\documentclass{aa}  
\usepackage{color}
\usepackage{natbib}
\usepackage{array}    % for better arrays (eg matrices) in maths
\usepackage{amsmath}  % Advanced maths commands
\usepackage{graphicx} % Including figure files
\usepackage{lscape} % landscape page
\usepackage{longtable} % longtables
\usepackage{placeins} % prevent floats placement away from their context
\usepackage{subfig}   % include more than one captioned figure/table in a single float
\usepackage[varg]{txfonts}
\usepackage[breaklinks]{hyperref}
\definecolor{cobalt}{rgb}{0.06, 0.2, 0.65}
\hypersetup{
  colorlinks,
  citecolor=cobalt,
  linkcolor=[rgb]{0.8, 0.2, 1.0},
  urlcolor=cobalt,
}
\begin{document} 

   \title{NSVS 14256825: Period variation and orbital stability analysis of two possible substellar companions}
   \titlerunning{NSVS 14256825: ETV and stability analysis}

   \author{K. Zervas
          \inst{1}
          \and
          P.-E. Christopoulou\inst{1}
          }

   \institute{Department of Physics, University of Patras, 26500 Patra, Greece\\
              \email{konst.zervas@upatras.gr}
             }

   % \date{Received ****; accepted ****}

% \abstract{}{}{}{}{} 
% 5 {} token are mandatory
 
  \abstract
   {Recent period investigations of the post-common envelop binary (PCEB) NSVS 14256825 suggest that two circumbinary companions are necessary to explain the observed eclipse timing variations (ETVs).}
   {Our objective in this work was to search for the best-fitting curve of two LTTE terms of the ETV diagram by implementing a grid search optimization scheme of Keplerian (kinematic) and Newtonian (N-body) fits alongside a dynamical stability analysis of N-body simulations.}
   {We compiled two datasets of archival photometric data covering different timelines and updated them with new observations and with three new times of minima calculated from the Transiting Exoplanet Survey Satellite (TESS). A grid search optimization process was implemented, and the resulting solutions that fell within the $90\%$ confidence interval of the best-fitting curve of the ETV diagram were tested for orbital stability using N-body simulations and the MEGNO chaos indicator.}
   {The Keplerian and Netwonian fits are in close agreement, and hundreds of stable configurations were identified for both datasets reaching a lifetime of 1 Myr. Our results suggest that the ETV data can be explained by the presence of a circumbinary planet with mass  $m_b=11 M_{Jup.}$ in a nearly circular inner orbit of period $P_b = 7$ yr. The outer orbit is unconstrained with a period range $P_c=20-50$ yr (from 3:1 to 7:1 MMR) for a circumbinary body of substellar mass ($m_c=11-70 M_{Jup.}$). The stable solutions of the minimum- and maximum-reduced chi-square value were integrated for 100 Myr and confirmed a non-chaotic behavior. Their residuals in the ETV data could be explained by a spin-orbit coupling model (Applegate-Lanza). However, continuous monitoring of the system is required in order to refine and constrain the proposed solutions.}
   {}


   \maketitle
%
%-------------------------------------------------------------------

\section{Introduction}
\label{sec:intro}
A well-known post-common envelop binary (PCEB), NSVS 14256825 (V1828 Aql, TIC 404635917) was discovered as a 13.2 mag variable star in the Northern Sky Variability Survey (NSVS) \citep{2004AJ....128.2965W}, and it has been recognized as an eclipsing binary by \cite{2007IBVS.5800....1W} ($\Delta V\sim$0.8 mag) with a very short period of P=0.110374230(2) d  and an sdOB+dM type light curve (LC; OB subdwarf and a M dwarf companion). From high precision $BVR_{c}I_{c}$ LCs, \cite{2011ASPC..451..155Z} derived a photometric mass ratio of 0.22(0.02) and a primary mass around the value for the He core flash at $M_{1} = 0.46 M_{\sun}$ and $M_{2} = 0.1 M_{\sun} = 105 M_{Jup}$ for the fully convective secondary star.  \cite{Almeida2012} classified NSVS 14256825 as sdOB+dM and obtained the first detailed physical and geometrical parameters from $UBVR_{c}I_{c}JH$ photometric and spectroscopic observations (inclination of the system $i= 82.5\degr \pm0.3\degr, M_{2}/M_{1}=0.260\pm 0.0012$) as $M_{1}=0.419\pm 0.07 M_{\sun}$,  $M_{2}=0.109\pm 0.023 M_{\sun}$, $\alpha=0.80\pm 0.04 R_{\sun}$ (separation between the components). \cite{NehirBulut2022}, using new $BVRI$ photometry (five nights, 2018-2019) and the old spectroscopy, presented the following updated parameters: i= 82.2$\degr \pm0.1\degr$, $M_{1}=0.351\pm 0.04 M_{\sun}$,  $M_{2}=0.095\pm 0.01 M_{\sun}$, and $\alpha=0.74\pm 0.03 R_{\sun}$. 

A large number of PCEBs show apparent period variations, which are frequently attributed to giant circumbinary planets \citep{Zorotovic2013, Heber2016} through eclipse timing variation (ETV) caused by the light-travel time effect (LTTE) \citep{Irwin1952, Borkovits2015}. \citet{Marsh2018} reports at least ten PCEBs with proposed circumbinary brown dwarfs or planets detected with ETV modeling.

\citet{Zorotovic2013} investigated the origin of these planets with binary population synthesis simulations and concluded that it is very unlikely that they formed in the primordial circumbinary disk (first-generation scenario) and survived the energetic common envelope (CE) evolution. The same authors proposed the formation of the planets from the CE material (second-generation scenario) as being the most likely for PCEBs, although their planetary formation is still an open question, with \citet{Bear2014} arguing in favor of the first-generation hypothesis.

\citet{Pulley2022} evaluated seven PCEBs with claimed circumbinary planets detected using the ETV method. Remarkably, they revealed that none of the more than 30 circumbinary models that have been proposed through the years have accurately predicted eclipse times within a year. Additionally, most of these models have failed the tests of dynamic stability, making the existence of these circumbinary planets questionable.
We give, as an example, HW Vir, which is the prototype of PCEBs, with a subdwarf (sdB) primary star and an M dwarf (dM) secondary. \citet{Beuermann2012B} presented a two-planetary companion model that proved to be stable for at least 10 Myr. However, recent studies \citep{Esmer2021, Brown-Sevilla2021, Mai2022} have found dynamically unstable configurations on short timescales.

Notably, NSVS 14256825 has a long and conflicting history of reports of possibly having circumbinary companions, with five circumbinary models proposed in the past. \citet{Beuermann2012A} proposed an eccentric ($e=0.5$) circumbinary planet of mass 12 $M_{J}$ and a period of 20 yr, whereas \citet{Almeida2013} suggested a two-planetary model with masses 2.9 $M_{J}$and 8.1 $M_{J}$; periods of 3.5 yr and 6.9 yr; and eccentricities of 0.0 and 0.52 for datasets covering the time span from 1999 to 2018 and 2007 to 2012, respectively.
\citet{Wittenmyer2013} ran N-body simulations for the two-planetary model and proved that it is unstable on a short timescale. Also, \citet{Hinse2014} could not confirm the two-planetary model of \citet{Almeida2013}, while their single LTTE solution was unconstrained due to a short time baseline. 
The ETV analysis of \citet{Nasiroglu2017} resulted in a single circumbinary brown dwarf with a of mass 15 $M_{J}$, a period of 9.9 yr, and an eccentricity of 0.175 for the time span 2007–2016. This solution is in close agreement with the results of \citet{Zhu2019} ($m_3$ = 14.2 $M_{J}$, $P_3= 8.83$ yr, $e_3 = 0.12$) and \citet{NehirBulut2022} ($m_3$ = 13.2 $M_{J}$, $P_3= 8.83$ yr, $e_3 = 0.13$) for datasets covering the time span 2008-2018 and 2007-2019, respectively. New data for the following years have indicated a divergence from these single LTTE term solutions. \citet{Wolf2021} concluded that at least two circumbinary bodies were necessary to explain the ETV data, as a single LTTE term with a period of 14 years was insufficient to fit the data. Similarly, \citet{Pulley2022} reached the same conclusion, finding that an updated ETV diagram (2007-2021) could not be adequately explained by a quadratic ephemeris and a circumbinary companion in a near circular orbit ($e=0.02$) with a period of 7.65 years.

In this study, we collect all available ETV data updated with new times of minima (TOMs) calculated from the Transiting Exoplanet Survey Satellite (TESS; Ricker et al. \citeyear{TESS2015}) LC data and new observations. Our objective was to search for the best-fitting curve of two LTTE terms in the ETV diagram by implementing a grid search optimization scheme alongside a dynamical stability analysis of N-body simulations, which we believe has not been attempted yet for this PCEB in this extent.

In the following section, we review alternative explanations for the ETVs in PCEBs. Then, we describe the new TOMs and the compilation of all available ETV data in Section~\ref{sec:data}. In Section~\ref{sec:analysis_results}, we analyze our search for the best-fitting curve of two LTTE terms in the ETV diagram and provide the results for different models, alongside a dynamical stability analysis and alternative mechanisms. Our discussion of the results can be found in Section~\ref{sec:conclusions}.

\section{Possible explanations for the ETVs in PCEBs}

The LTTE concept faces challenges not only in terms of dynamical models for hypothetical planetary systems but also with respect to optimization techniques and phenomena inherent in the binary. Specifically, the main datasets used for ETV analyses (i.e., eclipse timings) originate from diverse sources and methods, involve various software packages and algorithms, and are often accompanied by underestimated errors \citep{2012AN....333..754P}.
 
Alternative mechanisms have been proposed as explanations for the apparent ETV variations, such as mass transfer, apsidal motion \citep{Lacy1992}, magnetic effects \citep{Applegate1992, Volschow2016, Lanza2020}, and angular momentum loss through gravitational radiation \citep{Paczynski1967} or magnetic breaking \citep{Rappaport1983} in one or both stars. Since sdB binaries are detached systems, mass transfer cannot occur. Apsidal motion would not be expected in PCEBs with $\alpha<1 R_{\sun}$; thus the orbit can be considered circular (e = 0), and the rotation of the components synchronized with the orbit. \cite{2014MNRAS.438L..91P} suggest that orbital eccentricities much less than 0.001 can introduce a maximum  ETV of 5.5 s (for $P\sim0.1$ d), and thus it is considered a minor contributing factor to the observed ETVs. However, apsidal motion is the combined result of perturbations of a Keplerian orbit that include 
tidal and rotational contributions, a general relativity contribution, and possibly a third-body contribution. Analytical formulae can be found in  \cite{ 2014MNRAS.438L..91P, 2017MNRAS.468.3342B, 2020MNRAS.497.4022A, 2023AJ....166..114D}. 
Due to their proximity, PCEBs frequently undergo complex gravitational effects that depart from idealized spherical models. Due to dynamic interactions between the stars as they evolve and lose their spherical symmetry, caused by factors such as rotation, magnetic fields, and tidal forces, precession of the orbits may result. In situations, where the gravitational term is not taken into account, the rate of apsidal motion can be used as a probe of the internal structure of the stars. At the same time, although it was not ruled out, \cite{2020MNRAS.497.4022A} discovered that apsidal motion due to tidal and rotation effects was less plausible than substellar component models in the case of the PCEB GK Vir. Therefore, we do not take this effect into further consideration in the current investigation.
  
Because of the small masses, barely any recorded ETV signals may be attributed to gravitational radiation despite the short orbital periods. Considering the high surface gravities of the components, star winds should also be extremely weak and make minimal contributions. Given that the secular variations can occur at a timescale significantly longer than the baseline of data, it is very easy to confuse periodic changes with the former. The overall cyclical behavior of ETV cannot be explained by angular momentum loss (AML) from magnetic braking or gravitational radiation, which drives the binary to a constant reduction in the period without a quasi-cyclical component.

Magnetic activity cycles of the secondary components can imitate ETV behavior provided that the star has a sufficient energy budget (Applegate mechanism). According to the Applegate model \citep{Applegate1992} a magnetic active star in a binary can transfer angular momentum toward the surface from its interior and away from it during a complete magnetic cycle, 
leading to oscillations of the orbital period. 
The magnetic star's energy budget must be sufficient to adjust its quadrupole moment for this mechanism to work, resulting in a variable luminosity. \cite{Volschow2016} proposed a more accurate and realistic two-zone model where they treated a stellar density profile of the inner and outer regions for two different densities, including the change of the quadrupole moment both in the shell and the core, adding to previous improvements of the mechanism \citep{LanzaRodono2004,Brinkworth2006,Lanza2006,Tian2009}. A novel model has been developed recently by \citet{ Lanza2020}. It is based on the exchange of angular momentum between the binary orbital motion and the spin of the magnetically active secondary. This mechanism could help explain the observed ETVs and potentially get over the original Applegate mechanism's energy constraints. 
Although these alternatives often fail to explain the magnitude of the observed ETVs or to satisfy the required observational evidence of each mechanism \citep{Pulley2022} in most cases, their contribution can become significant \citep[RR Cae, DE CVn;][]{Pulley2022}.

\section{Data acquisition}
\label{sec:data}

We compiled 305 eclipse minimum times from the literature as follows: 152 from Table 4 of \cite{Nasiroglu2017} (these include 18 by \cite{2007IBVS.5800....1W}, 32 by \citet{Beuermann2012A}, nine by \cite{2012MNRAS.421.3238K}, and ten by\cite{Almeida2012}); 84 by \citet{Zhu2019},  41 by \citet{Wolf2021}, 20 by \cite{Pulley2022}, and eight by \cite{NehirBulut2022} (only the primaries). All TOMs are in BJD-TDB. Details of the observations (error, eclipse type, observatory) can be found in the papers mentioned. In addition, we updated the TOMs with our own observations carried out in 2017 using the 14-inch Schmidt-Cassegrain telescope at the Patras University “Mythodea” Observatory \citep{PapagChris2015}. Three new TOMs were calculated from TESS sector 51 with a 2-min cadence for the year 2022, available from the Mikulski Archive for Space Telescopes (MAST),\footnote{\url{https://archive.stsci.edu/}} and two new TOMs were collected from the online B.R.N.O.\footnote{\url{http://var2.astro.cz/brno/protokoly.php?lang=en}} database during the years 2023 and 2024 (Table~\ref{tab:TOMs}).

Our own observations were taken with an $R_{C}$ filter. The standard processes of bias subtraction, flat fielding, and dark subtraction were used to reduce the CCD frames. Differential magnitudes were obtained by choosing a non-variable star in the field of view comparable to the target in brightness (GSC 00504 00639) and a comparison star (2MASS 1084555518). Image reduction and differential photometry were performed using a fully automated pipeline \citep{2015ASPC..496..181P} that incorporates Pyraf \citep{2012ascl.soft07011S} and the Astrometry.net packages \citep{2010AJ....139.1782L}. 

\begin{figure*}
    % \centering
    \subfloat[]{\includegraphics[width=8cm, height=5cm, keepaspectratio]{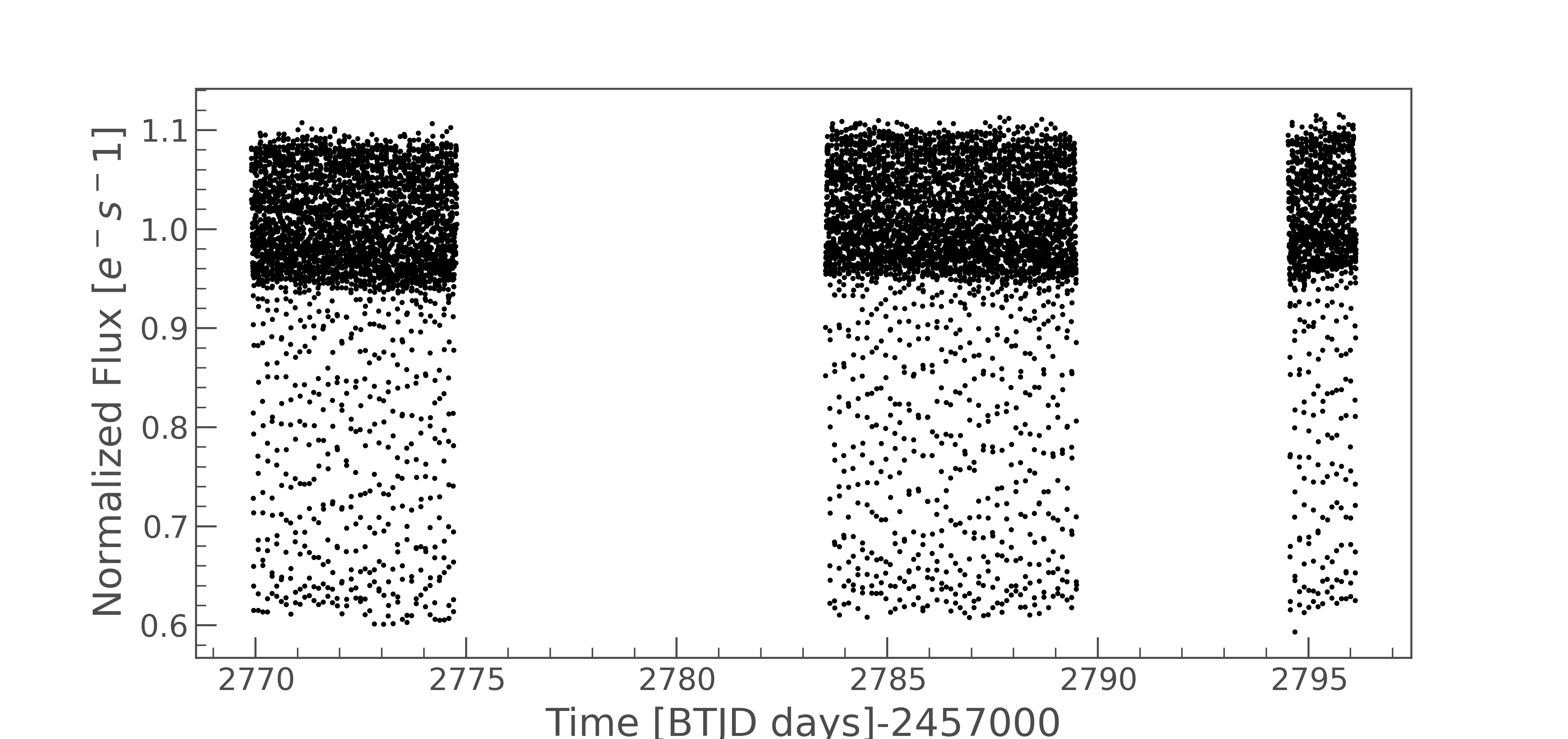}\label{fig:sbf_LC}}
    \hspace{-0.2cm}\subfloat[]{\includegraphics[width=10cm, height=7cm, keepaspectratio]{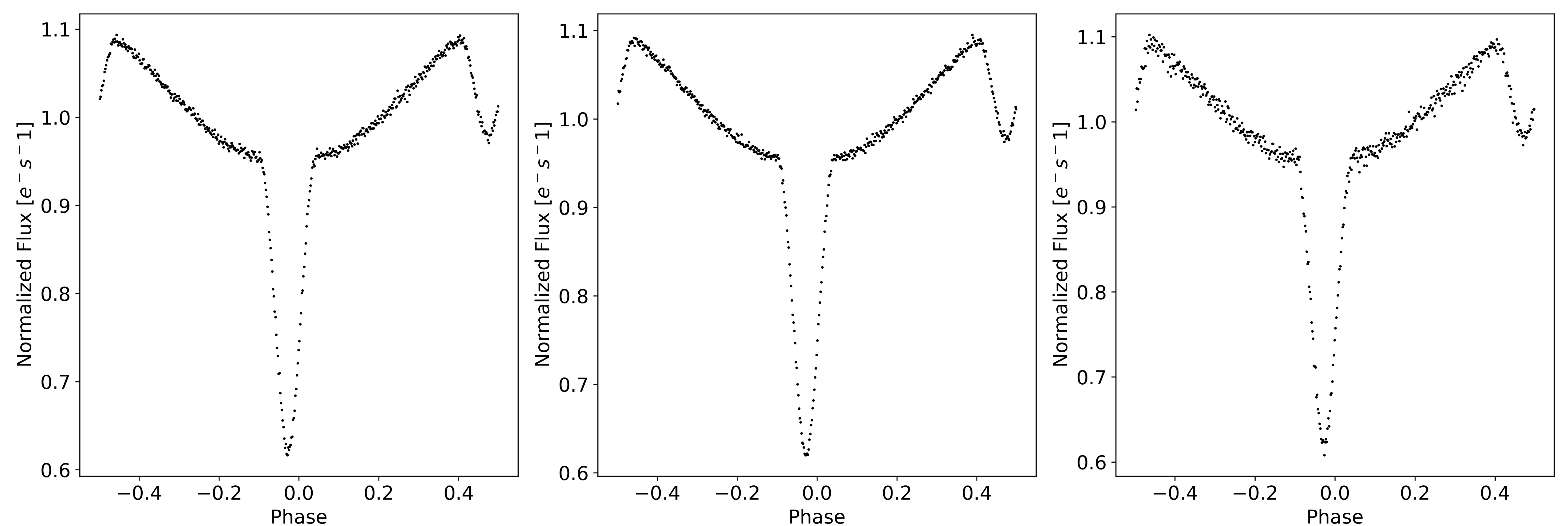}\label{fig:sbf_FBLCs}}
\caption{(a) TESS LC time series composed of three segments (sector 51) (b) phase-folded binned LCs for each segment.}
\label{fig:TESS}
\end{figure*}

The TESS LC in Fig.~\ref{fig:sbf_LC} was plotted with normalized SAP fluxes of the best quality (quality flag zero) data points using the Lightkurve Python package
for Kepler and TESS data analysis \citet{Lightkurve}, from which we generated a phased-folded and binned LC (FBLC) for each time series segment (Fig.~\ref{fig:sbf_FBLCs}). Employing a modified Kwee–Van Woerden \citep{Deeg2020} method for eclipse minimum timing with reliable error estimates, one primary TOM was calculated from each FBLC following the process of TOM determination described in \citet{Zervas2024ApJ}. All the HJD times were transformed
into TDB-based BJD ones by using the publicly available code of \cite{2010PASP..122..935E}.

\begin{figure*}
    \centering
    \subfloat[]{\includegraphics[width=9cm, height=6cm, keepaspectratio]{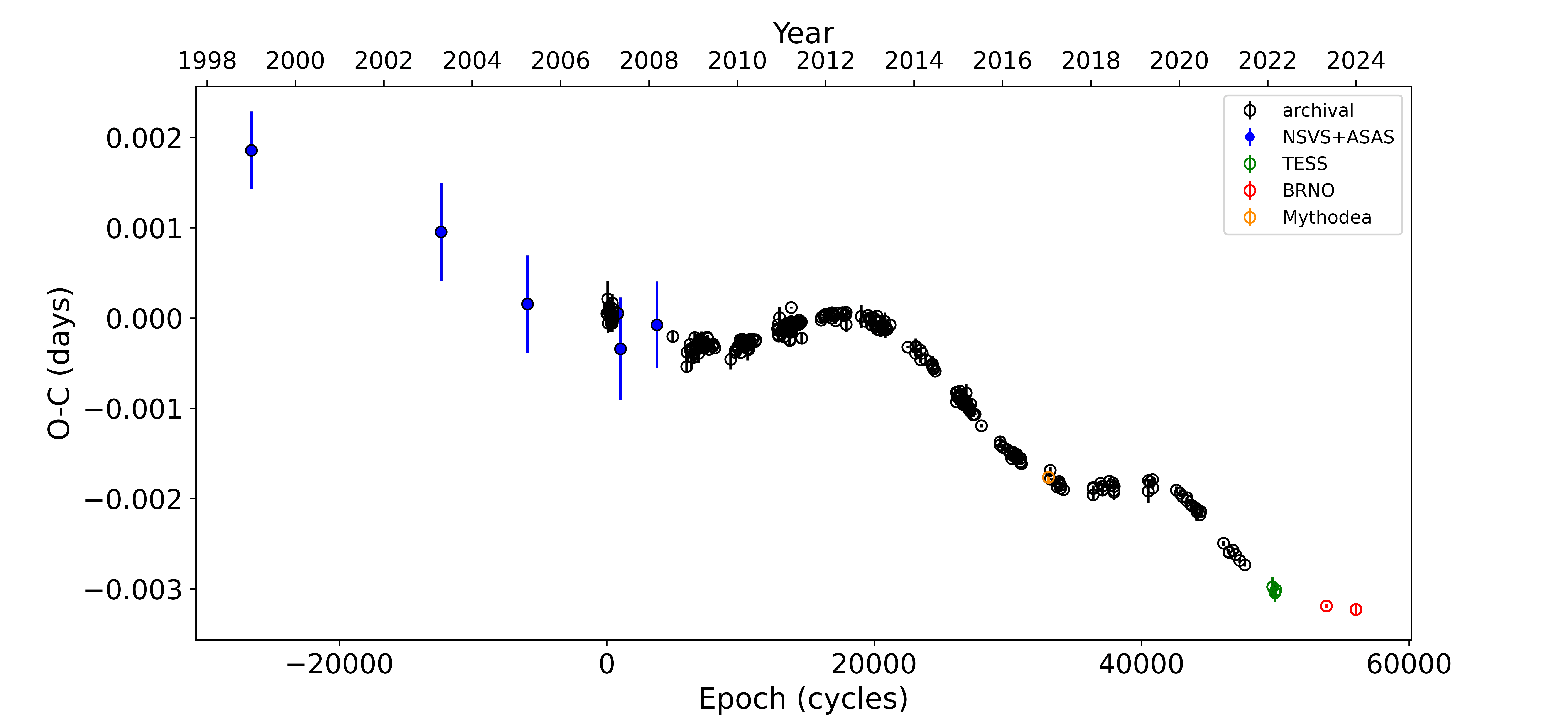}\label{fig:sbf_OC_A}}
    \subfloat[]{\includegraphics[width=9cm, height=6cm, keepaspectratio]{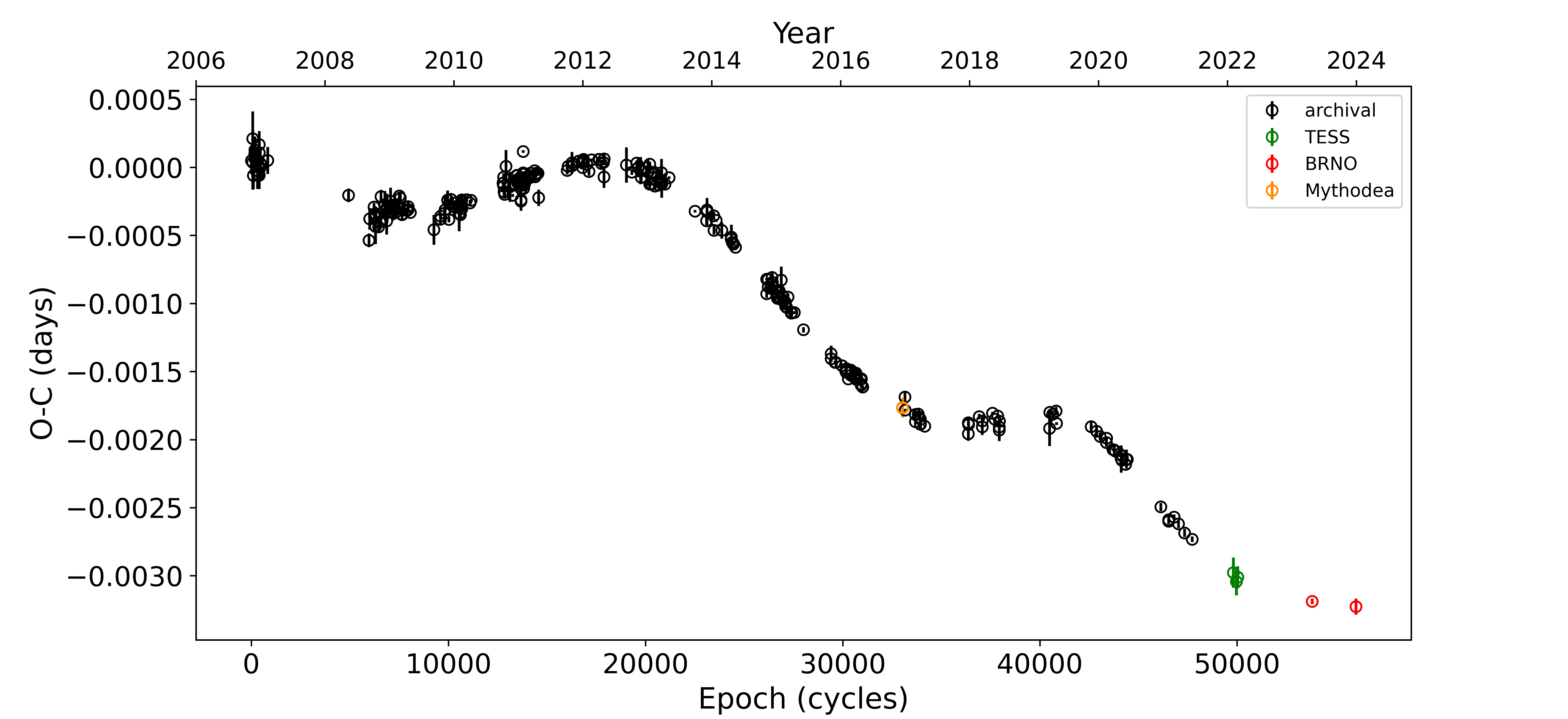}\label{fig:sbf_OC_B}}
\caption{ETV diagrams of NSVS 14256825. (a) ETV diagram of dataset A spanning 26 years (1999-2024). Archival data $\sim$ black open circles; NSVS+ASAS data $\sim$ blue open circles; Mythodea data $\sim$ orange open circles; TESS data $\sim$ green open circles; B.R.N.O. data $\sim$ red open circles. (b) ETV diagram of dataset B spanning 18 years (2007-2024). Color indexing is the same as in (a).}
\label{fig:ETV}
\end{figure*}

A total of 312 TOMs were used in the ETV analysis (Table~\ref{tab:long}), excluding low-precision timings and using only the timings with errors smaller than $\sim$17sec (0.0002 d). Consistent with previous data categorizations \citep{Beuermann2012A, Nasiroglu2017, Pulley2022}, we established two datasets. The first one includes the less precise NSVS and All Sky Automated Survey (ASAS) data, offering a broader time frame from 1999 to 2024 (dataset A), while the second one with 307 TOMs excludes these data, narrowing down the analysis to the years 2007 to 2024 (dataset B). Using the linear part of ephemeris from \citet{Pulley2022},
\begin{equation}
BJD = 2454274.20875(5) + 0.110374157(4)E
\label{eq:ephemeris},
\end{equation}
we calculated and plotted the ETV diagram (Fig.~\ref{fig:ETV}) of each dataset.

\begin{table}
    \centering
    \caption{New times of minimum light for NSVS 14256825.}
    \label{tab:TOMs}
    \resizebox{\hsize}{!}{%
    \begin{tabular}{ccccc}
        \hline\hline
        BJD-240000 (d) & Epoch (cycles) & O-C (d) & Error (d) & Source \\
        \hline
        57919.47908 & 33026.5 & -0.00177 & 0.00007 & Mythodea \\
        57919.53427 & 33027.0 & -0.00176 & 0.00007 & Mythodea \\
        59771.05954 & 49802.0 & -0.00298 & 0.00011  & TESS \\
        59789.27121 & 49967.0 & -0.00304 & 0.00010  & TESS \\
        59796.11444 & 50029.0 & -0.00301 & 0.00008  & TESS \\
        60213.43895 & 53810.0  & -0.00319 & 0.00002  & B.R.N.O. \\
        60458.46954 & 56030.0  & -0.00323 & 0.00006  & B.R.N.O. \\
        \hline
    \end{tabular}%
    }
\end{table}

\section{Analysis and results}
\label{sec:analysis_results}

\subsection{Analysis outline}
\label{sec:analysis_outline}

We adopted the formulation of \citet{Gozdziewski2012} for the LTTE term:

\begin{equation}
    LTTE_{p}= K_p \left[ \sin\omega_{p} \cos(E_p(t)-e_p)+cos\omega_{p} \sqrt{1-e_{p}^2} \sin E_p(t) \right]
\label{eq:lite},
\end{equation}

where $K_p$, $e_p$, $\omega_{p}$ are the semi-amplitude of the LTTE signal, eccentricity, and argument of the pericenter, respectively, for body $p$ ($p=b, c$ for inner and outer orbit, in accordance with the common convention). The orbital period $P_p$ and the pericenter passage $T_p$ depend on the eccentric anomaly $E_p(t)$, and hence they do not explicitly appear in equation \ref{eq:lite}. This is a revised kinematic (Keplerian) formulation of the LTTE effect \citep{Irwin1952} for multiple circumbinary  companions, where the eclipse ephemerides is expressed with regard to Jacobi coordinates, with the origin at the center of mass (CM) of the binary.

According to \citet{Gozdziewski2012}, the osculating orbital elements and masses derived with this revised formulation best match the true N-body initial condition of the system with mutually interacting planets. However, these Keplerian orbits are only an approximation of the true (Newtonian) orbits in the case of more than two bodies, and as a result, one needs to determine whether the degree of approximation is significant. This discrepancy has been highlighted first in radial velocity studies of extrasolar planets \citep{LaughlinChambers2001,Gozdziewski2003} and becomes substantial in cases of massive, strongly interacting planets close to low-order mean motion resonances (MMRs).

In total, our least-squares fit involves twelve parameters: two for the linear ephemeris of the binary (the epoch $T_0$ and period $P_{bin}$) and five orbital elements for each body ($e_p$, $P_p$, $\omega_p$, $T_p$, $\alpha_{p}$), where $\alpha_{p}$ is the semi-major axis and $p=b, c$ for each LTTE orbit. We followed a grid search approach similar to that of \citet{Beurmann2013}. Our optimization process consisted of two sets of runs in the $e_b$, $e_c$ plane utilizing a combination of the Nelder-Mead Simplex and Levenberg-Marquardt algorithms using the \textsc{lmfit} Python package \citep{LMFIT2015}. The goal was to identify the best-fitting curve in the ETV diagram of models featuring two LTTE terms.

Additionally, each solution that achieved a reduced chi-square ($\chi_{\nu}^2$) value below a preset limit (listed below) was integrated with the \textsc{rebound} N-body code \citep{Rein2012} using the \textsc{ias15} integrator \citep{Rein2015}, requiring a lifetime $\tau=1$Myr and a MEGNO chaos indicator \citep{MEGNO2003} value $\langle Y \rangle\sim2$ as an additional acceptance criterion. Therefore, an acceptable model can be considered one that provides a good fit to the data and is secularly stable with non-chaotic orbital behavior.

This optimization process was independently applied to two distinct ETV datasets (dataset A, dataset B), each covering a different time span. The initialization of each integration was set up with the transformation of the initial condition from the Jacobian kinematic frame to the N-body Cartesian osculating frame, with the origin at the CM, at the osculating epoch $T_0$. Furthermore, we used these initial conditions and refined them in terms of the exact N-body model. An iterative Levenberg-Marquardt process was implemented, and we examined how the $\chi_{\nu}^2$ value of the fit depends on variations of all orbital elements \citep{LaughlinChambers2001}. In that way, a fully self-consistent Newtonian fit to the ETV data could be made.

\begin{figure}
	\resizebox{\hsize}{!}{\includegraphics{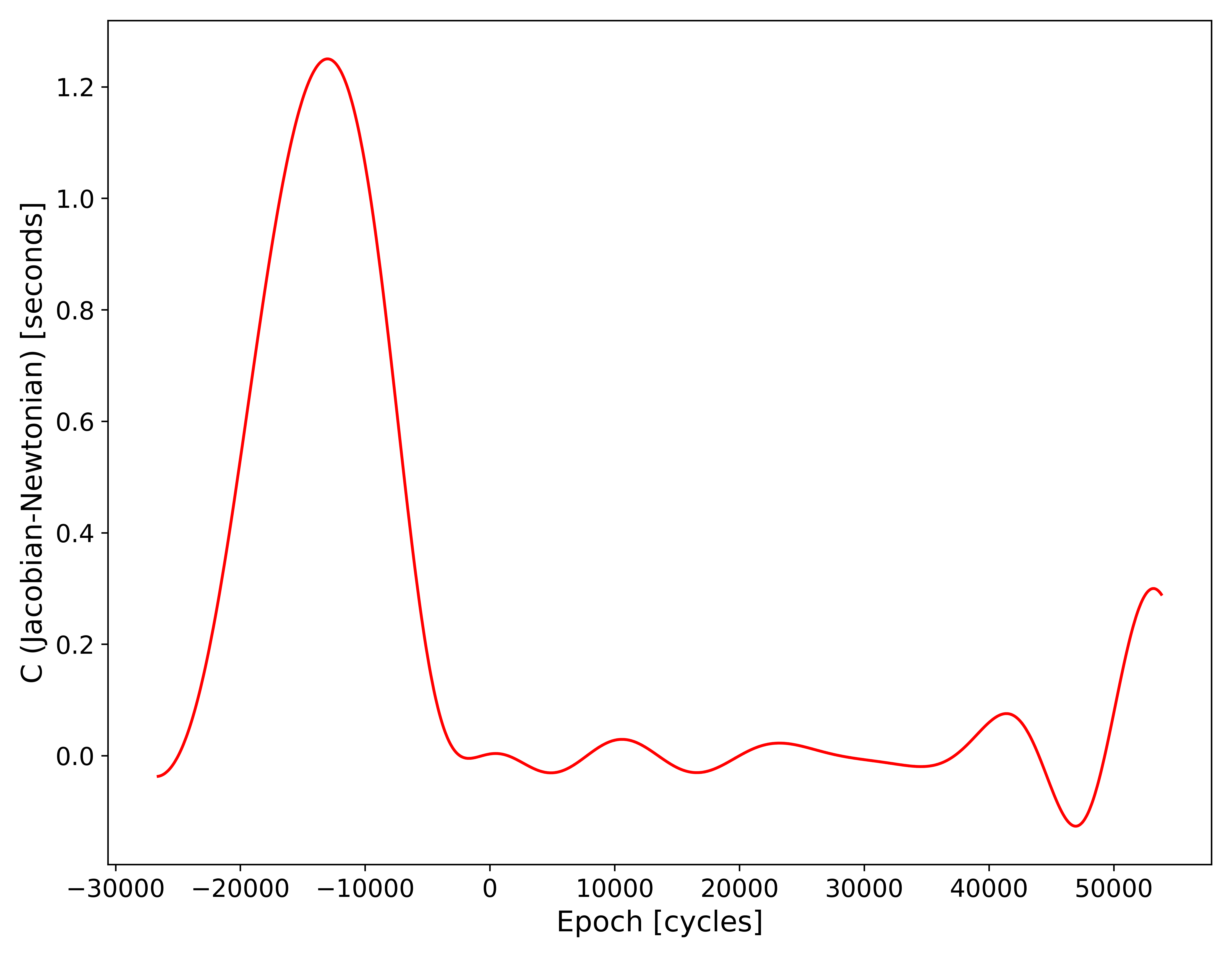}}
	\caption{Differences of the LTTE signals derived from a typical Jacobian fit and from respective osculating N-body models integrated numerically with the inferred initial condition at the osculating epoch $T_0 = 54274.20875$ BJD of the zeroth cycle, where the two models seem to agree.}
	\label{fig:Jacobian-Newtonian}
\end{figure}

Figure \ref{fig:Jacobian-Newtonian} illustrates the difference between the synthetic LTTE signals of a Jacobian (Keplerian) and an N-body (Newtonian) fit for dataset A. The maximum time difference of 1.2 sec is comparable to the discrepancy of Keplerian and Newtonian two-planet models for the ETV diagram of the post-common envelope binary NN Ser, as has been shown in \citet{Marsh2014}, while these differences are about ten times larger for the cataclysmic binary HU Aqr \citep{Gozdziewski2015}.

The steps of the fitting process are outlined as follows:
\begin{enumerate}
	\item Run 1: Grid search optimization  in the $e_b$, $e_c$ plane with fixed eccentricity values in the range [0, 0.98] and a step size of 0.01 (9801 models). The rest of the parameters were subsequently optimized, initiating from values chosen randomly within physically accepted limits.
	\item Run 2: The resulting solutions of Run 1 served as the initial points for a new optimization run. However, this time all the parameters, including eccentricities, were adjusted during the fitting process.
	\item N-body simulations for 1 Myr of the obtained Keplerian solutions of both runs while selectively excluding models that deviate beyond the $90\%$ confidence level of the reduced chi-square ($\chi_{\nu}^2$) for two degrees of freedom in order to acquire well-fitted curves. Specifically, we considered solutions with $\chi_{\nu}^2 \leq \chi_{\nu,best}^2 + \Delta\chi$, where $\chi_{\nu,best}^2$ corresponds to the chi-square value of the best-fitting curve and $\Delta\chi$ = 4.6.
	\item Self-consistent N-body (Newtonian) orbital fits initialized from each Keplerian fit of step 3 and subsequent integrations for 1Myr.
\end{enumerate}

In the first optimization run (Run 1), the eccentricities are fixed, and the rest of the parameters are free to adjust, while in the second run (Run 2), all parameters are allowed to vary. Throughout all orbital simulations, we treated the central binary as a single object and considered only co-planar, edge-on cases. This is the simplest scenario, and it is commonly assumed in stability analyses of exoplanets, as it minimizes the masses of the planets relative to the binary, thereby promoting stability \citep{Beuermann2012B, Horner2012, Marsh2014, Mai2022, Esmer2023}. However, there are cases, such as the putative HU Aqr planetary system, that appear to be more complex (highly non-coplanar), resulting in stable orbits in contrast to the co-planar scenario \citep{Gozdziewski2015}. Fundamental parameters of the binary are from \cite{NehirBulut2022} and were used for the rest of the analysis when needed.

\subsection{Period variation and orbital stability}
\label{sec:results}

Optimization results of Keplerian models are shown in Table~\ref{tab:Keplerian_run1} and Table~\ref{tab:Keplerian_run2} for each dataset and Runs 1 and 2, respectively. Each optimization run is characterized by three critical values: the best reduced chi-square value ($\chi_{\nu,best}^2$) of the best-fitting curve to ETV data and the minimum- ($\chi_{\nu,min}^2$) and the maximum-reduced chi-square ($\chi_{\nu,max}^2$) values among all stable solutions. The same description is followed in Tables \ref{tab:Newtonian_run1}, \ref{tab:Newtonian_run2} for the Newtonian models. The reported parameter uncertainties correspond to the standard 1-$\sigma$ error, and their calculation resulted from the covariance matrix.

\begin{table*}
         \renewcommand{\arraystretch}{1.1}
	\caption{Grid search optimization results of Keplerian two-planet LTTE fit after Run 1 of the fitting process for both datasets (A and B).}
	 % \hspace{-0.39cm}
          \begin{tabular*}{0.95\textwidth}{l>{\centering}m{2.1cm}>{\centering}m{2cm}>{\centering}m{2.1cm}>{\centering}m{2.1cm}>{\centering}m{2.1cm}>{\centering\arraybackslash}m{2cm}}
		\hline
                \multicolumn{1}{c}{Parameter} & \multicolumn{3}{c}{Dataset A (Run 1)} & \multicolumn{3}{c}{Dataset B (Run 1)} \\
		\multicolumn{1}{c}{} & \multicolumn{1}{c}{$\chi_{\nu,best}^2$} & \multicolumn{1}{c}{$\chi_{\nu,min}^2$} & \multicolumn{1}{c}{$\chi_{\nu,max}^2$} & \multicolumn{1}{c}{$\chi_{\nu,best}^2$} & \multicolumn{1}{c}{$\chi_{\nu,min}^2$} & \multicolumn{1}{c}{$\chi_{\nu,max}^2$} \\
		\hline
		e$_{b}$ & 0.11 & 0.09 & 0.25 & 0.09 & 0.06 & 0.29 \\
                K$_b$ (d) & 0.00043(2) & 0.00034(2) & 0.00031(3) & 0.00047(5) & 0.00035(2) & 0.00035(3) \\
                $\omega_b$ (rad) & 0.00(5) & 5.52(3) & 3.42(11) & 5.64(4) & 5.45(5) & 0.00(7) \\
                P$_b$ (yr) & 7.67(3) & 7.02(4) & 6.68(5) & 8.11(4) & 7.05(3) & 7.02(4) \\
                T$_b$ (BJD-2400000) & 52963.22(5.27) & 58083.25(11.23) & 54801.26(5.15) & 58344.17(8.32) & 58071.62(10.53) & 50697.31(30.14) \\
                e$_{c}$ & 0.84 & 0.07 & 0.15 & 0.90 & 0.25 & 0.10 \\
                K$_{c}$ (d) & 0.00043(1) & 0.00072(3) & 0.00069(3) & 0.00055(2) & 0.00523(4) & 0.00094(42) \\
                $\omega_c$ (rad) & 4.78(3) & 0.23(5) & 1.01(5) & 4.45(7) & 5.41(5) & 6.27(15) \\
		P$_c$ (yr) & 14.45(2) & 21.70(8) & 20.42(7) & 16.09(5) & 38.04(8) & 23.01(4.32) \\
                T$_c$ (BJD-2400000) & 54316.21(10.35) & 54318.59(12.22) & 55444.47(23.65) & 53745.16(22.24) & 48325.17(20.44) & 53710.02(51.53) \\
                P$_{bin}$ (d) & 0.11037409(2) & 0.11037412(2) & 0.11037412(3) & 0.11037409(2) & 0.11037428(4) & 0.11037413(4) \\
                T$_0$ (BJD-2400000) & 54274.20881(4) & 54274.20846(5) & 54274.20864(6) & 54274.20847(5) & 54274.20261(7) & 54274.20811(5) \\
		$m_{b}$ ($M_{Jup}$) & 12.22 & 9.94 & 10.04 & 12.57  & 10.39 & 10.88  \\
		$m_{c}$ ($M_{Jup}$) & 7.91 & 10.15 & 9.86  & 9.5 & 53.82  & 12.73  \\
	 	$\chi_{\nu}^2$ & 5.26 & 7.06 & 9.90 & 5.28 & 6.89 & 9.73  \\
		\hline
	\end{tabular*}
    \tablefoot{Columns $\chi_{\nu,best}^2$ display the best-fitting curve parameters, whereas columns $\chi_{\nu,min}^2$ and $\chi_{\nu,max}^2$ contain parameters corresponding to the minimum- and maximum-reduced chi-square values of stable configurations, respectively. Digits in parentheses are for the uncertainty at the last significant place.}
\label{tab:Keplerian_run1}
\end{table*}

\begin{table*}
    \renewcommand{\arraystretch}{1.1}
	\caption{Grid search optimization results of Keplerian two-planet LTTE fit after Run 2 of the fitting process for both datasets (A and B).}
	% \hspace{-0.45cm}
    \begin{tabular*}{0.95\textwidth}{l>{\centering}m{2.1cm}>{\centering}m{2.1cm}>{\centering}m{2.2cm}>{\centering}m{2.1cm}>{\centering}m{2.2cm}>{\centering\arraybackslash}m{2cm}}
	\hline
    \multicolumn{1}{c}{Parameter} & \multicolumn{3}{c}{Dataset A (Run 2)} & \multicolumn{3}{c}{Dataset B (Run 2)} \\
	\multicolumn{1}{c}{} & \multicolumn{1}{c}{$\chi_{\nu,best}^2$} & \multicolumn{1}{c}{$\chi_{\nu,min}^2$} & \multicolumn{1}{c}{$\chi_{\nu,max}^2$} & \multicolumn{1}{c}{$\chi_{\nu,best}^2$} & \multicolumn{1}{c}{$\chi_{\nu,min}^2$} & \multicolumn{1}{c}{$\chi_{\nu,max}^2$} \\
	\hline
	e$_{b}$ & 0.09(2) & 0.09(3) & 0.06(4) & 0.09(2) & 0.09(2) & 0.00(1) \\
    K$_b$ (d) & 0.00046(5) & 0.00034(2) & 0.00034(2) & 0.00047(7) & 0.00035(4) & 0.00034(4) \\
    $\omega_b$ (rad) & 5.58(3) & 5.34(4) & 0.00(6) & 5.64(5) & 5.53(5) & 3.33(6) \\
    P$_b$ (yr) & 8.09(4) & 6.97(5) & 7.02(6) & 8.10(6) & 7.02(5) & 6.99(04) \\
    T$_b$ (BJD-2400000) & 55366.53(12.34) & 58096.21(45.32) & 55841.14(22.37) & 55386.28(33.22) & 58092.21(40.72) & 57189.16(32.47) \\
    e$_{c}$ & 0.89(3) & 0.15(3) & 0.0(5) & 0.90(2) & 0.14(3) & 0.00(3) \\
    K$_{c}$ (d) & 0.00056(4) & 0.00093(5) & 0.00507(2) & 0.00054(4) & 0.00679(9) & 0.00096(6) \\
    $\omega_c$ (rad) & 4.61(8) & 0.00(58) & 6.04(9) & 4.45(8) & 5.06(22) & 2.78(9) \\
	P$_c$ (yr) & 16.25(5) & 22.64(5) & 47.40(8) & 16.07(7) & 47.91(28) & 25.00(35) \\
    T$_c$ (BJD-2400000) & 59660.53(10.42) & 45591.75(83.24) & 48715.53(112.27) & 53749.47(22.41) & 45174.62(115.43) & 48319.52(108.45) \\
    P$_{bin}$ (d) & 0.11037410(2) & 0.11037413(3) & 0.11037426(3) & 0.11037410(2) & 0.11037432(6) & 0.11037413(3) \\
    T$_0$ (BJD-2400000) & 54274.20840(6) & 54274.20814(9) & 54274.20353(8) & 54274.20847(5) & 54274.20155(6) & 54274.20807(4) \\
	$m_{b}$ ($M_{Jup}$) & 12.36 & 10.05 & 10.11 & 9.53  & 10.23 & 12.19  \\
	$m_{c}$ ($M_{Jup}$) & 9.41 & 12.72 & 43.90  & 12.57 & 59.64  & 9.95  \\
	$\chi_{\nu}^2$ & 5.20 & 7.02 & 7.97 & 5.28 & 6.97 & 7.47  \\
	\hline
	\end{tabular*}
    \tablefoot{Columns $\chi_{\nu,best}^2$ display the best-fitting curve parameters, whereas columns $\chi_{\nu,min}^2$ and $\chi_{\nu,max}^2$ contain parameters corresponding to the minimum- and maximum-reduced chi-square values of stable configurations, respectively. Digits in parentheses are for the uncertainty at the last significant place.}
\label{tab:Keplerian_run2}
\end{table*}

\begin{table*}
	\renewcommand{\arraystretch}{1.1}
	\caption{Self-consistent Newtonian two-planet orbital fits initialized from the results of optimization Run 1 for both datasets (A and B).}
	% \hspace{-0.39cm}
	\begin{tabular*}{0.95\textwidth}{l>{\centering}m{2.1cm}>{\centering}m{2cm}>{\centering}m{2.1cm}>{\centering}m{2.1cm}>{\centering}m{2.1cm}>{\centering\arraybackslash}m{2cm}}
		\hline
		\multicolumn{1}{c}{Parameter} & \multicolumn{3}{c}{Dataset A (Run 1)} & \multicolumn{3}{c}{Dataset B (Run 1)} \\
		\multicolumn{1}{c}{} & \multicolumn{1}{c}{$\chi_{\nu,best}^2$} & \multicolumn{1}{c}{$\chi_{\nu,min}^2$} & \multicolumn{1}{c}{$\chi_{\nu,max}^2$} & \multicolumn{1}{c}{$\chi_{\nu,best}^2$} & \multicolumn{1}{c}{$\chi_{\nu,min}^2$} & \multicolumn{1}{c}{$\chi_{\nu,max}^2$} \\
		\hline
		e$_{b}$ & 0.18(3) & 0.10(4) & 0.29(5) & 0.08(3) & 0.03(5) & 0.30(5) \\
		K$_b$ (d) & 0.00041(3) & 0.00034(3) & 0.00034(3) & 0.00047(4) & 0.00035(3) & 0.00035(4) \\
		$\omega_b$ (rad) & 0.13(6) & 5.85(4) & 0.08(8) & 5.65(3) & 4.54(5) & 5.74(6) \\
		P$_b$ (yr) & 6.96(6) & 7.03(4) & 6.91(5) & 8.14(3) & 6.90(3) & 7.06(6) \\
		M$_b$ (rad) & 4.53(5) & 0.10(33) & 5.92(3) & 1.39(8) & 2.32(7) & 4.56(8) \\
		e$_{c}$ & 0.85(5) & 0.15(3) & 0.12(4) & 0.88(5) & 0.26(4) & 0.08(6) \\
		K$_{c}$ (d) & 0.00046(4) & 0.00071(5) & 0.00078(4) & 0.00058(5) & 0.00343(6) & 0.00146(9) \\
		$\omega_c$ (rad) & 4.79(8) & 6.19(4) & 0.52(9) & 4.44(7) & 5.38(5) & 3.98(7) \\
		P$_c$ (yr) & 15.67(8) & 20.51(6) & 22.26(7) & 17.57(6) & 33.92(7) & 27.19(6) \\
		M$_c$ (rad) & 0.75(7) & 3.43(6) & 2.32(8) & 5.33(5) & 6.20(4) & 1.82(5) \\
		P$_{bin}$ (d) & 0.11037410(2) & 0.11037411(3) & 0.11037412(2) & 0.11037409(2) & 0.11037428(3) & 0.11037413(3) \\
		T$_0$ (BJD-2400000) & 54274.20851(5) & 54274.20851(4) & 54274.20834(5) & 54274.20819(4) & 54274.20500(9) & 54274.20712(6) \\
		$m_{b}$ ($M_{Jup}$) & 12.32 & 10.09 & 10.60 & 12.68  & 10.33 & 10.69  \\
		$m_{c}$ ($M_{Jup}$) & 7.97 & 10.68 & 10.56  & 8.98 & 38.12  & 17.66  \\
		$\chi_{\nu}^2$ & 5.76 & 7.07 & 7.38 & 5.53 & 6.97 & 7.31  \\
		\hline
	\end{tabular*}
	\tablefoot{Columns $\chi_{\nu,best}^2$ display the best-fitting curve parameters, whereas columns $\chi_{\nu,min}^2$ and $\chi_{\nu,max}^2$ contain parameters corresponding to the minimum- and maximum-reduced chi-square values of stable configurations, respectively. The term $M_b$, $M_c$ is the resulting mean anomaly of the inner and outer orbits, with each configuration being set up in the N-body osculating frame at the osculating epoch $T_0$. Digits in parentheses are for the uncertainty at the last significant place.}
	\label{tab:Newtonian_run1}
\end{table*}

\begin{table*}
	\renewcommand{\arraystretch}{1.1}
	\caption{Self-consistent Newtonian two-planet orbital fits initialized from the results of optimization Run 2 for both datasets (A and B).}
	% \hspace{-0.45cm}
	\begin{tabular*}{0.95\textwidth}{l>{\centering}m{2.1cm}>{\centering}m{2.1cm}>{\centering}m{2.2cm}>{\centering}m{2.1cm}>{\centering}m{2.2cm}>{\centering\arraybackslash}m{2cm}}
		\hline
		\multicolumn{1}{c}{Parameter} & \multicolumn{3}{c}{Dataset A (Run 2)} & \multicolumn{3}{c}{Dataset B (Run 2)} \\
		\multicolumn{1}{c}{} & \multicolumn{1}{c}{$\chi_{\nu,best}^2$} & \multicolumn{1}{c}{$\chi_{\nu,min}^2$} & \multicolumn{1}{c}{$\chi_{\nu,max}^2$} & \multicolumn{1}{c}{$\chi_{\nu,best}^2$} & \multicolumn{1}{c}{$\chi_{\nu,min}^2$} & \multicolumn{1}{c}{$\chi_{\nu,max}^2$} \\
		\hline
		e$_{b}$ & 0.09(3) & 0.08(4) & 0.06(4) & 0.09(3) & 0.10(4) & 0.06(3) \\
		K$_b$ (d) & 0.00046(3) & 0.00034(3) & 0.00035(4) & 0.00047(3) & 0.00036(5) & 0.00034(6) \\
		$\omega_b$ (rad) & 5.56(6) & 5.45(4) & 6.18(3) & 5.63(4) & 5.46(7) & 6.21(4) \\
		P$_b$ (yr) & 8.14(4) & 7.06(5) & 7.07(4) & 8.13(5) & 6.99(7) & 7.04(6) \\
		M$_b$ (rad) & 5.42(7) & 5.91(6) & 6.03(6) & 4.59(6) & 3.22(7) & 6.06(4) \\
		e$_{c}$ & 0.91(4) & 0.13(5) & 0.17(4) & 0.91(5) & 0.33(5) & 0.10(6) \\
		K$_{c}$ (d) & 0.00058(5) & 0.00091(2) & 0.00530(5) & 0.00058(4) & 0.00714(7) & 0.00067(7) \\
		$\omega_c$ (rad) & 4.65(6) & 5.86(9) & 5.44(5) & 4.48(9) & 5.20(8) & 5.71(7) \\
		P$_c$ (yr) & 16.94(4) & 21.73(7) & 50.45(6) & 17.09(6) & 41.78(7) & 24.91(8) \\
		M$_c$ (rad) & 4.21(8) & 1.27(6) & 0.32(8) & 3.86(8) & 5.30(9) & 5.14(9) \\
		P$_{bin}$ (d) & 0.11037410(2) & 0.11037413(2) & 0.11037426(4) & 0.11037410(3) & 0.11037432(5) & 0.11037413(4) \\
		T$_0$ (BJD-2400000) & 54274.20824(4) & 54274.20834(8) & 54274.20294(4) & 54274.20815(4) & 54274.20349(7) & 54274.20819(5) \\
		$m_{b}$ ($M_{Jup}$) & 12.33 & 10.12 & 10.20 & 12.69  & 10.73 & 11.12  \\
		$m_{c}$ ($M_{Jup}$) & 9.50 & 12.75 & 44.31  & 9.61 & 70.43  & 8.57  \\
		$\chi_{\nu}^2$ & 5.45 & 7.14 & 7.67 & 5.52 & 6.89 & 7.44  \\
		\hline
	\end{tabular*}
	\tablefoot{Columns $\chi_{\nu,best}^2$ display the best-fitting curve parameters, whereas columns $\chi_{\nu,min}^2$ and $\chi_{\nu,max}^2$ contain parameters corresponding to the minimum- and maximum-reduced chi-square values of stable configurations, respectively. The term $M_b$, $M_c$ is the resulting mean anomaly of the inner and outer orbits, with each configuration being set up in the N-body osculating frame at the osculating epoch $T_0$. Digits in parentheses are for the uncertainty at the last significant place.}
	\label{tab:Newtonian_run2}
\end{table*}

\begin{figure*}
    \centering
    \subfloat[]{\includegraphics[width=0.4\textwidth, keepaspectratio]{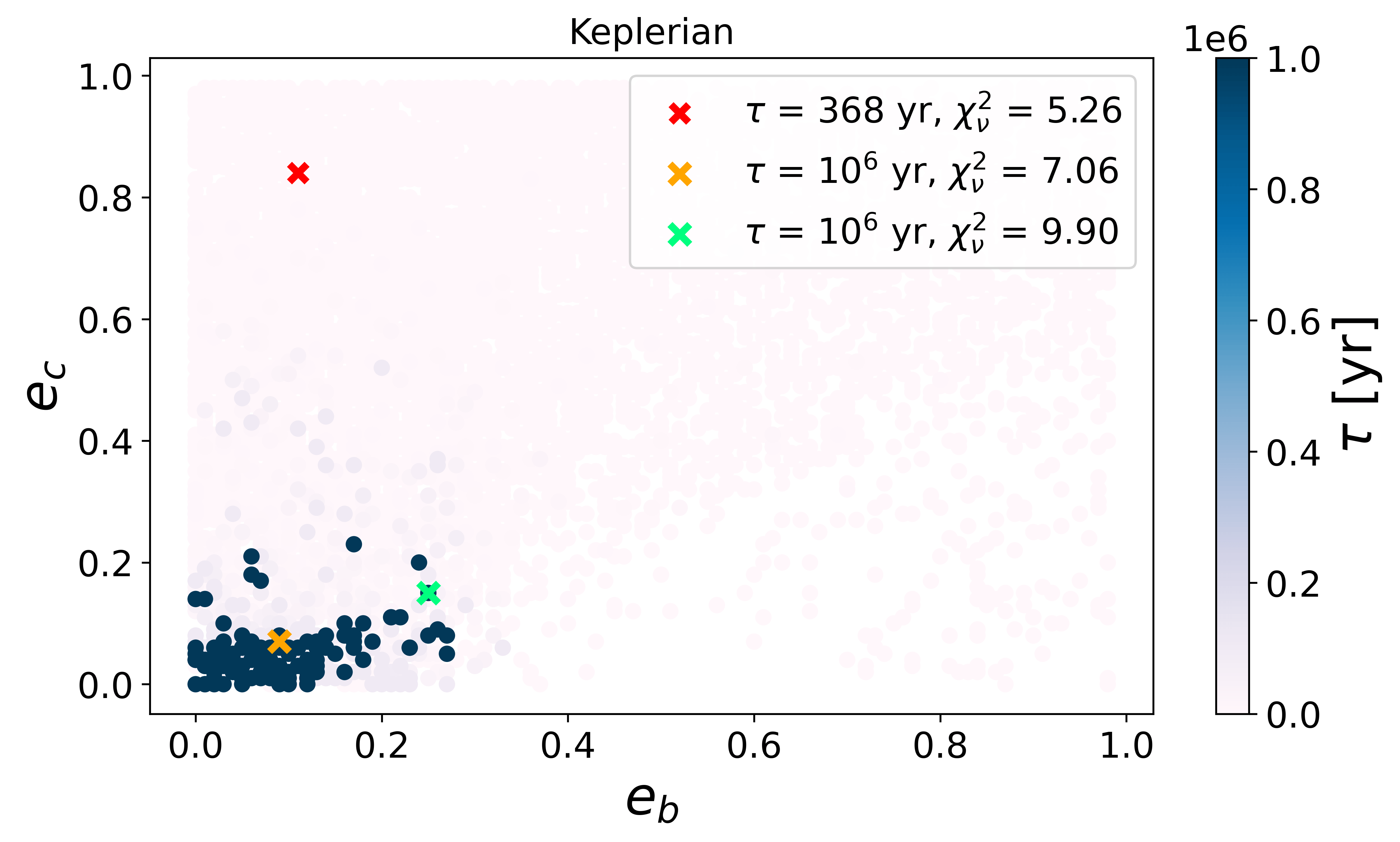}\label{fig:sbf_run1A_eieolt_Keplerian}}
    \subfloat[]{\includegraphics[width=0.4\textwidth, keepaspectratio]{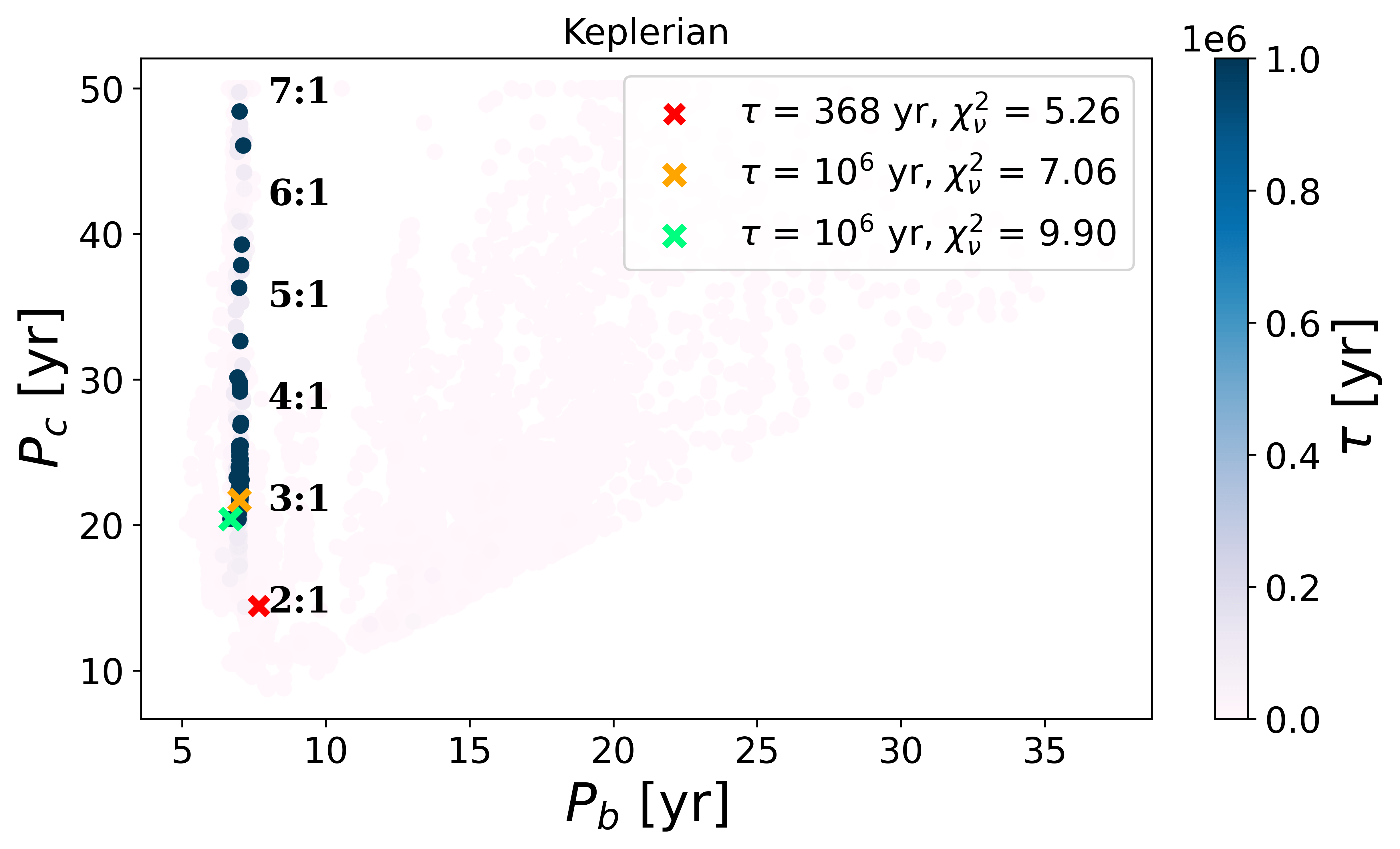}\label{fig:sbf_run1A_PiPolt_Keplerian}}\\
    \subfloat[]{\includegraphics[width=0.4\textwidth, keepaspectratio]{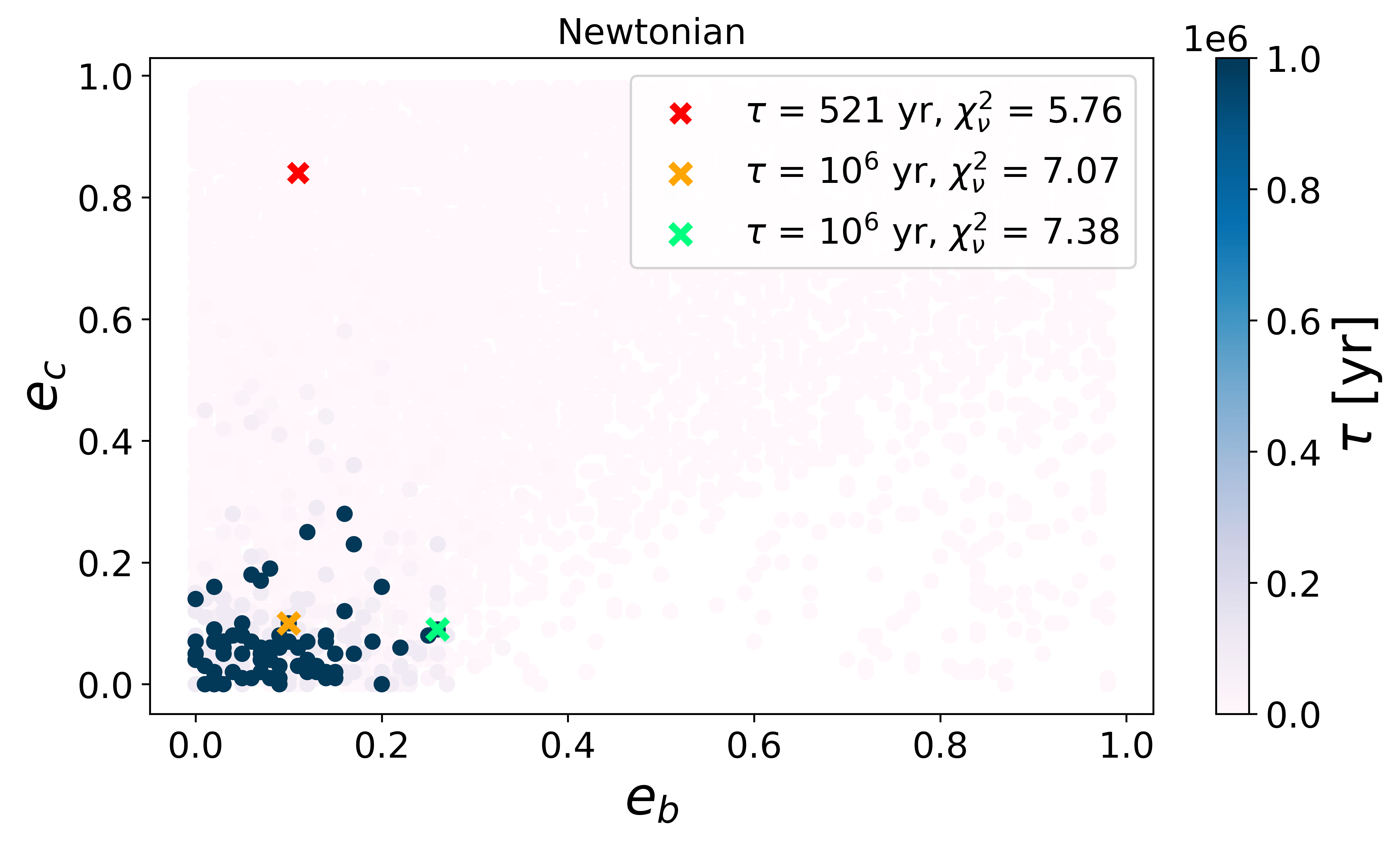}\label{fig:sbf_run1A_eieolt_Newtonian}}
    \subfloat[]{\includegraphics[width=0.4\textwidth, keepaspectratio]{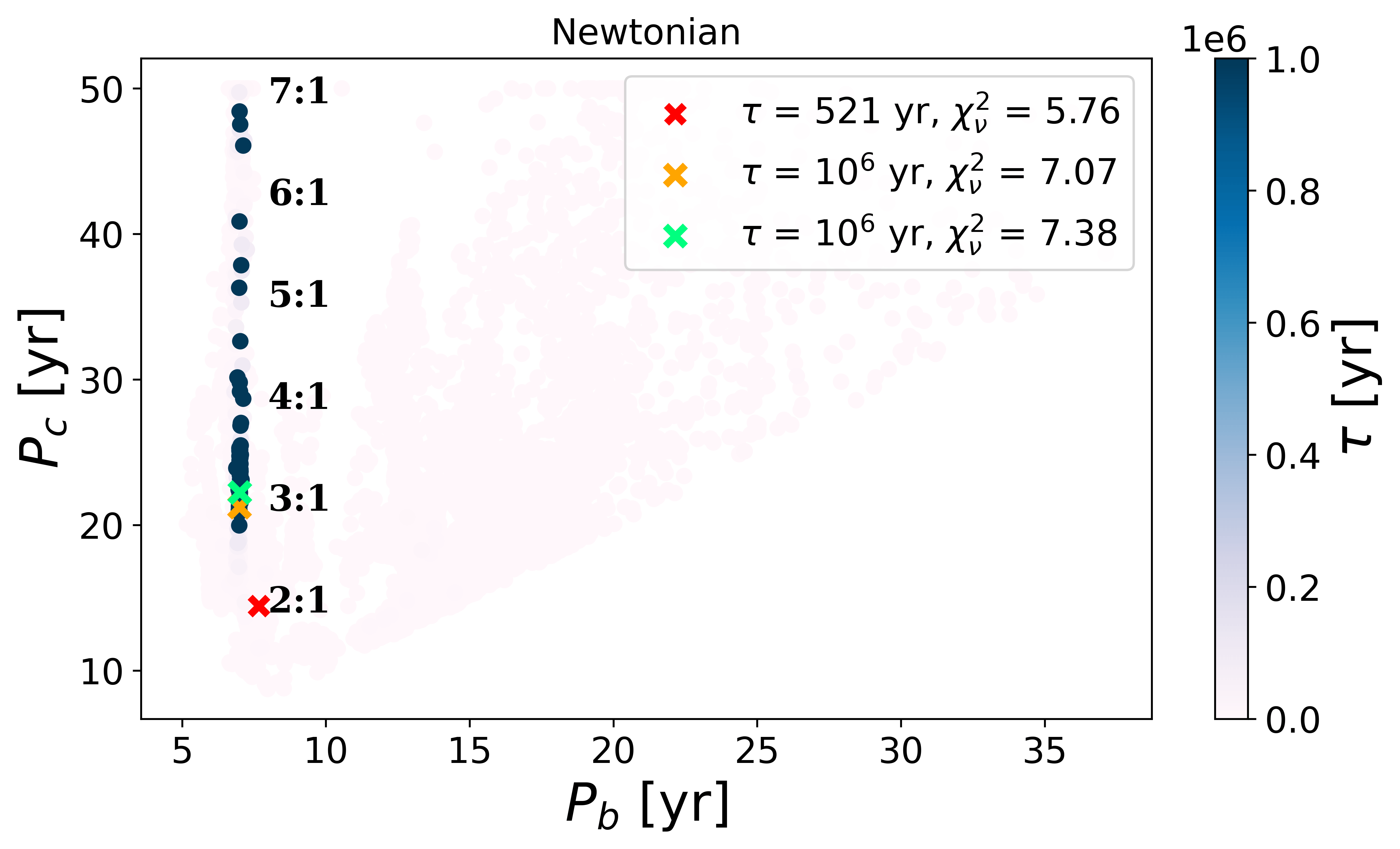}\label{fig:sbf_run1A_PiPolt_Newtonian}}

    \caption{Top: Lifetime colormap plots in the (a) $e_b,e_c$ and (b) $P_b,P_c$ planes for 5838 Keplerian models (99 stable) within the $90\%$ confidence level of $\chi_{\nu,best}^2$ as they resulted from optimization Run 1 for dataset A. Bottom: Same as top but for 5838 Newtonian models (74 stable). The color coding displays the range of lifetimes from white to dark blue, while the orange and green crosses correspond to the positions of the minimum- and maximum-reduced chi-square value of the stable configurations. The best-fitting curve to the data is identified by the red cross, which is unstable in all cases. The MMR solutions have been labeled in the $P_b,P_c$ plane for the nominal $P_b = 7$ yr.}
    \label{fig:lifetime_distributions_rA1}
\end{figure*}

\begin{figure*}
    \hspace{1cm}\subfloat[$\chi_{\nu,best}^2$ (dynamically unstable)]{\includegraphics[width=0.3\textwidth, keepaspectratio]{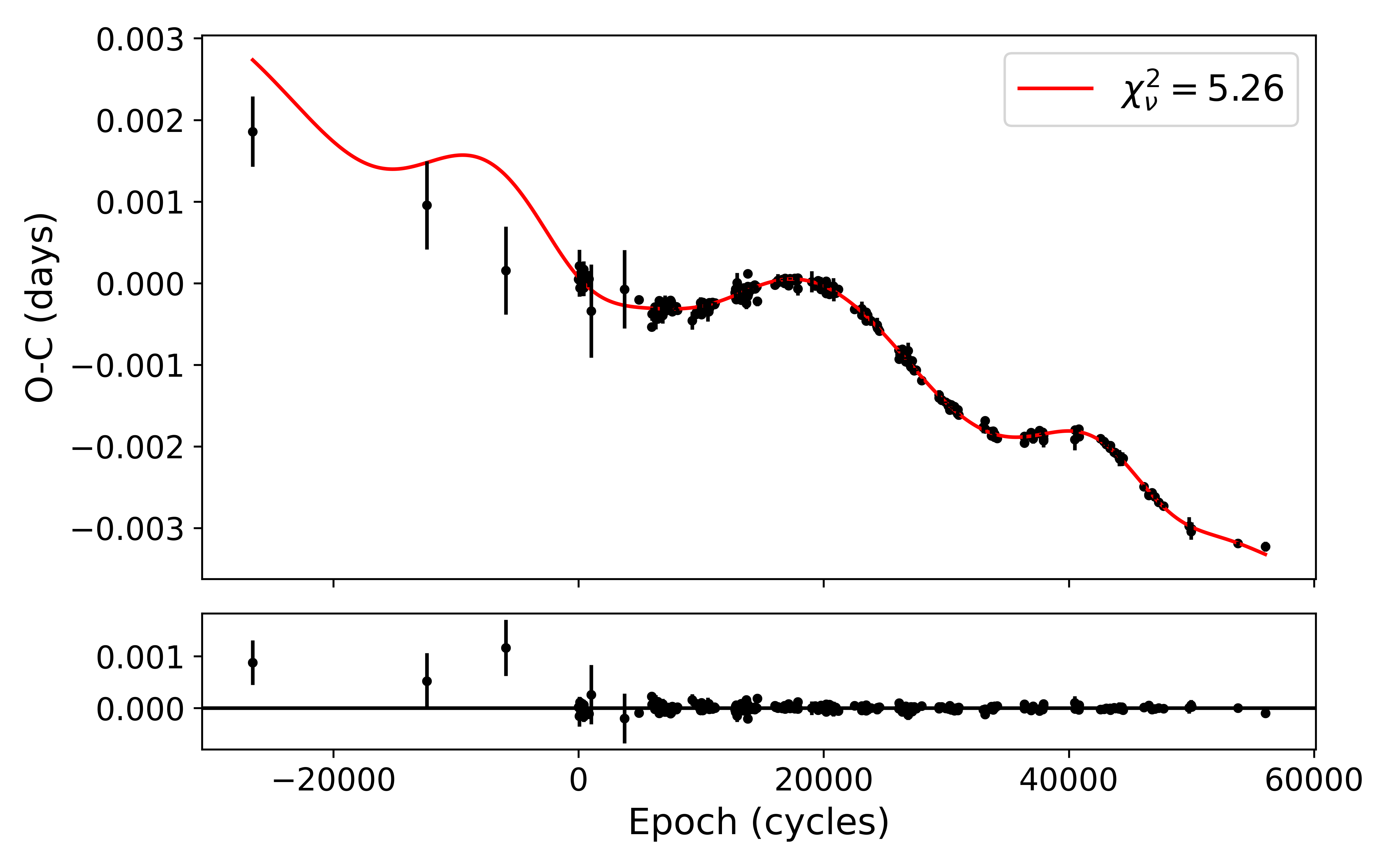}}
    \subfloat[$\chi_{\nu,min}^2$ (dynamically stable)]{\includegraphics[width=0.3\textwidth, keepaspectratio]{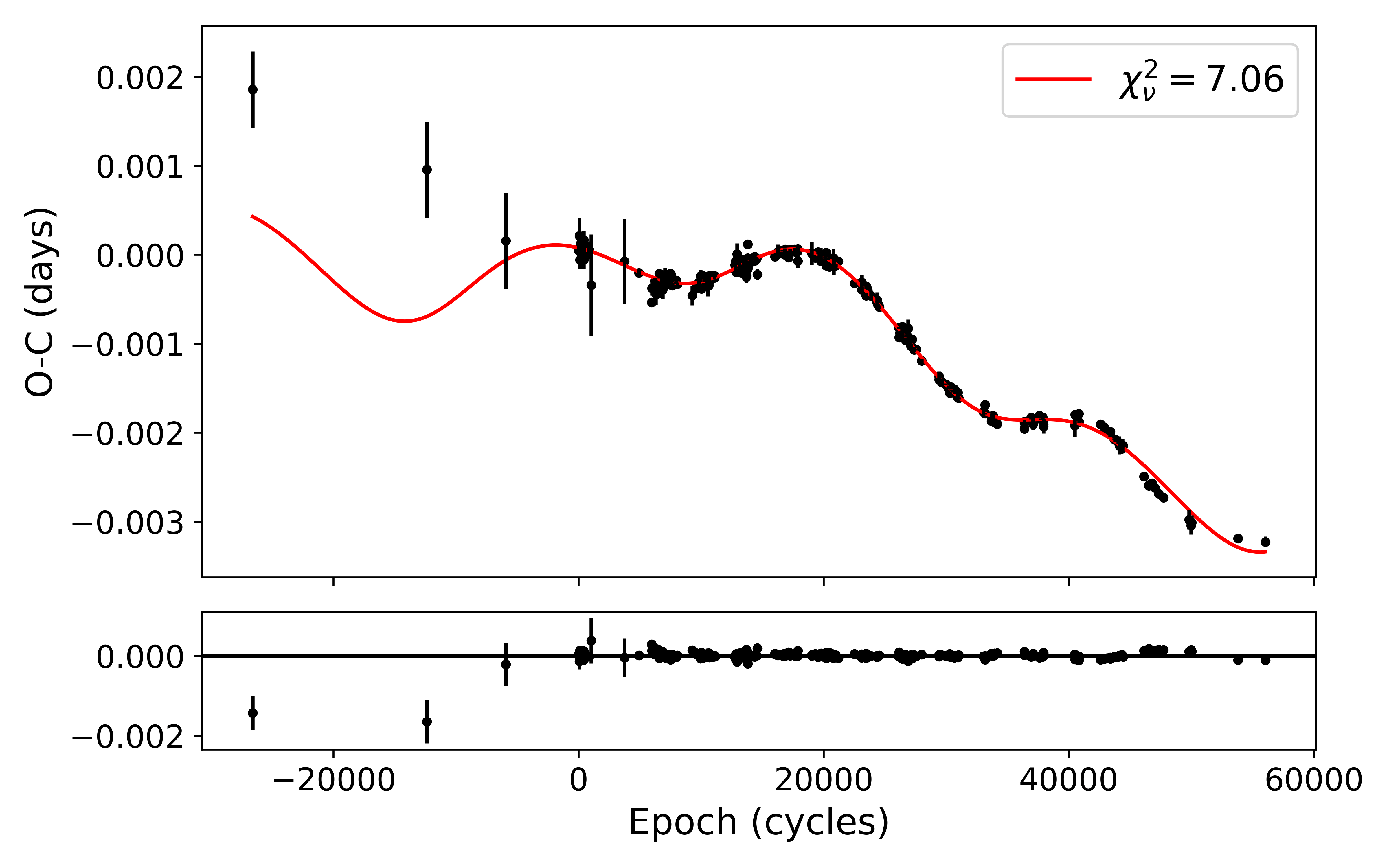}}
    \subfloat[$\chi_{\nu,max}^2$ (dynamically stable)]{\includegraphics[width=0.3\textwidth, keepaspectratio]{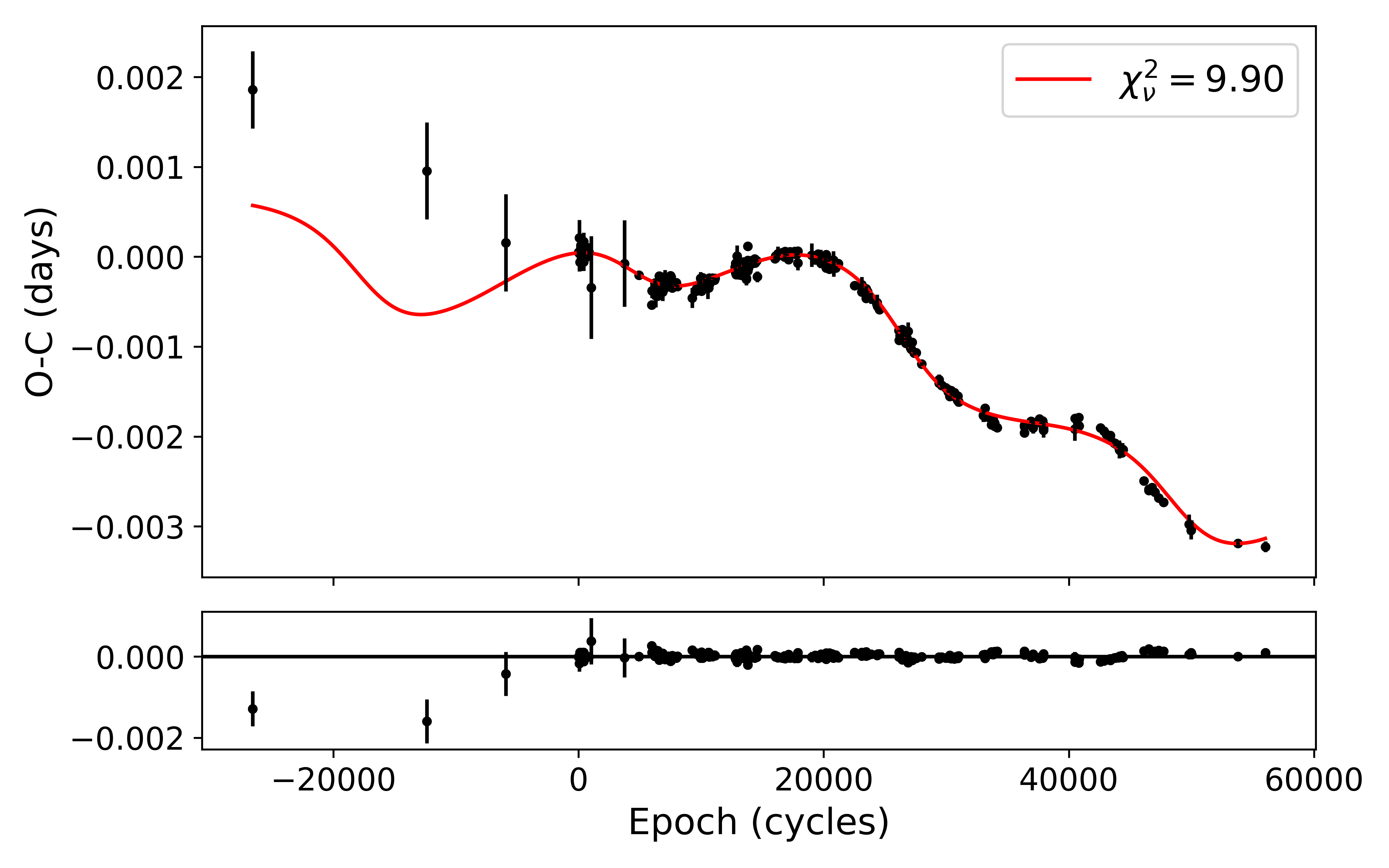}}\\
    
    \hspace{1cm}\subfloat[$\chi_{\nu,best}^2$ (dynamically unstable)]{\includegraphics[width=0.3\textwidth, keepaspectratio]{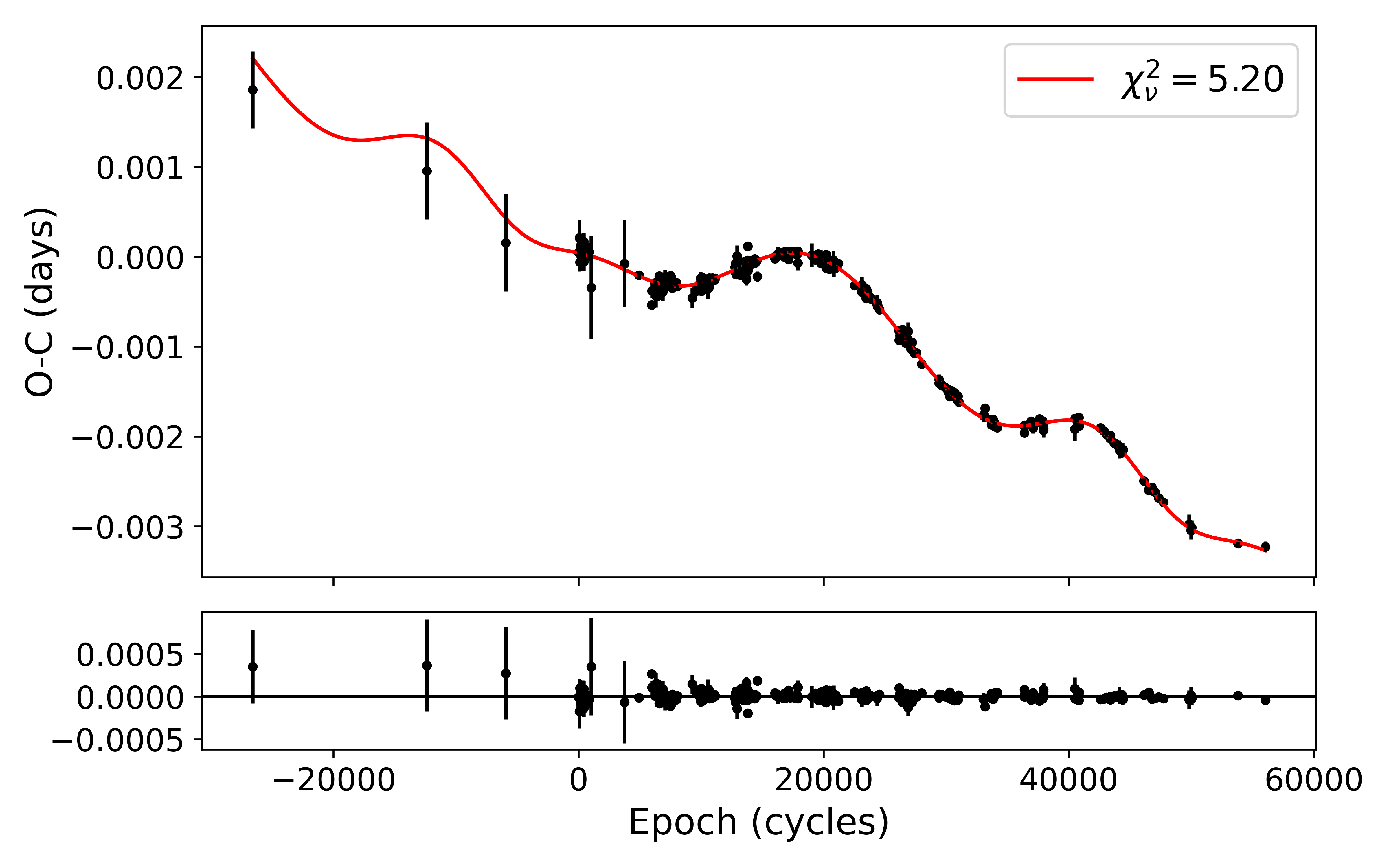}}
    \subfloat[$\chi_{\nu,min}^2$ (dynamically stable)]{\includegraphics[width=0.3\textwidth, keepaspectratio]{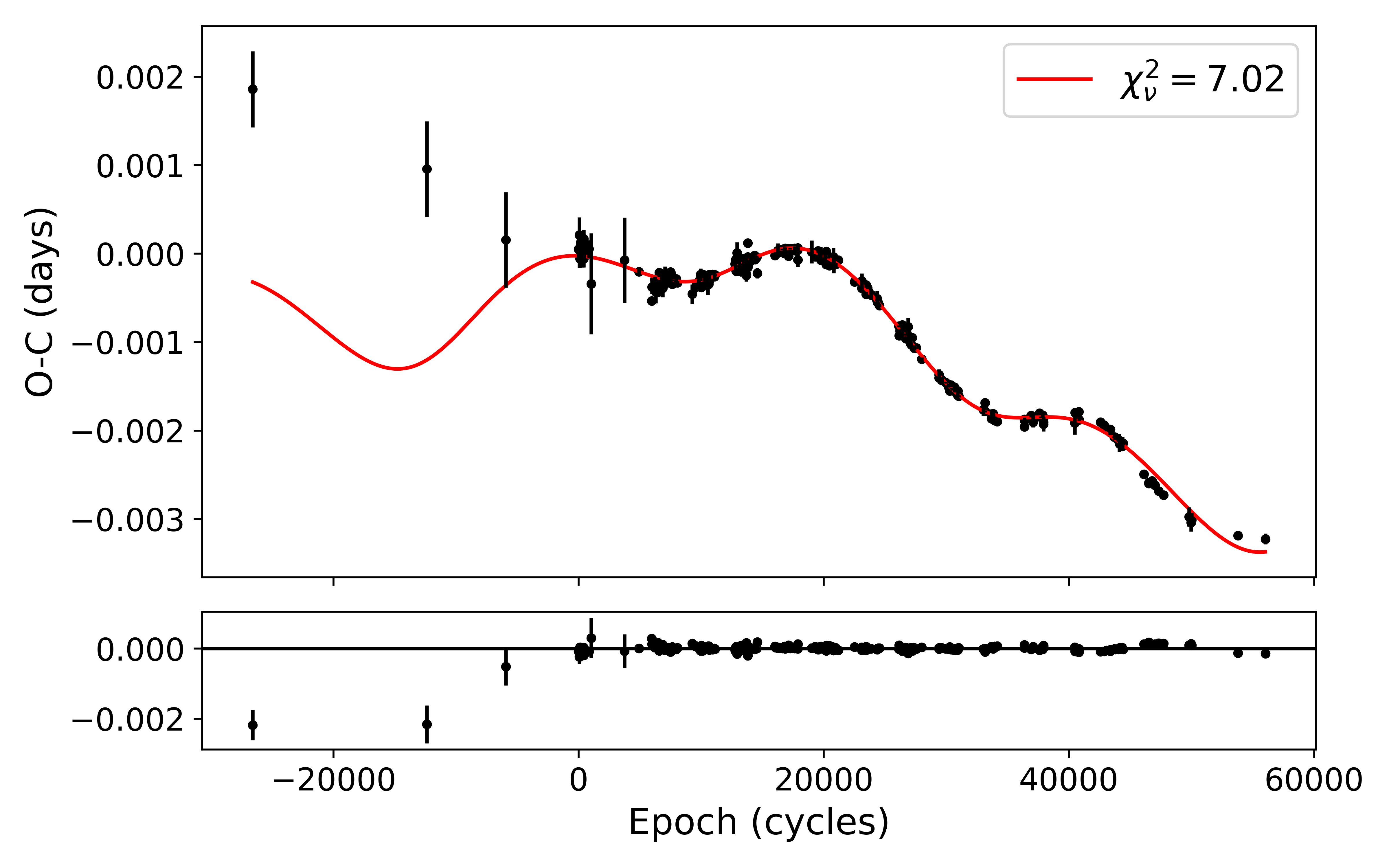}}
    \subfloat[$\chi_{\nu,max}^2$ (dynamically stable)]{\includegraphics[width=0.3\textwidth, keepaspectratio]{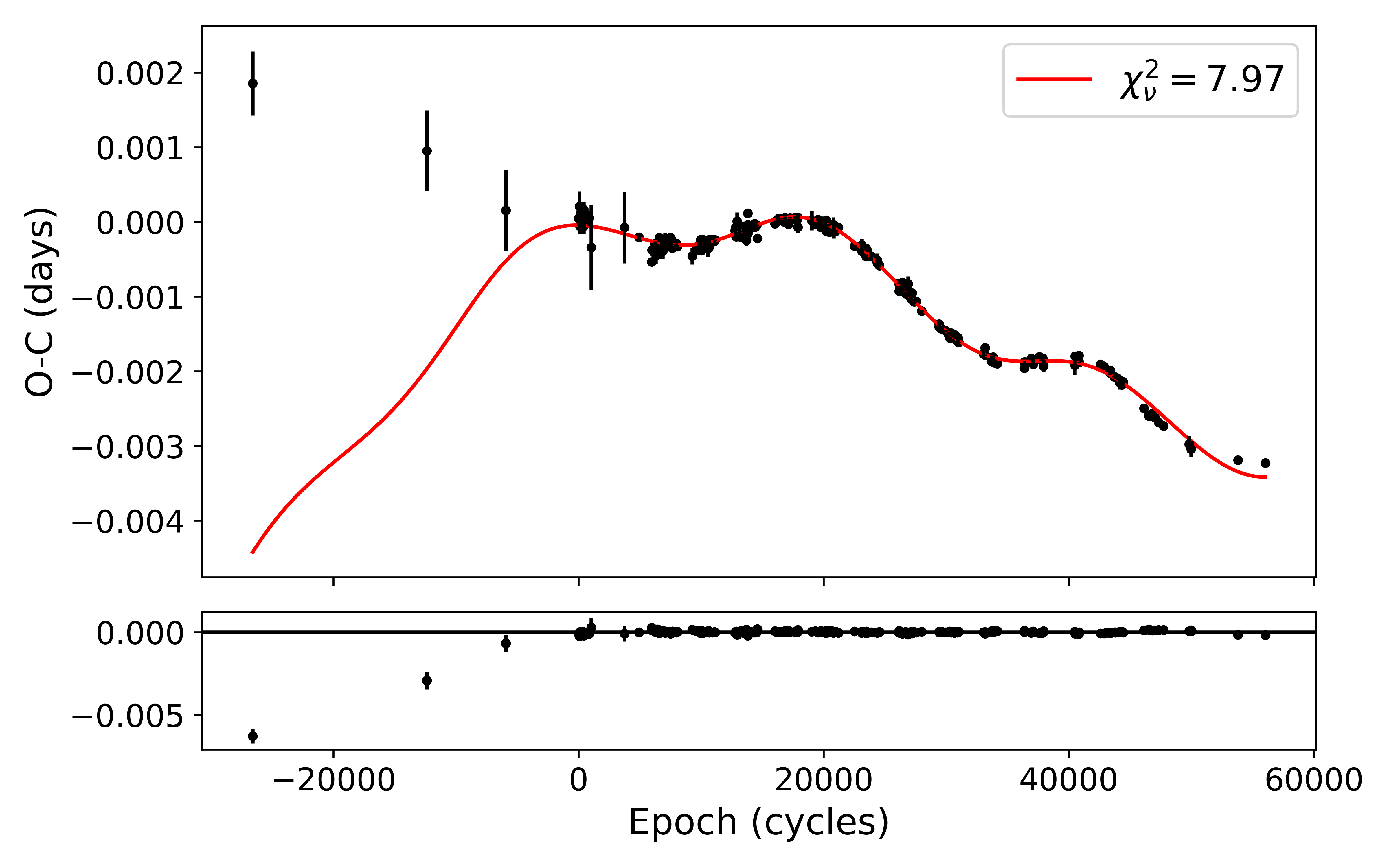}}\\
    
    \hspace{1cm}\subfloat[$\chi_{\nu,best}^2$ (dynamically unstable)]{\includegraphics[width=0.3\textwidth, keepaspectratio]{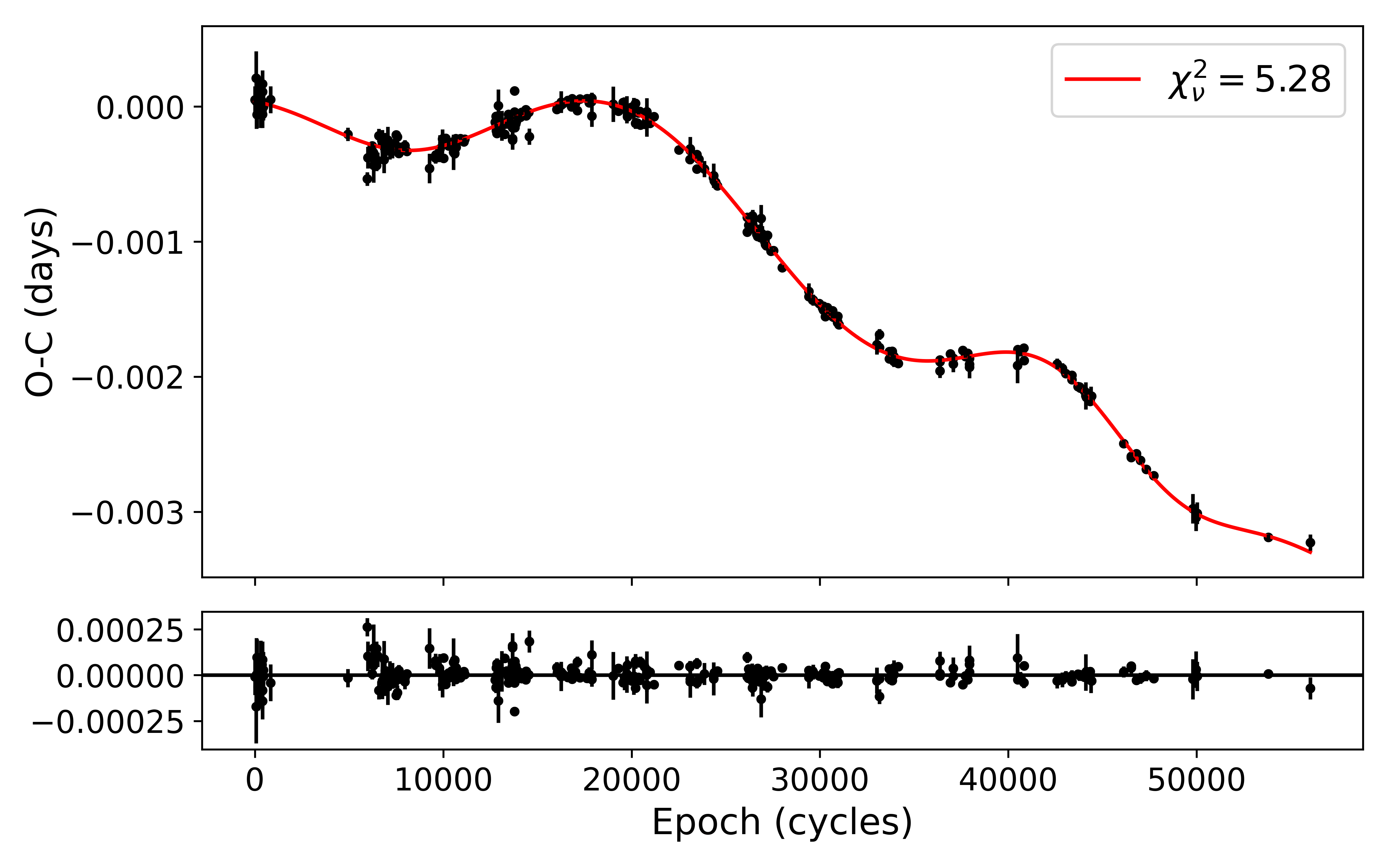}}
    \subfloat[$\chi_{\nu,min}^2$ (dynamically stable)]{\includegraphics[width=0.3\textwidth, keepaspectratio]{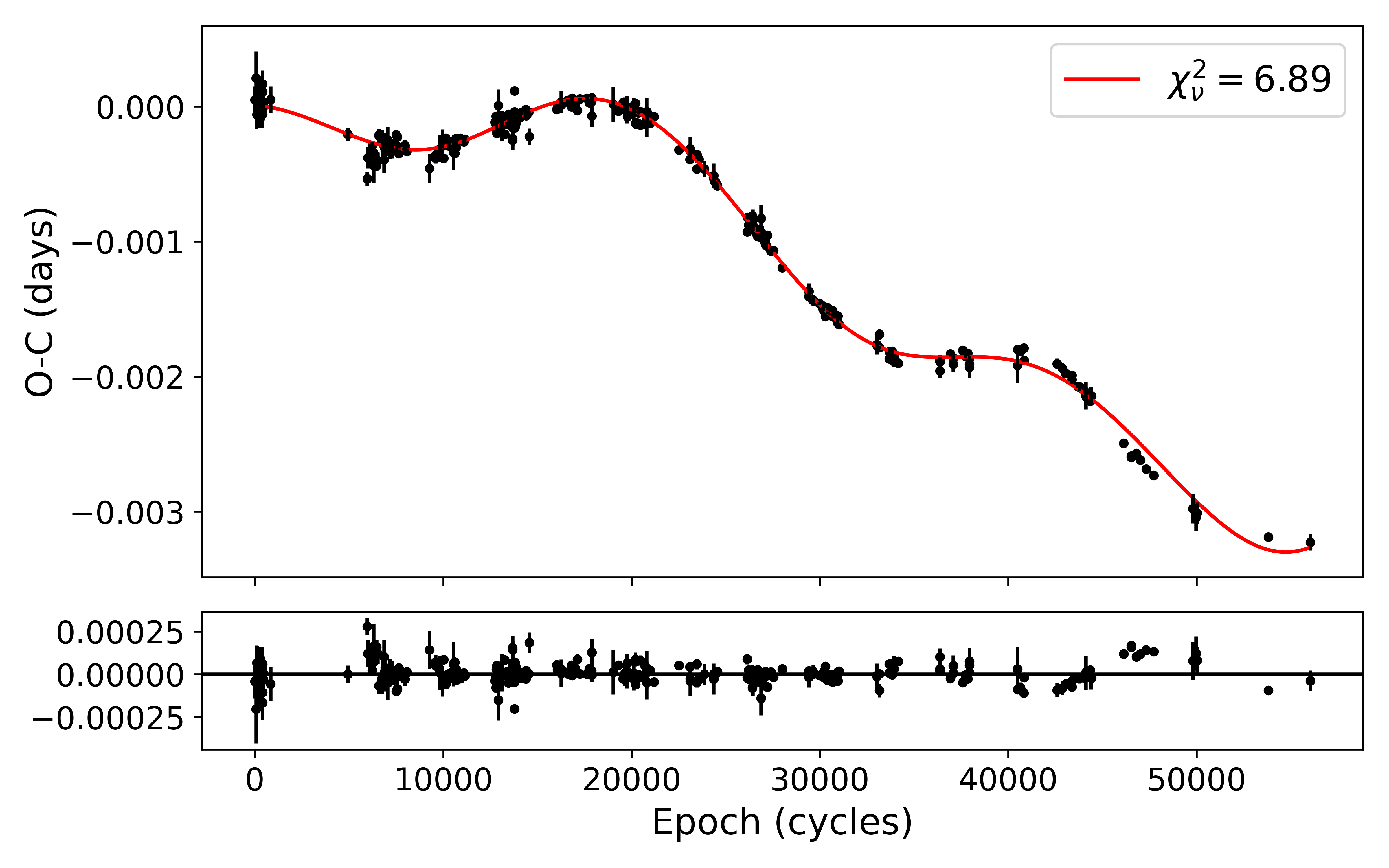}}
    \subfloat[$\chi_{\nu,max}^2$ (dynamically stable)]{\includegraphics[width=0.3\textwidth, keepaspectratio]{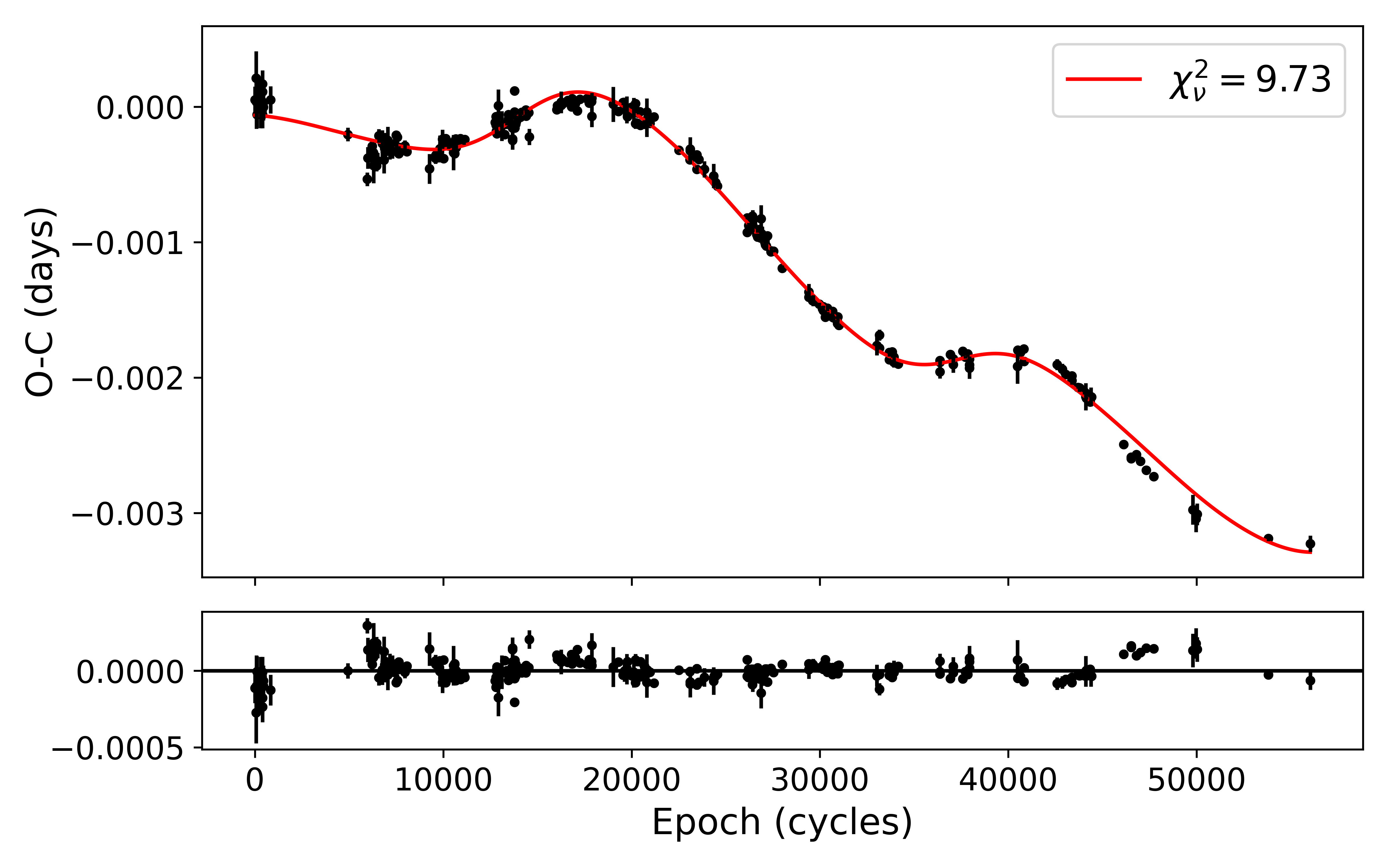}}\\
    
    \hspace{1cm}\subfloat[$\chi_{\nu,best}^2$ (dynamically unstable)]{\includegraphics[width=0.3\textwidth, keepaspectratio]{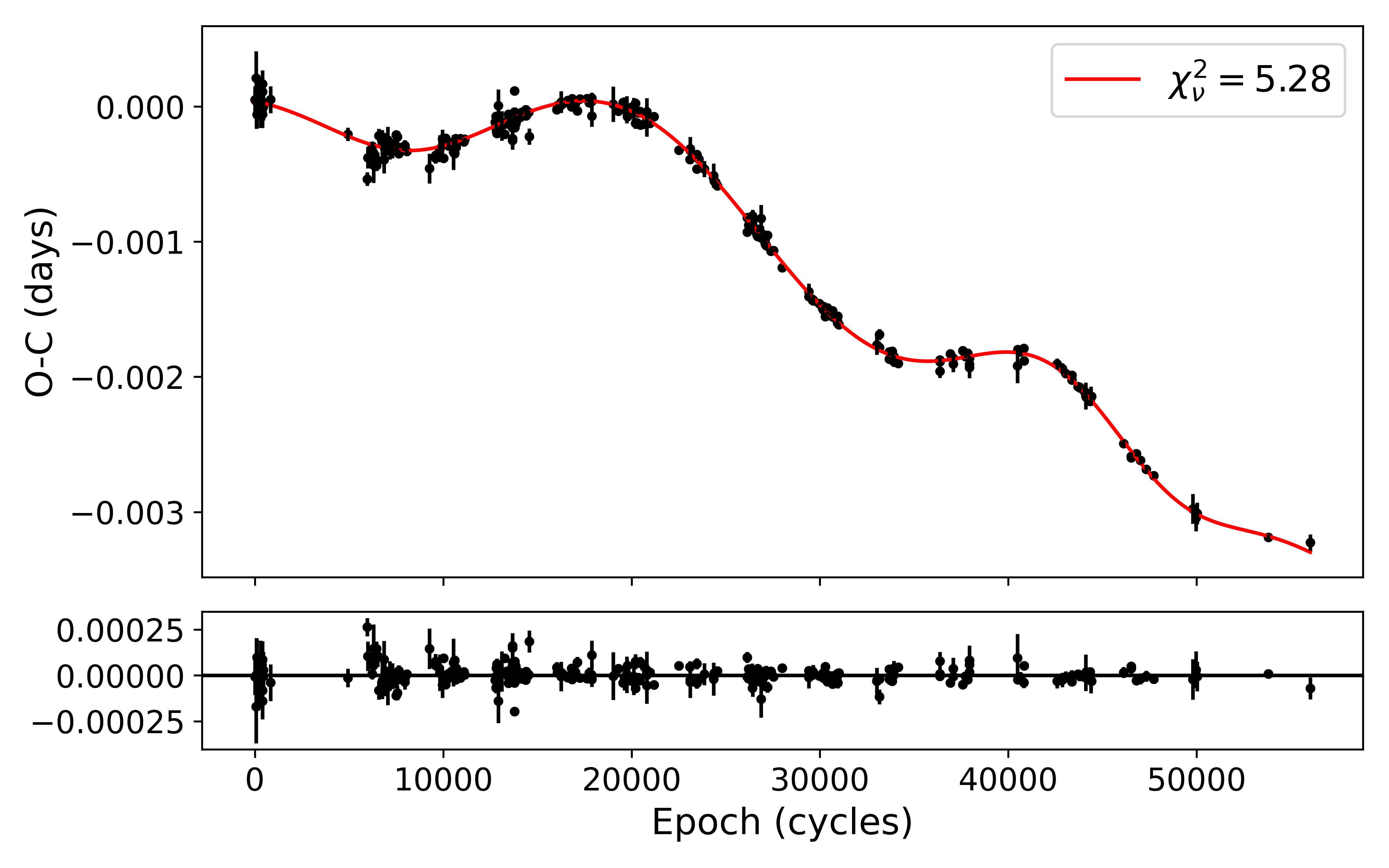}}
    \subfloat[$\chi_{\nu,min}^2$ (dynamically stable)]{\includegraphics[width=0.3\textwidth, keepaspectratio]{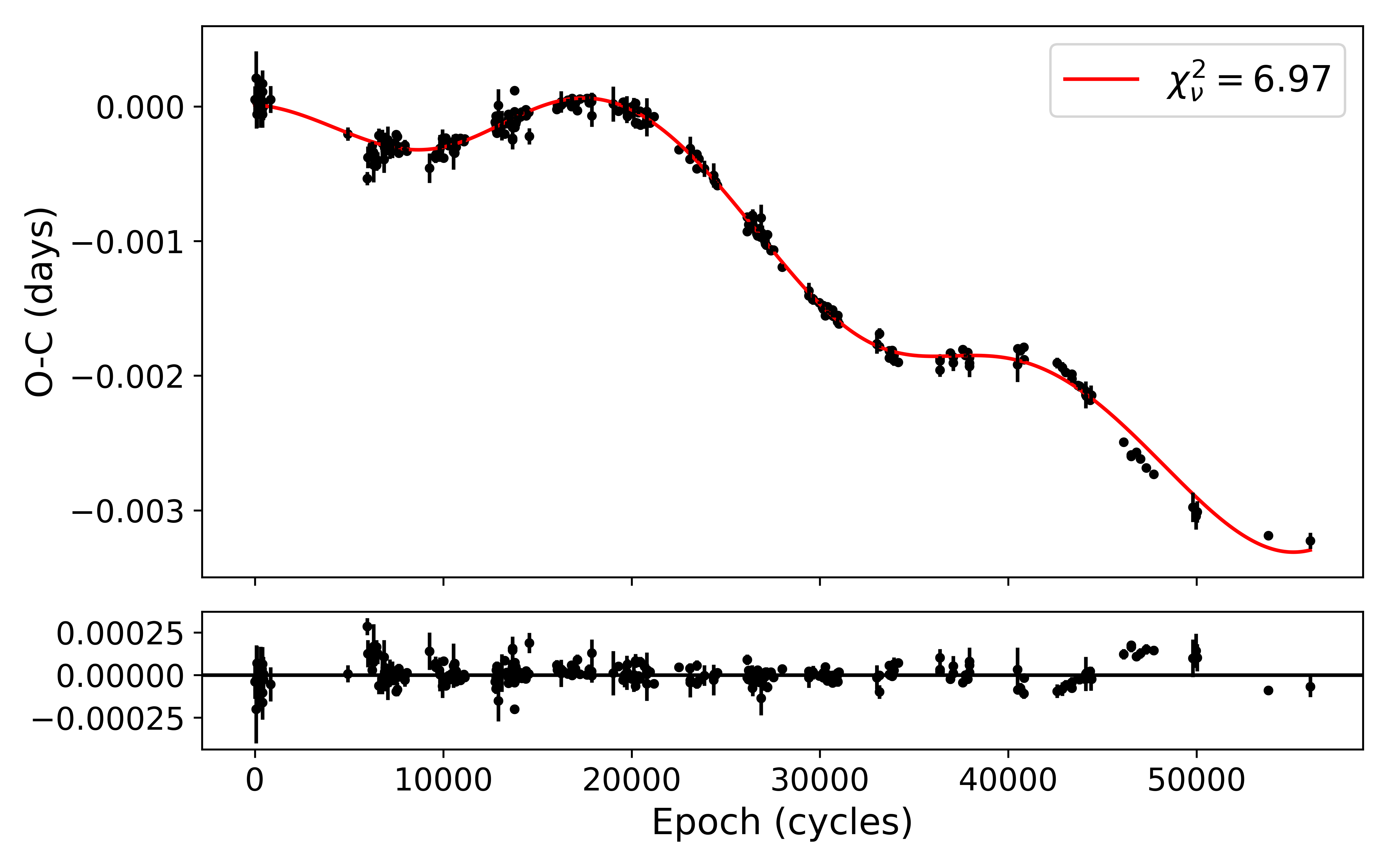}}
    \subfloat[$\chi_{\nu,max}^2$ (dynamically stable)]{\includegraphics[width=0.3\textwidth, keepaspectratio]{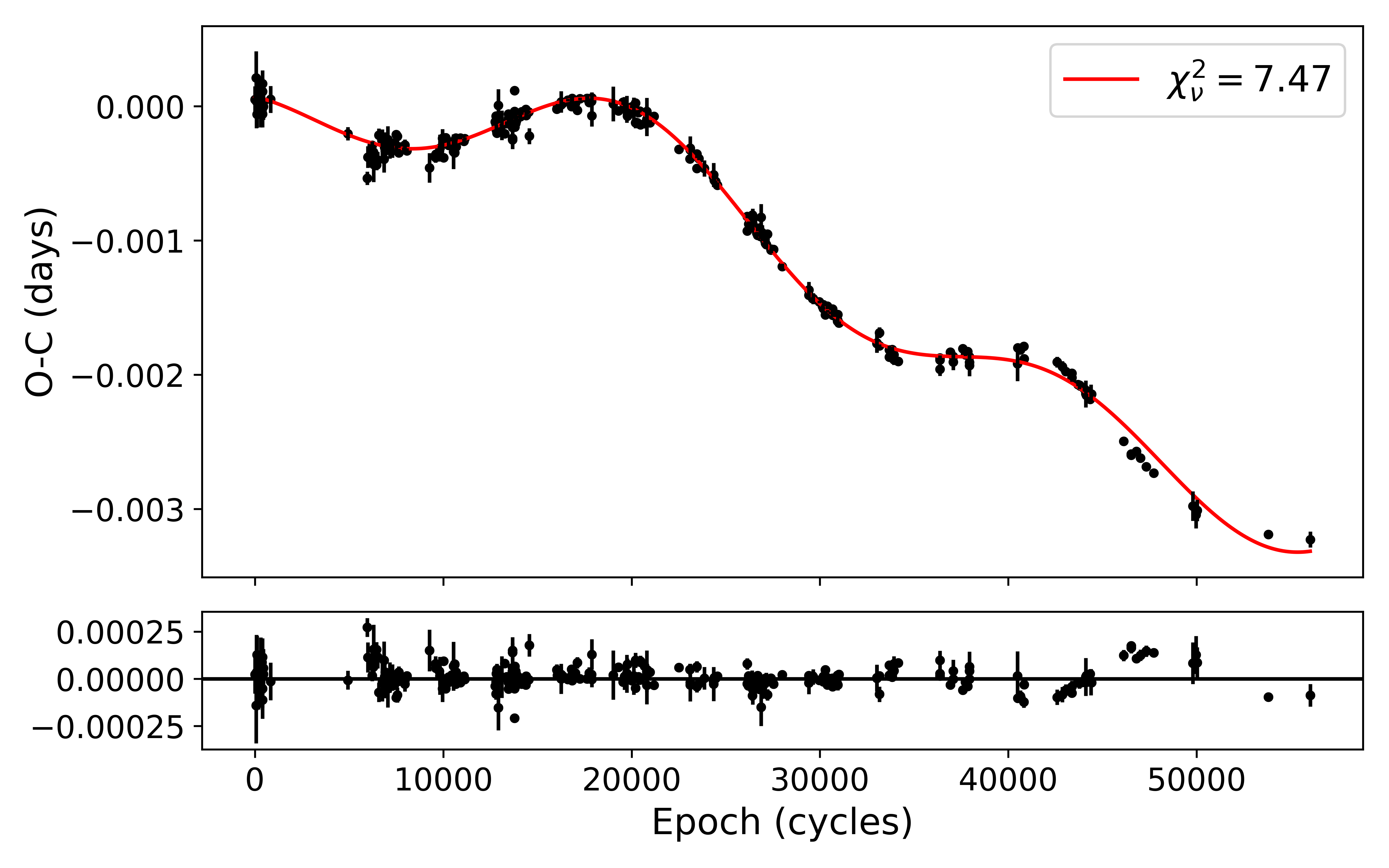}}
 
\caption{Keplerian fits for both datasets [A: (a)-(f), B: (g)-(l)] and optimization runs [1: (a)-(c), (g)-(i) ,2: (d)-(f), (j)-(l)]. The best-fitting curves ($\chi_{\nu,best}^2$) result in dynamically unstable configurations, while the fitting curves of the minimum- and maximum-reduced chi-square value ($\chi_{\nu,min}^2$, $\chi_{\nu,max}^2$) define the $\chi_{\nu}^2$ limits among all stable configurations. The parameters of each case can be found in Table~\ref{tab:Keplerian_run1} and Table~\ref{tab:Keplerian_run2}.}
\label{fig:Keplerian_fits}
\end{figure*}

\begin{figure*}
	\centering
	\subfloat[]{\includegraphics[width=0.45\textwidth, keepaspectratio]{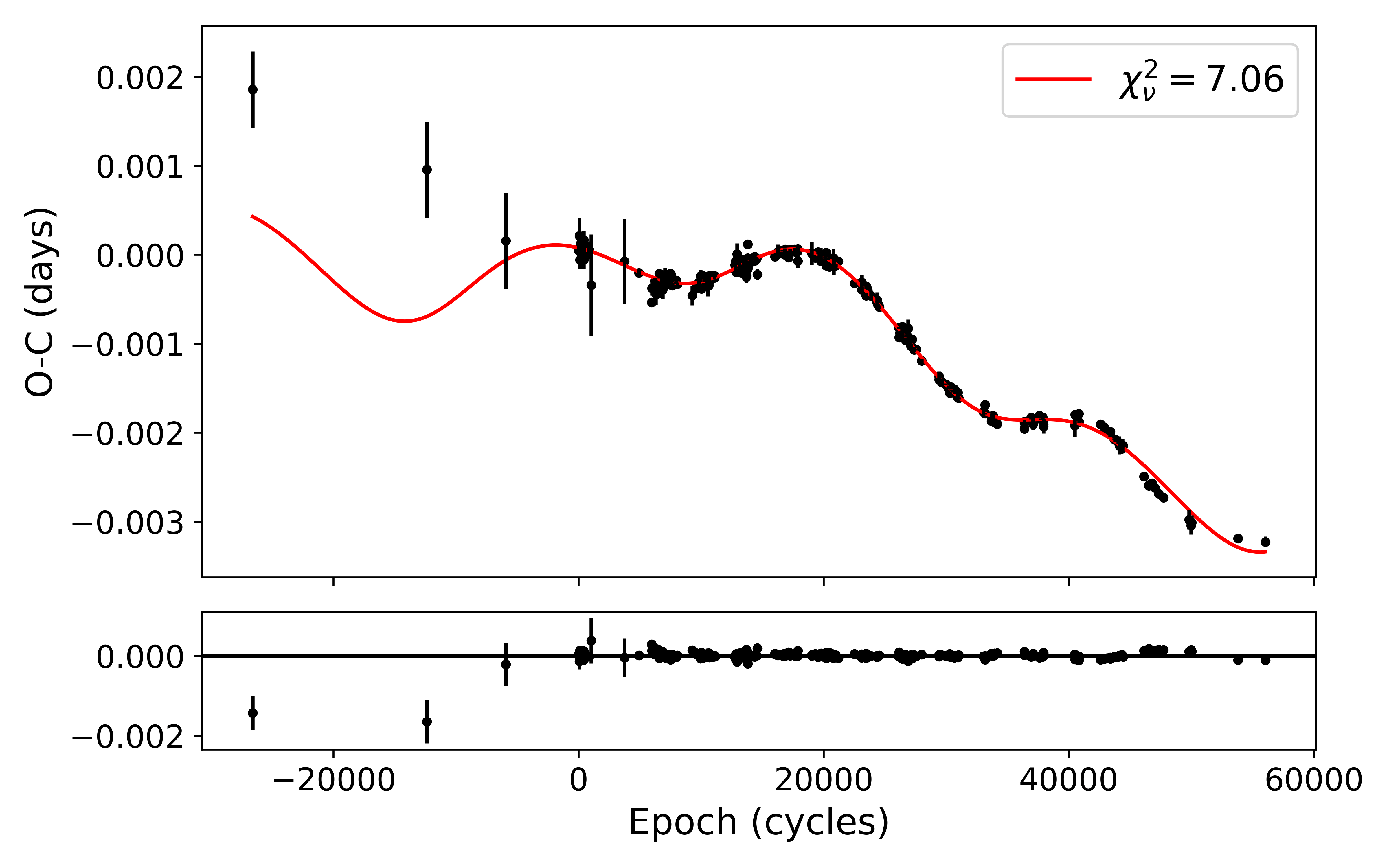}\label{fig:Kfit}}
	\subfloat[]{\includegraphics[width=0.45\textwidth, keepaspectratio]{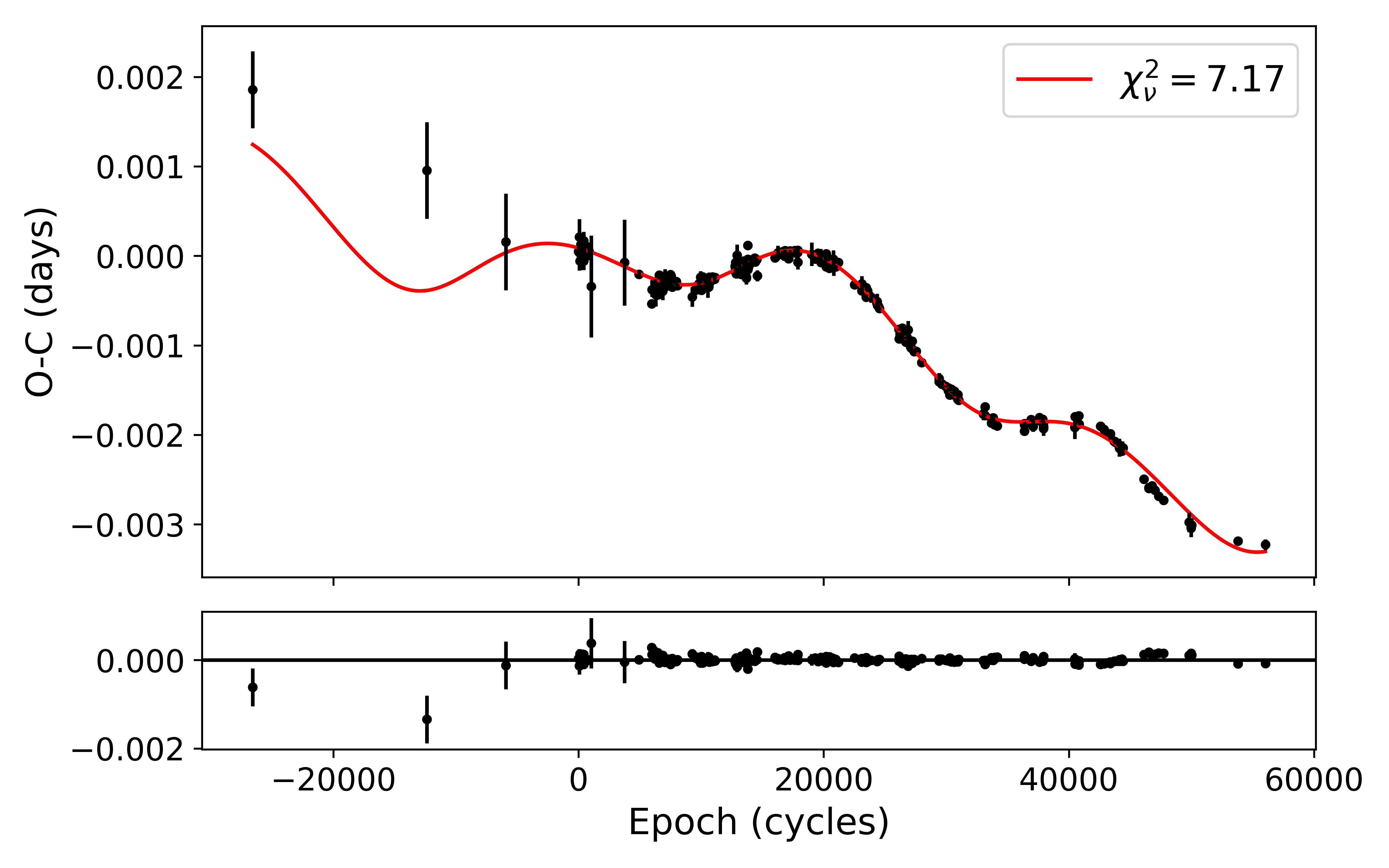}\label{fig:Nfit}}
	
	\caption{ETV diagrams with synthetic curves. Left: Keplerian fit of two-planet Jacobian solution (Table \ref{tab:Keplerian_run1}, $\chi_{\nu,min}^2$, dataset A). Right: N-body synthetic curve for the osculating initial condition derived through the formal transformation of the Jacobian elements (Keplerian fit) to the N-body Cartesian osculating frame, centered at the CM of the binary.}
	\label{fig:Kfit_Nfit}
\end{figure*}

\begin{figure*}
    \subfloat[]{\includegraphics[width=0.33\textwidth, keepaspectratio]{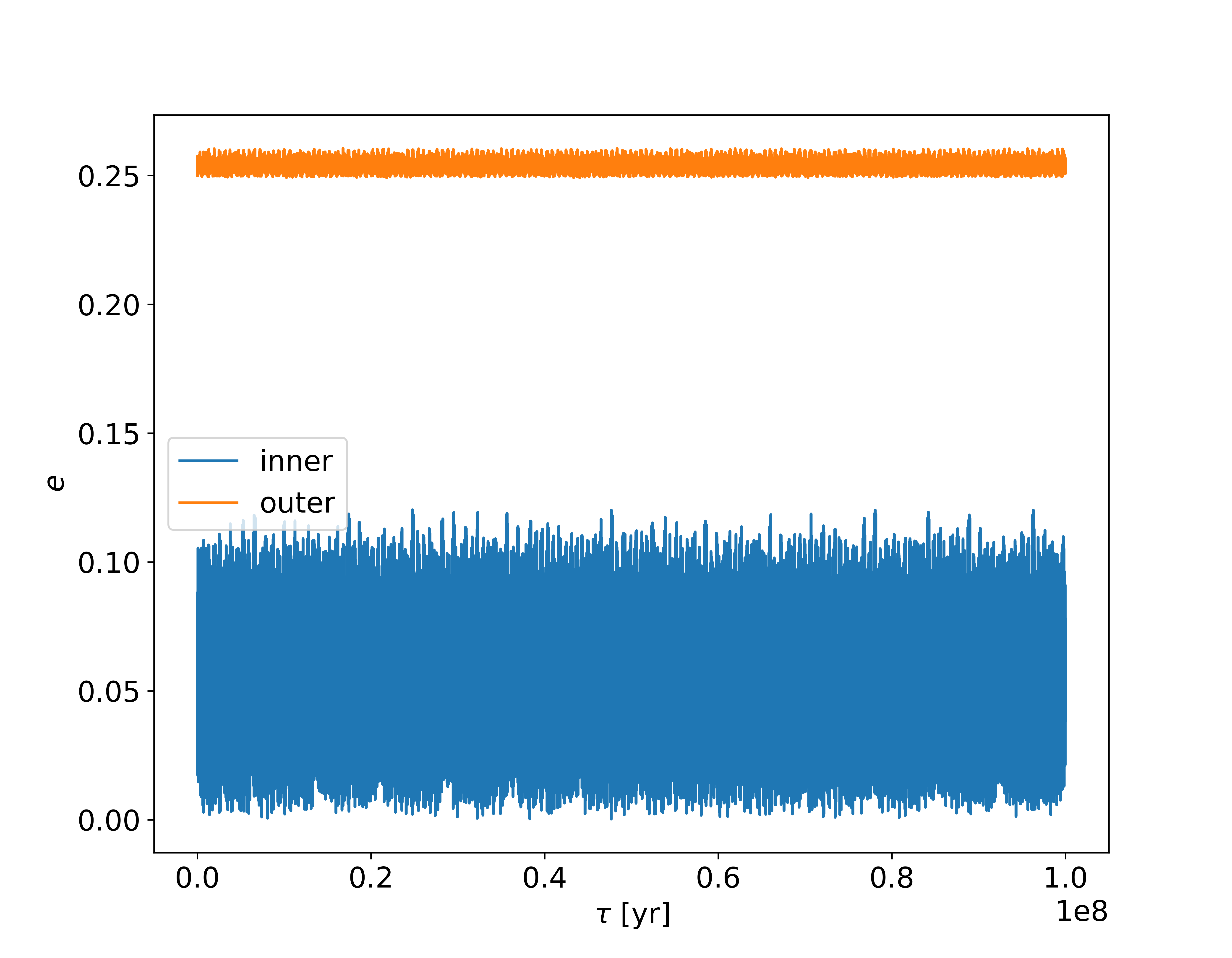}\label{fig:dsA_e3e4}}
    \subfloat[]{\includegraphics[width=0.33\textwidth, keepaspectratio]{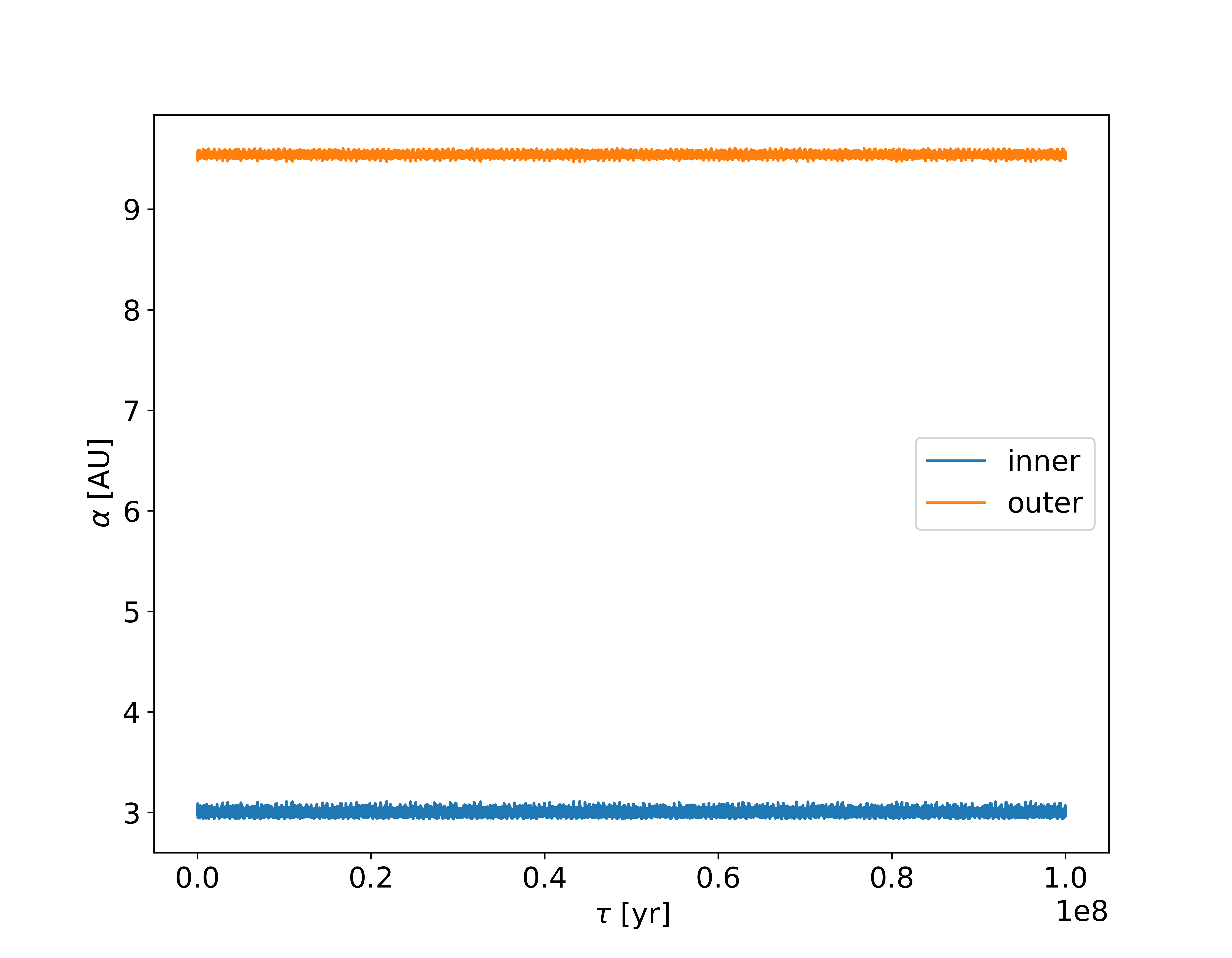}\label{fig:dsA_a3a4}}
    \subfloat[]{\includegraphics[width=0.33\textwidth, keepaspectratio]{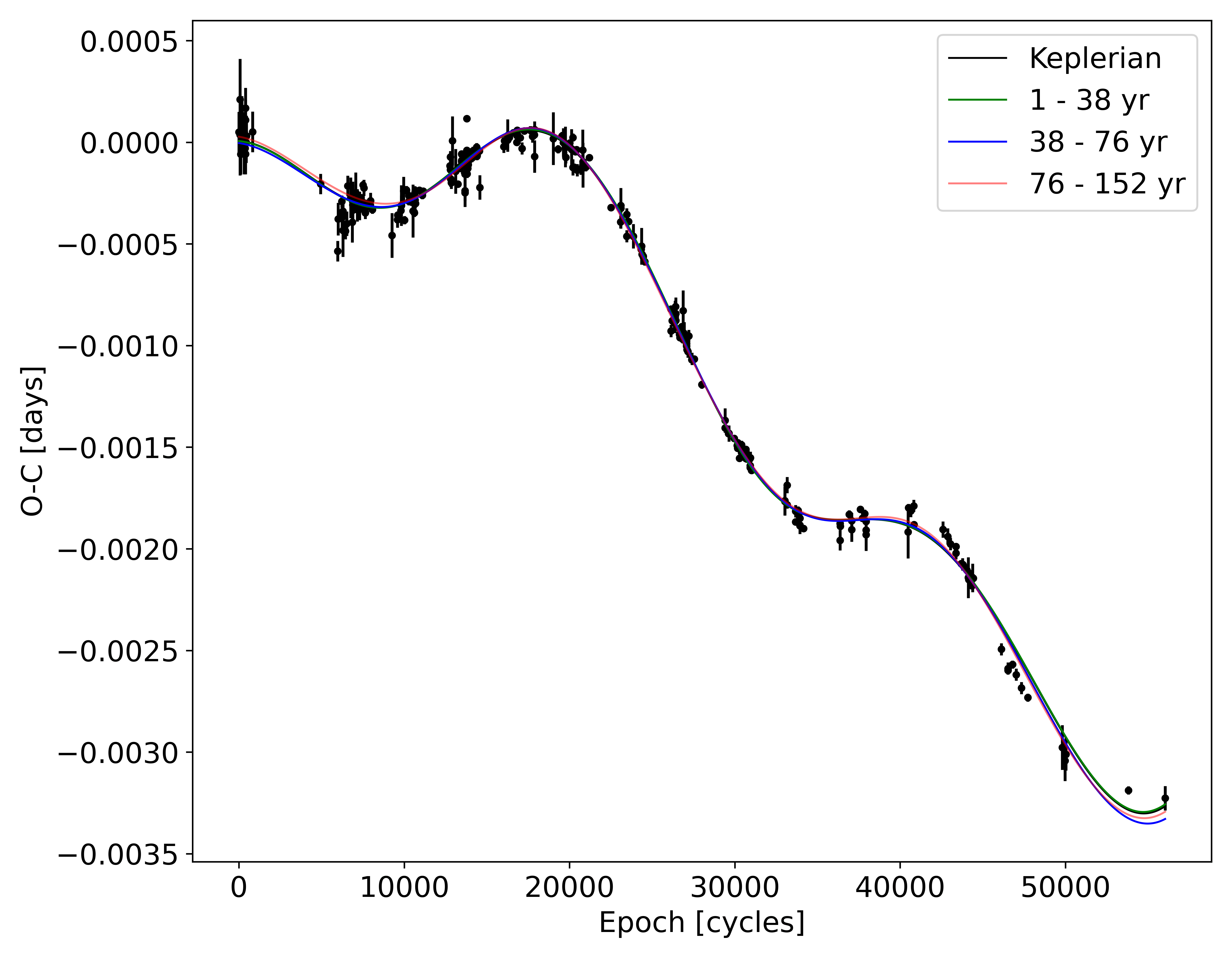}\label{fig:dsB-run1-Kfit_fold}}
\caption{Orbital evolution of (a) eccentricities and (b) semi-major axes for the inner and outer planets over 100 Myr (dataset B, Run 1, $\chi_{\nu,min}^2$). (c) The temporal variation of the $\chi_{\nu,min}^2$ fitting curve (dataset B, Run 1) over the first 152 years is represented by five sets of parameters, each corresponding to a complete outer orbit revolution (38 years). The black synthetic curve is the Keplerian (kinematic) fit from which this dynamical model was initialized, and it appears to be in absolute agreement with the first 38 years of orbital evolution (green curve) and in close agreement with the subsequent fits (blue and red curves).}
\label{fig:orbital_evolution}
\end{figure*}

\begin{figure*}
    \subfloat[]{\includegraphics[width=0.33\textwidth, keepaspectratio]{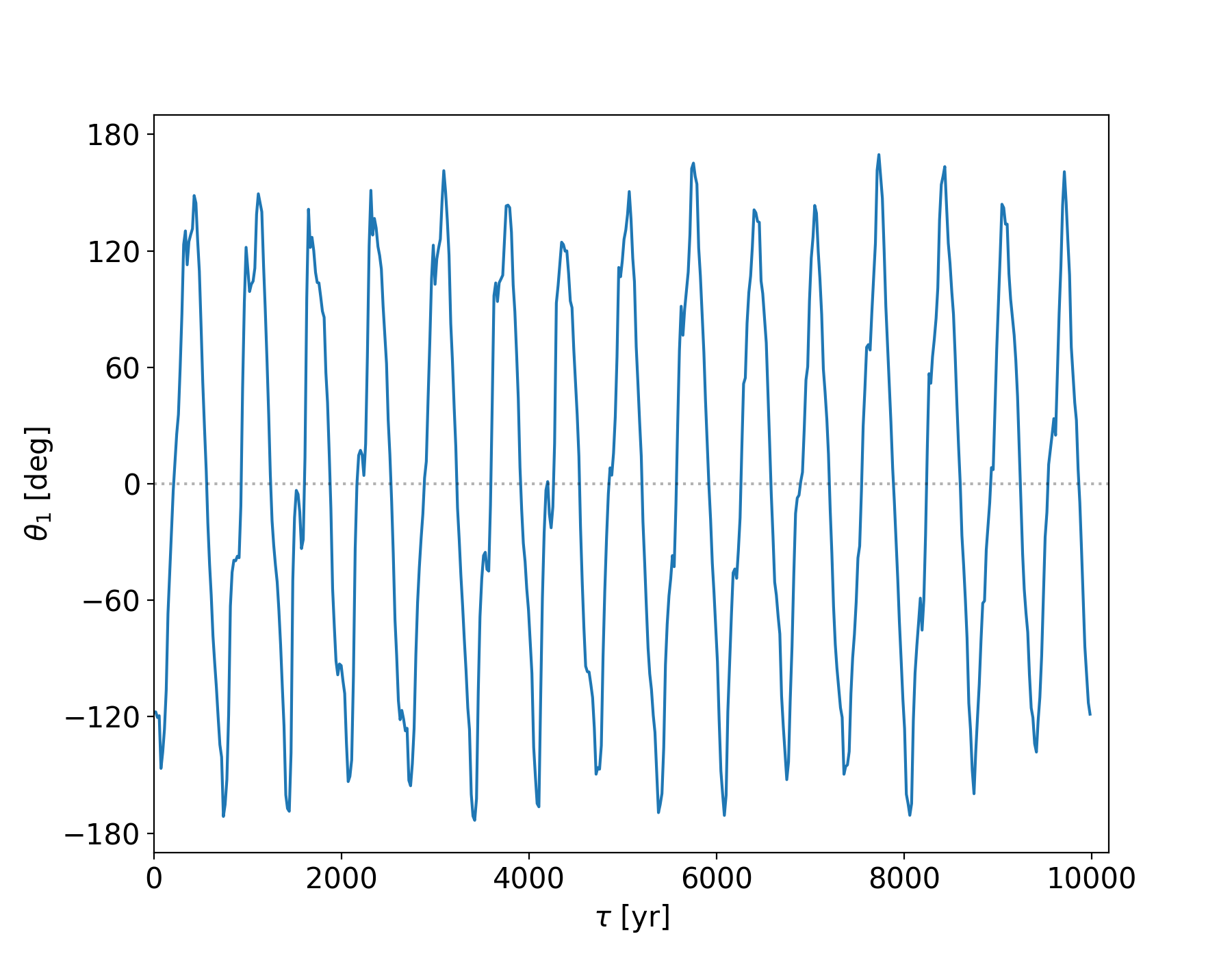}\label{fig:theta1}}
    \subfloat[]{\includegraphics[width=0.33\textwidth, keepaspectratio]{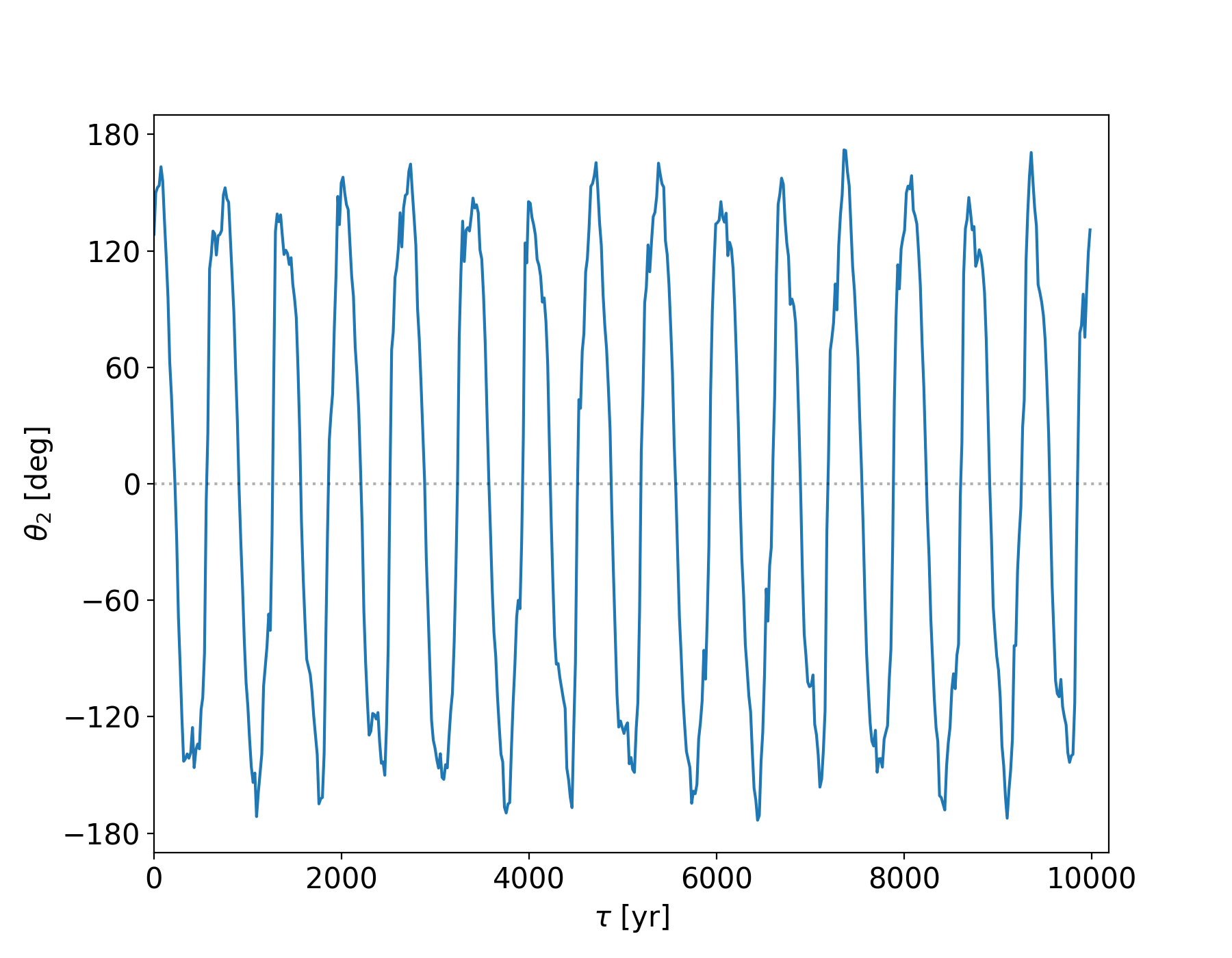}\label{fig:theta2}}
    \subfloat[]{\includegraphics[width=0.33\textwidth, keepaspectratio]{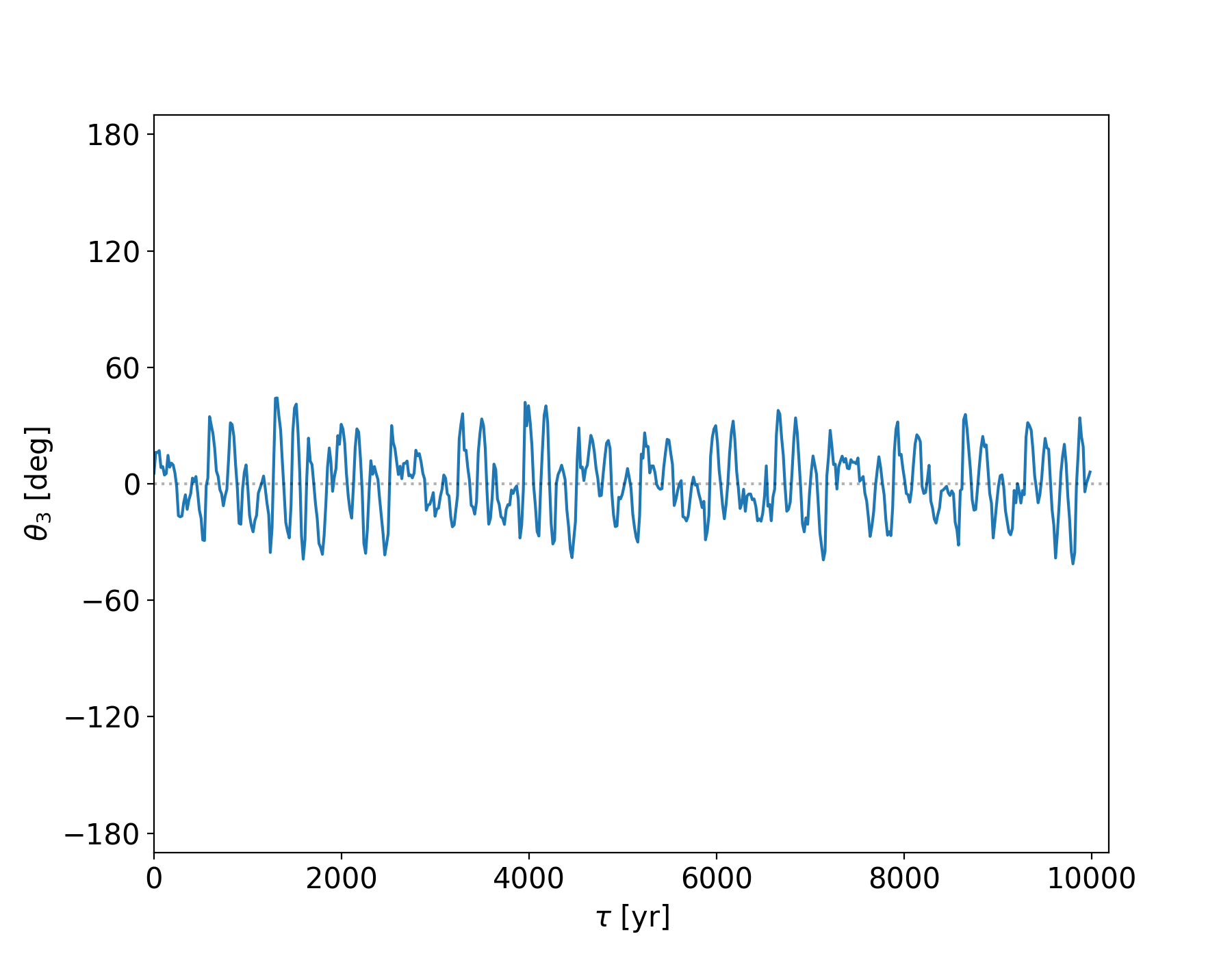}\label{fig:theta3}}
\caption{Temporal evolution of resonant angles (a) $\theta_{1}$, (b) $\theta_{2}$, and (c) $\theta_{3}$ of a stable solution close to the 3:1 MMR (Table \ref{tab:Keplerian_run1}, $\chi_{\nu,min}^2$, dataset A) over the first 10000 years of integration.}
\label{fig:mmr_evolution}
\end{figure*}

It is notable that none of the best-fitting curves ($\chi_{\nu,best}^2$) are stable for more than a few hundred years. However, there are hundred of stable orbits ($\tau=1$Myr, $\langle Y \rangle\sim2$)  within the $90\%$ confidence level of each of the best reduced chi-square values.  Illustrated in Fig.~\ref{fig:lifetime_distributions_rA1}-Fig.~\ref{fig:Newtonian_fits} are the lifetime distributions of eccentricities as well as the periods and the fitting curves corresponding to the three critical cases of Keplerian and Newtonian models for each dataset.

For dataset A, 99 stable Keplerian orbits and 74 stable Newtonian orbits were found among 5838 models, each with a chi-square value falling within the $90\%$ confidence level of the best-fitting curve ($\chi_{\nu,best}^2 = 5.26$) during optimization Run 1. Subsequently, following optimization Run 2, 153 stable Keplerian orbits and 93 stable Newtonian orbits were identified out of 8123 models, all meeting the same criteria.

Similarly, in dataset B, after optimization Run 1, we found 198 stable Keplerian orbits and 171 stable Newtonian orbits among 6055 models, each with a chi-square value lying within the $90\%$ confidence level of the best-fitting curve ($\chi_{\nu,best}^2 = 5.76$). The greatest number of stable solutions was discovered after optimization Run 2 of dataset B, revealing 454 stable Keplerian orbits and 423 stable Newtonian orbits among 8216 models, all having reduced chi-square values within the $90\%$ confidence level of $\chi_{\nu,best}^2 = 5.26$.

The lifetime distributions of $P_b$, $P_c$ (subplots b and d of Figs. \ref{fig:lifetime_distributions_rA1}-\ref{fig:lifetime_distributions_rB2}) reveal the existence of a constrained inner orbit with period $P_b$ = 7 yr and an unconstrained outer orbit with periods in the range of 3:1 to 7:1 MMR. The same conclusion about the constrained (unconstrained) inner (outer) orbits can be deduced by the histograms of the Keplerian (Figure \ref{fig:hist_Keplerian}) and Newtonian (Figure \ref{fig:hist_Newtonian}) models for both datasets.

Both datasets seem to constrain the inner planet to a nearly circular orbit with a mean period of $P_{b}\sim7$ yr, while the outer companion has a circular orbit for either dataset and optimization run. When comparing the mean outer periods across datasets and optimization runs, a noticeable deviation of almost 27 years was observed between the two datasets of optimization Run 1 (Figure \ref{fig:hist1-Po-Keplerian}, Figure \ref{fig:hist1-Po-Newtonian}), while this discrepancy decreases to a deviation of at least seven years for the results of optimization Run 2 (Figure \ref{fig:hist2-Po-Keplerian}, Figure \ref{fig:hist2-Po-Newtonian}). Nevertheless, the well-defined peak in the outer period histograms of dataset A suggests a more constrained outer period with a mean value of $P_{c}\sim23$ yr, in contrast to the broader distributions of dataset B with a mean value of $P_{c}\sim40$ yr.

The N-body osculating initial conditions derived from the kinematic (Keplerian) fits provide $\chi_{\nu}^2$ values that are comparable with the Jacobian solutions. Figure \ref{fig:Kfit_Nfit} illustrates a kinematic model and its formal transformation to the N-body Cartesian osculating frame, from which Figure \ref{fig:Jacobian-Newtonian} was produced. Similarly, the refined self-consistent N-body solutions result in $\chi_{\nu}^2$ values comparable to those of the Keplerian fits. The only exception is the $\chi_{\nu,max}^2$ solutions of optimization Run 1 (Tables \ref{tab:Keplerian_run1}, \ref{tab:Newtonian_run1}), which is justified because the eccentricities of self-consistent N-body models oscillate, in contrast to their fixed values in Run 1. Therefore, the kinematic solutions and the transformed kinematic initial conditions are as consistent with the observations as the self-consistent Newtonian fits. 

We investigated further the orbital stability of the eight critical solutions ($\chi_{\nu,min}^2$, $\chi_{\nu,max}^2$ of both datasets and all optimization runs) by integrating each one for 100 Myr. All orbits were found to be stable in this timescale, which is considered the lifetime of the PCEB phase of sdB binaries \citep{King2002, Schreiber2003}. In Fig.~\ref{fig:orbital_evolution}, we plot indicatively the orbital evolution of eccentricities and semi-major axes, initialized by a Keplerian solution (Table \ref{tab:Keplerian_run1},  dataset B,  $\chi_{\nu,min}^2$), over 100 Myr.

Additionally, in Fig.~\ref{fig:dsB-run1-Kfit_fold}, we display the LTTE effect of this indicative solution for the first 152 years, phase folded in the ETV diagram. Each synthetic curve corresponds to an integration time equal to the outer period (38 years). The black synthetic curve is the Keplerian (kinematic) fit, which appears to be in absolute agreement with the dynamical fit for the first 38 years (green curve). The rest of the synthetic curves are also in close agreement with the initial Keplerian fit. This allowed us to test the validity of the Keplerian fits to the data.

Finally, in Fig. \ref{fig:mmr_evolution} we plot the temporal evolution of the mean-motion resonant angles $\theta_{1} = 3\lambda_{o} - \lambda_{i} - 2\varpi_{i}$, $\theta_{2} = 3\lambda_{o} - \lambda_{i} - 2\varpi_{o}$, and $\theta_{3} = 3\lambda_{o} - \lambda_{i} - \varpi_{i} - \varpi_{o}$ \citep{BeaugeFerrazMicht2003,Michtchenko2008} for the first 10000 years of a stable solution close to 3:1 MMR (Table \ref{tab:Keplerian_run1}, $\chi_{\nu,min}^2$, dataset A). It is apparent that these critical angles librate around zero degrees, indicating a dynamical resonant system. 

\subsection{Alternative explanations for ETV}

We have investigated if the cyclic period modulation can also be driven by magnetic activity cycles \citep{Applegate1992} of the secondary component of NSVS 14256825, as it is a convective red dwarf star with an effective temperature of 3077 K and a mass of $0.095 M_{\odot}$ \citep{NehirBulut2022}. Using the online tool of \citet{Volschow2016}$\footnote{\url{http://theory-starformation-group.cl/applegate/index.php}}$, we calculated the energy demand ($\Delta E / E_2$) for two different models: (i) the finite-shell two-zone model  \citep{Volschow2016} and (ii) the thin-shell model \citep{Tian2009}. We also calculated the requirements for the spin-orbit coupling model \citep{Lanza2020}. To explain the observed ETV modulation, the magnetic activity of the secondary must meet the criterion $(\Delta E/E_2<<1)$ for the energy demand. All the calculated energy ratios for a period modulation of 7 to 48 yrs 
were found to be larger than the threshold. 
Thus, we investigated the requirements for the spin-orbit coupling model (Applegate-Lanza) to account for residual changes in the O-C stable model fits. For the two datasets, A and B, these are 22s and 43s over T=18 yr and T=26 yr, respectively. Following the rationale of the equations of \cite{ Mai2022} and \citet{Lanza2020}, the change in rotational energy $\Delta E_{rot} \sim -(2-3) \times 10^{32} J$ is smaller than the maximum energy available $(4.4-6.3) \times 10^{32} J$ from luminosity changes over the timescale for O-C variability in the secondary star $E_{max}= L_{2} \times T$. Therefore,  $\Delta E_{rot}/ E_{max}$ is $\sim 0.5$, and it could account for the residual changes in the ETV model fits. Given the spectral type M5-M8 for the secondary of NSVS 14256825, this is in agreement with the conclusion of \cite{2016MNRAS.460.3873B} that as magnetic activity decreases toward a later spectral type, we expect to drive much smaller O-C variations. Nevertheless it is still a high number, meaning that half of the energy generated by the secondary star should account for the non-axisymmetric gravitational quadrupole moment. We also calculated the tidal synchronization timescale following the assumptions and equations of \citet{Lanza2020} and found $t_{s}=3.4-34 $ yr for the dimensionless tidal quality factor $Q=10^5-10^6$ that respectively characterizes the efficiency of tidal energy dissipation in the active (secondary) component. As these times are similar to the observed O-C changes, the timescale for spin-orbit coupling may be determined by tidal synchronization instead of by magnetic activity timescales. 

\section{Discussion and conclusions}
\label{sec:conclusions}

We implemented a grid search approach consisting of two sets of optimization runs (Run 1: for fixed eccentricities, Run 2: adjusting all resulting parameters from Run 1) in the $e_b, e_c$ plane utilizing a combination of the Nelder-Mead Simplex and Levenberg-Marquardt algorithms. The goal was to find the best-fitting curve ($\chi_{\nu,best}^2$) in the ETV diagram of two datasets (dataset A, dataset B), implementing a revised kinematic (Keplerian) formulation of the LTTE effect and self-consistent N-body (Newtonian) models.
Additionally, we ran three-body simulations of the resulting solutions that lie within the $90\%$ confidence interval of the best-fitting curve, searching for stable ($\tau$=1Myr) and non-chaotic ($\langle Y \rangle\sim2$) orbits.

As a result, hundreds of stable orbits, with a lifetime of at least 1 Myr, were found for two circumbinary bodies of substellar mass within the $90\%$ confidence interval of the best-fitting curve. However, none of the Keplerian and Newtonian $\chi_{\nu,best}^2$ solutions were dynamically stable for more than a few hundred years. The histograms and lifetime distributions of inner and outer periods revealed the existence of a constrained inner orbit with period $P_b$ = 7 yr, mass $m_b = 11 M_{Jup}$, and an unconstrained outer orbit with a period in the range of 3:1 to 7:1 MMR ($P_c = 20-50$ yr) and a mass range of $m_c = 11-70 M_{Jup}$. Recent ETV studies of multiple-planet systems have reported that orbits close to low-order MMRs are strongly unstable \citep{Horner2012, Gozdziewski2015, HinseHorner2014, Esmer2021} for timescales as short as thousands of years. The only exception based on ETV analysis is NY Vir \citep{Esmer2023},  with a period variation attributed to two putative Jovian planets close to 2:1 MMR, although this is uncertain due to the short time interval of the observations.

Furthermore, the observational timeline of ETV diagrams are usually narrow relative to the putative orbital periods, and as a consequence, the inferred masses of hypothetical planets reach the brown dwarf and the red dwarf limits \citep{HinseGodz2012, Gozdziewski2015}. Such cases may introduce significant deviations between synthetic signals derived from the Keplerian and osculating Newtonian elements, leading to qualitatively different best-fitting configurations constrained by the available data. In the case of NN Ser, the Newtonian fits of \citet{Marsh2014} are in qualitative agreement with the Keplerian fits of \citet{Beurmann2013}. However, the following analyses \citep{2016MNRAS.460.3873B,Pulley2022,NNSer2023MNRAS} of the system revealed that the initially stable system in 2:1 MMR does not fit the updated time measurements. Here, both datasets confirm the presence of a circumbinary planet with $P_{b}\sim7$ yr, but deviations in the outer period were observed:  $P_{c}\sim23$ yr for dataset A and $P_{c}\sim40$ yr for dataset B. This discrepancy suggests that the extended timeline of dataset A, spanning 26 years, may provide better constraints on the orbital parameters compared to the narrower 18-year time frame of dataset B.

In any case, the fitting curves of stable orbits are not perfectly fitted to the complete datasets in comparison to the best-fitting curves. The low precision NSVS and ASAS data seem to follow the general trend of strong peak-to-peak amplitude reduction over time, although the fitting curves of the minimum- and maximum-reduced chi-square of stable solutions do not fit well within their respective time frames (1999-2007). Moreover, both datasets exhibit visually similar fits for the minimum- ($\chi_{\nu,min}^2$) and maximum-reduced ($\chi_{\nu,max}^2$) chi-square of stable solutions in the 2007-2024 interval, but they do not fit perfectly through the latest data of 2021-2024. In general, the self-consistent N-body models account for mutual gravitational interactions between the circumbinary bodies, which could affect the ETV trends. We showed that the temporal variation of the mid-eclipse times of the binary, as predicted by a best-fitting dynamical two-planet model, are in close agreement with the initial Keplerian fit. In addition, our Keplerian fits produce qualitatively the same results as the Newtonian ones. As a result, we attribute these ETV residuals to the sparse data of the corresponding time frames and/or additional mechanisms (Applegate-Lanza, spin-orbit coupling model). We applied the model of \citet{Lanza2020}, and although it depends on poorly constrained binary system parameters, such as the tidal energy dissipation and the internal mass distribution of the stars, we found that it may contribute to the observed ETV, apart from the potential circumbinary bodies, revealing the complex nature of the orbital structure of NSVS 14256825, as in the case of HS 0705+6700 \citep{Mai2022} and V471 Tau \citep{2022MNRAS.517.5358K}. 

The plentiful stable solutions that were found, coupled with their relatively well-fitted ETV curves, could serve as evidence supporting the existence of two circumbinary companions of substellar mass orbiting NSVS 14256825. We also tested the stability of the eight critical stable solutions ($\chi_{\nu,min}^2$, $\chi_{\nu,max}^2$ of both datasets and of both optimization runs) for an integration interval of 100 Myr. All of these solutions were found to be stable in this timescale, which is considered to be the lifetime of the PCEB phase of sdB binaries.

We could not suggest one conclusive model of two circumbinary planets since one cannot rely solely on the statistics of a reduced chi-square. Regarding the reduced chi-square for each dataset and optimization run, the eight critical stable solutions outline the extrema of a group of solutions. However, within these limits exist hundreds of solutions with various parameter values that are not centered around a best-fit solution. Dataset A seems to constrain the model parameters for both companions, but the fitting curves of the critical stable solutions are far from satisfactory for the earliest NSVS and ASAS data. On the other hand, dataset B provides more satisfactory fits, but there are cases where the mean outer period exceeds 30 years in both optimization runs and the inferred masses reach values greater than 30 $M_{Jup}$.

Among the ten PCEBs exhibiting LTTE signals \citep[for the complete list, refer to][]{Pulley2022}, half of them (HW Vir, NY Vir,  NN Ser, HS 0705+6700, Kepler-451) exhibit two or more cyclic variations attributed to circumbinary planets. Their ETV and dynamical stability analyses share some common features, among which are the continuously updated LTTE solutions, peak-to-peak amplitude reduction \citep{Pulley2022}, and stable regions of higher $\chi^2$ than the best-fit solution \citep{Potter2011, Dreizler2012, Esmer2022, NNSer2023MNRAS}. From our analysis, it seems that NSVS 14256825 is not an exception. 

We examined the potential for detecting the planets of ETV using Gaia \citep[DR3;][]{2023A&A...674A...1G} astrometry. The detection limit of an outer planet can be calculated by following the formalism of \cite{2015MNRAS.447..287S} and the NSVS 14256825 distance, 752.6 pc according to the Gaia DR3 parallax measurements. Among the stable models, using the masses and the periods of, for example, the $\chi_{\nu,max}^2$ dataset B (Run2) solution ($m_{b}=11.12 M_{J}$, $P_{b}=7.04$ yr, $m_{c}=8.57 M_{J}$, $P_{c}=24.91$ yr), the gravitational pull of the planet will displace the system’s barycenter, with a semi-major axis of $a_{b} \sim 19$ mas and $a_{c} \sim 34$ mas, respectively. This represents the planet's astrometric signature, which to be detectable must be greater than the scan length accuracy per optical field pass $\sigma_{fov}$ corresponding to the luminosity G \citep{2014ApJ...797...14P}. For the NSVS 14256825 (G = 13.22) $\sigma_{fov}\sim 42 \mu$as, and thus Gaia can detect the planets of NSVS 14256825.

As recently proposed for the case of HW Vir by \cite{Baycroft2023}, Gaia full-epoch astrometry (circa 2025) will provide an alternative method to ETVs for confirming circumbinary planets of PCEBs. In the meantime, it is essential to continue monitoring the system in order to expand the observational timeline with new data, providing further validation and refinement of the proposed solutions.

\begin{acknowledgements}
		We thank the anonymous referee for numerous constructive comments and suggestions that greatly improved this paper.
\end{acknowledgements}

\bibpunct{(}{)}{;}{a}{}{,} % to follow the A&A style
\bibliographystyle{aa_url} % style aa.bst
\bibliography{aa_NSVS1425} % your references Yourfile.bib

\begin{appendix}

\section{Additional figures}

	\begin{figure*}[h!]
		\centering
		\subfloat[example]{\includegraphics[width=0.4\textwidth, keepaspectratio]{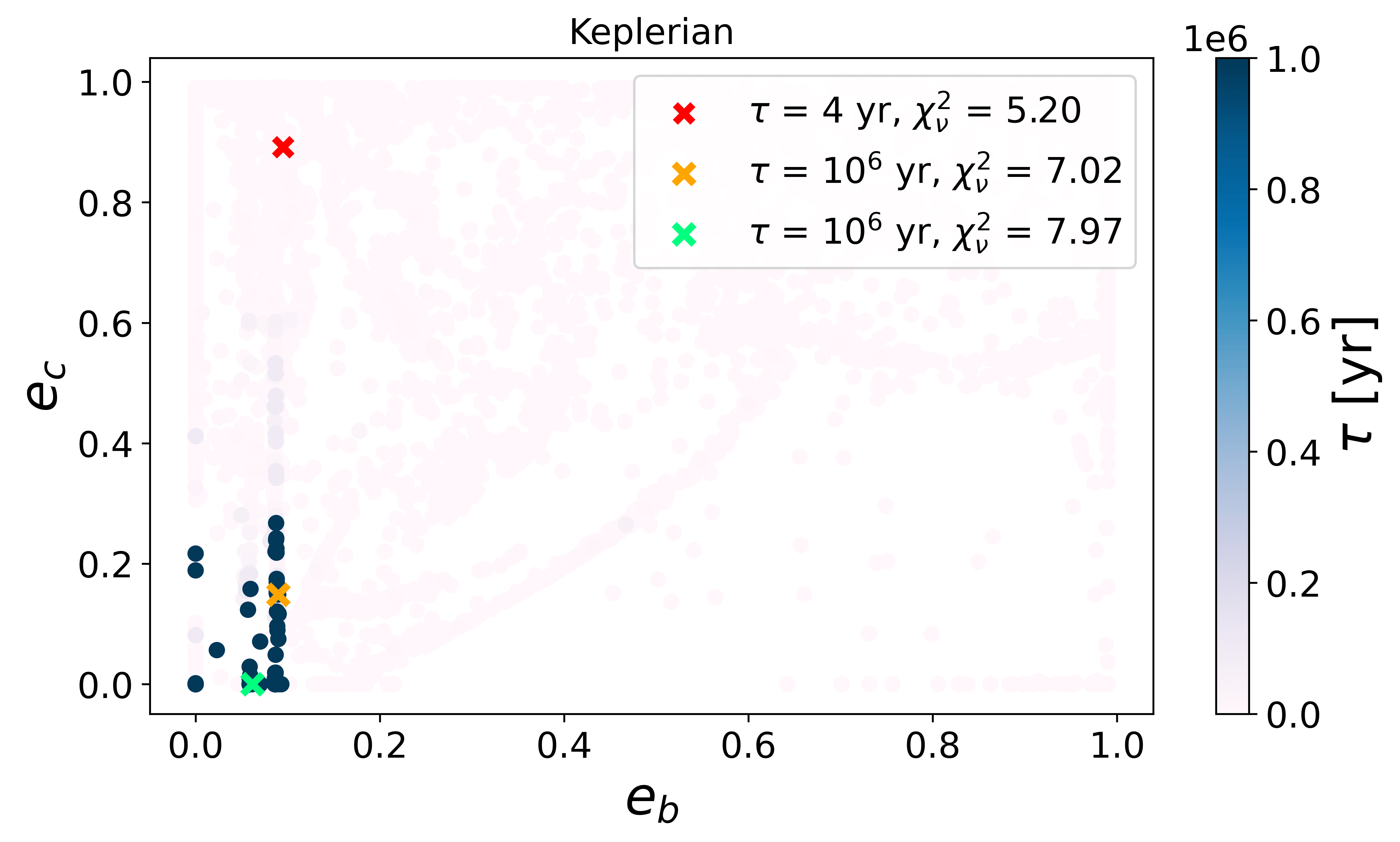}\label{fig:sbf_run2A_eieolt_Keplerian}}
		\subfloat[example]{\includegraphics[width=0.4\textwidth, keepaspectratio]{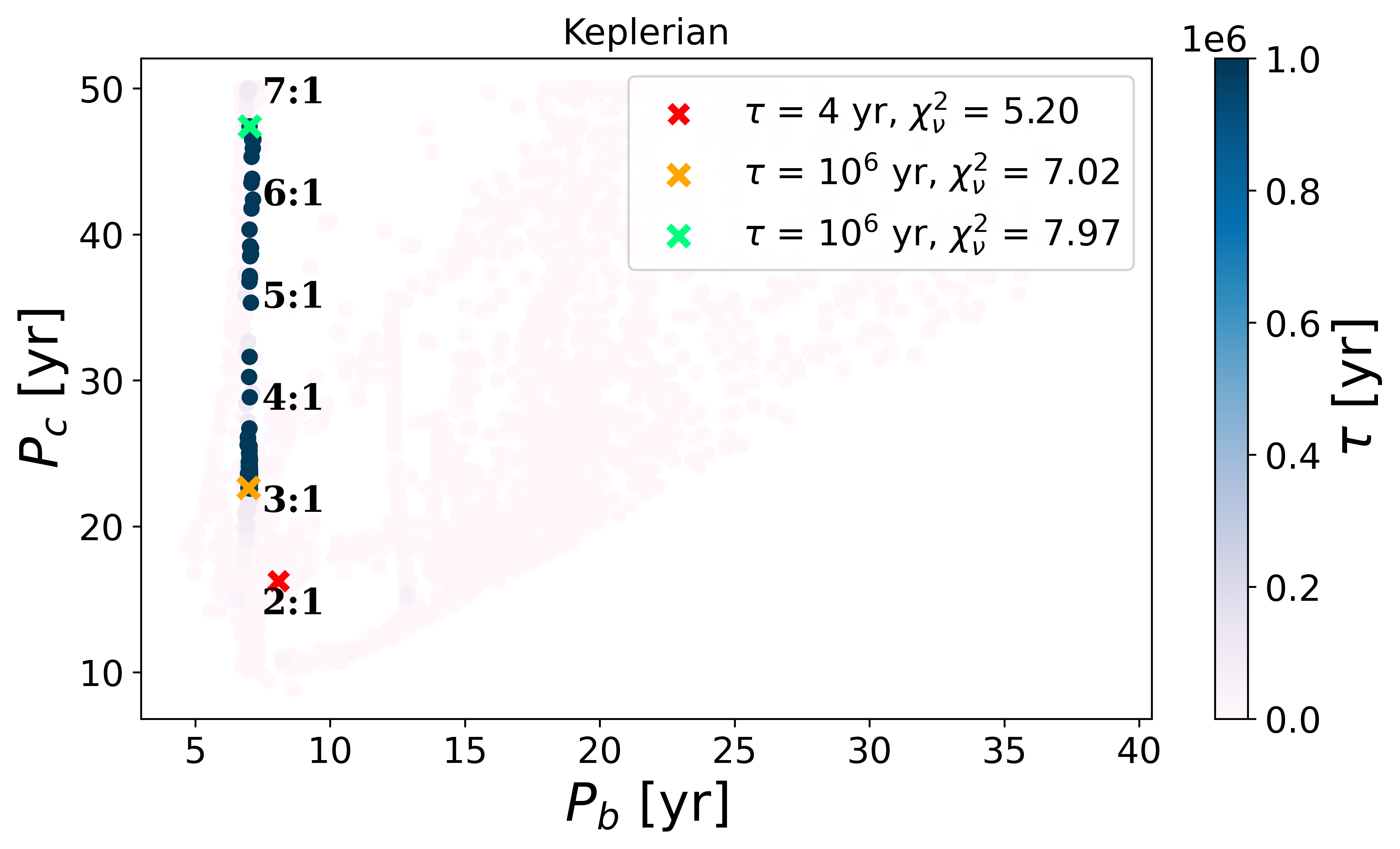}\label{fig:sbf_run2A_PiPolt_Keplerian}}\\
		\subfloat[example]{\includegraphics[width=0.4\textwidth, keepaspectratio]{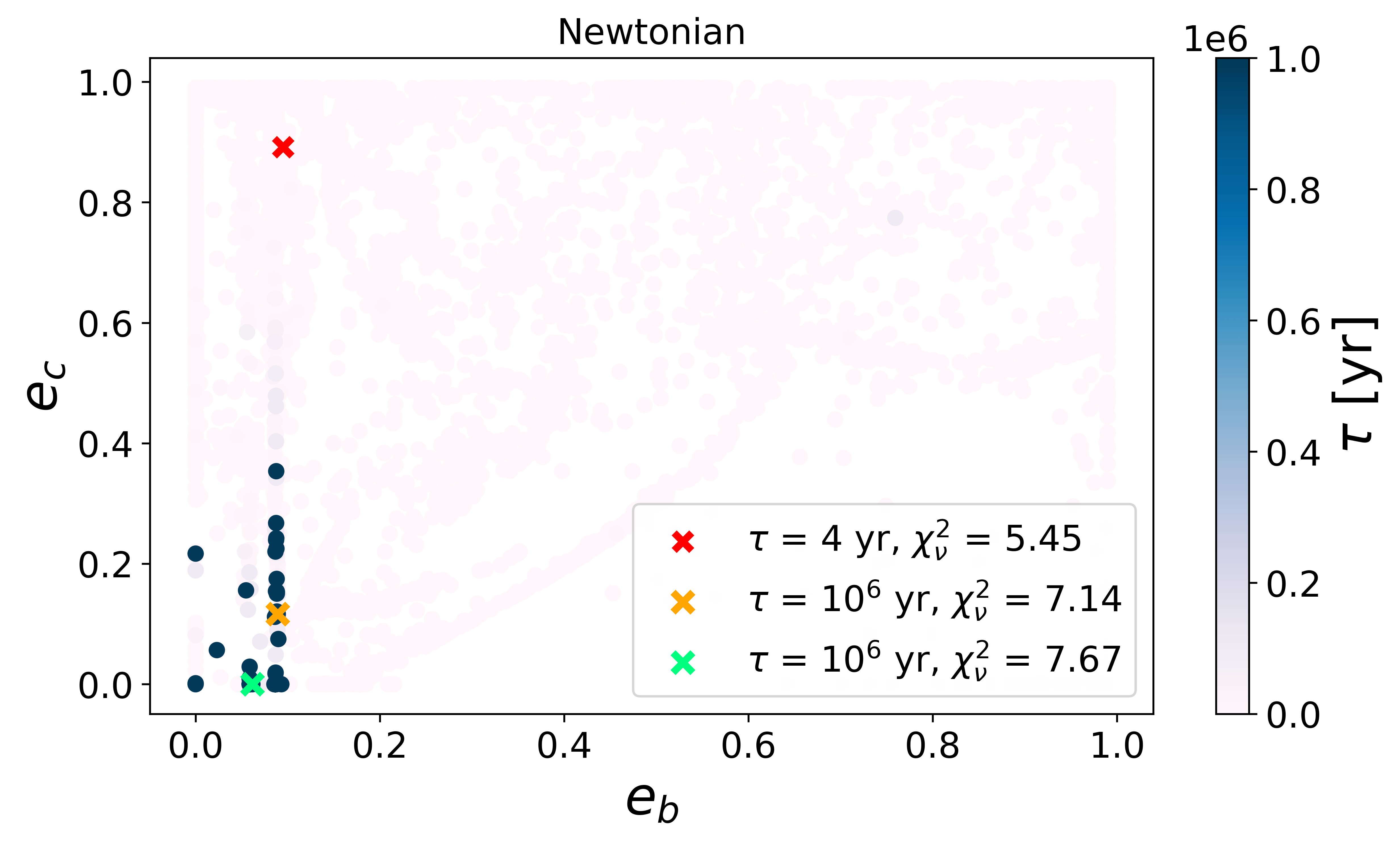}\label{fig:sbf_run2A_eieolt_Newtonian}}
		\subfloat[example]{\includegraphics[width=0.4\textwidth, keepaspectratio]{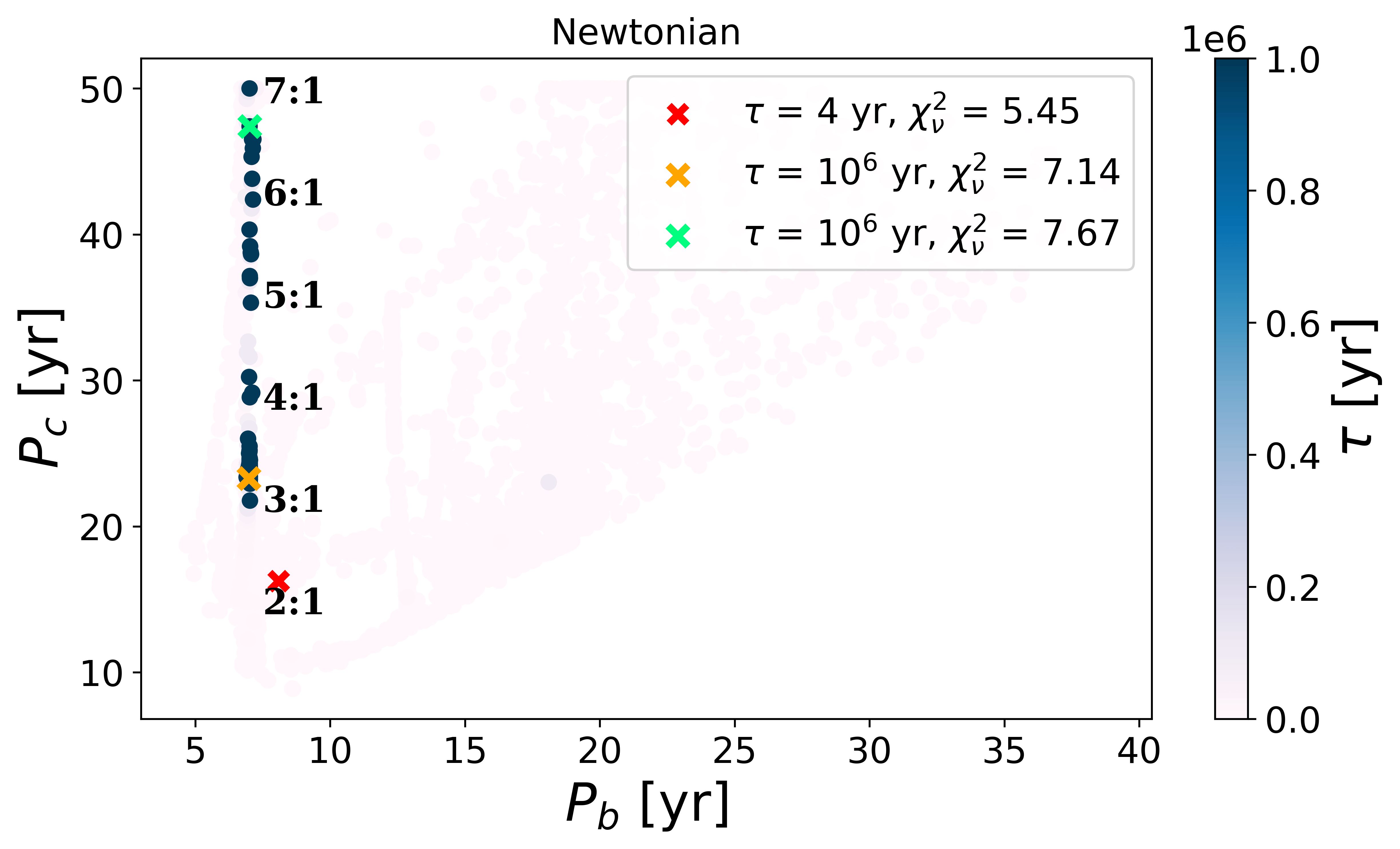}\label{fig:sbf_run2A_PiPolt_Newtonian}}
		
		\caption{Same as Figure \ref{fig:lifetime_distributions_rA1} but for 8123 Keplerian models (153 stable) within the $90\%$ confidence level of $\chi_{\nu,best}^2$ as they resulted from optimization Run 2 for dataset A (top) and for 8123 Newtonian models (93 stable, bottom).}
		\label{fig:lifetime_distributions_rA2}
	\end{figure*}
	
	\begin{figure*}[h!]
		\centering
		\subfloat[]{\includegraphics[width=0.4\textwidth, keepaspectratio]{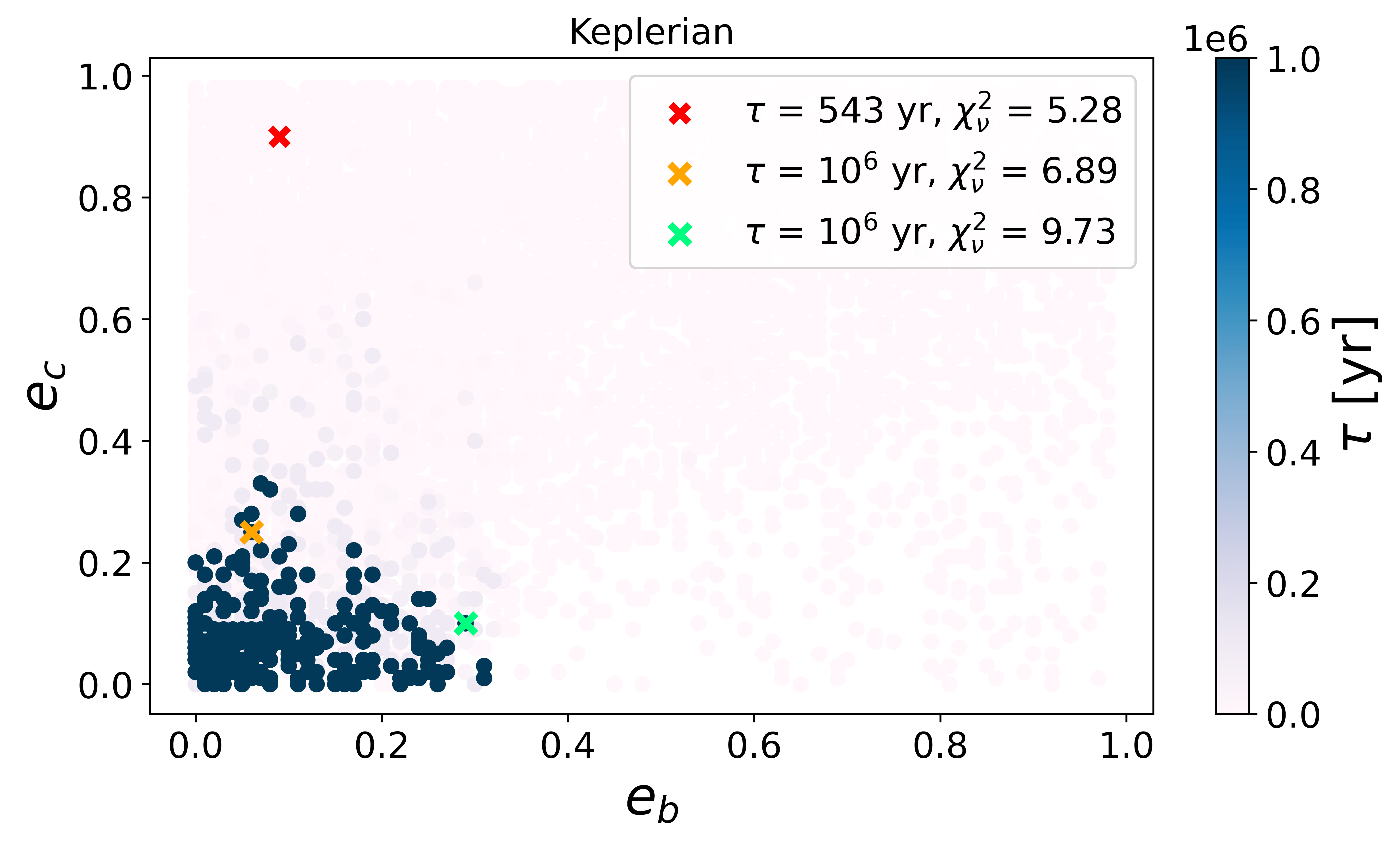}\label{fig:sbf_run1B_eieolt_Keplerian}}
		\subfloat[]{\includegraphics[width=0.4\textwidth, keepaspectratio]{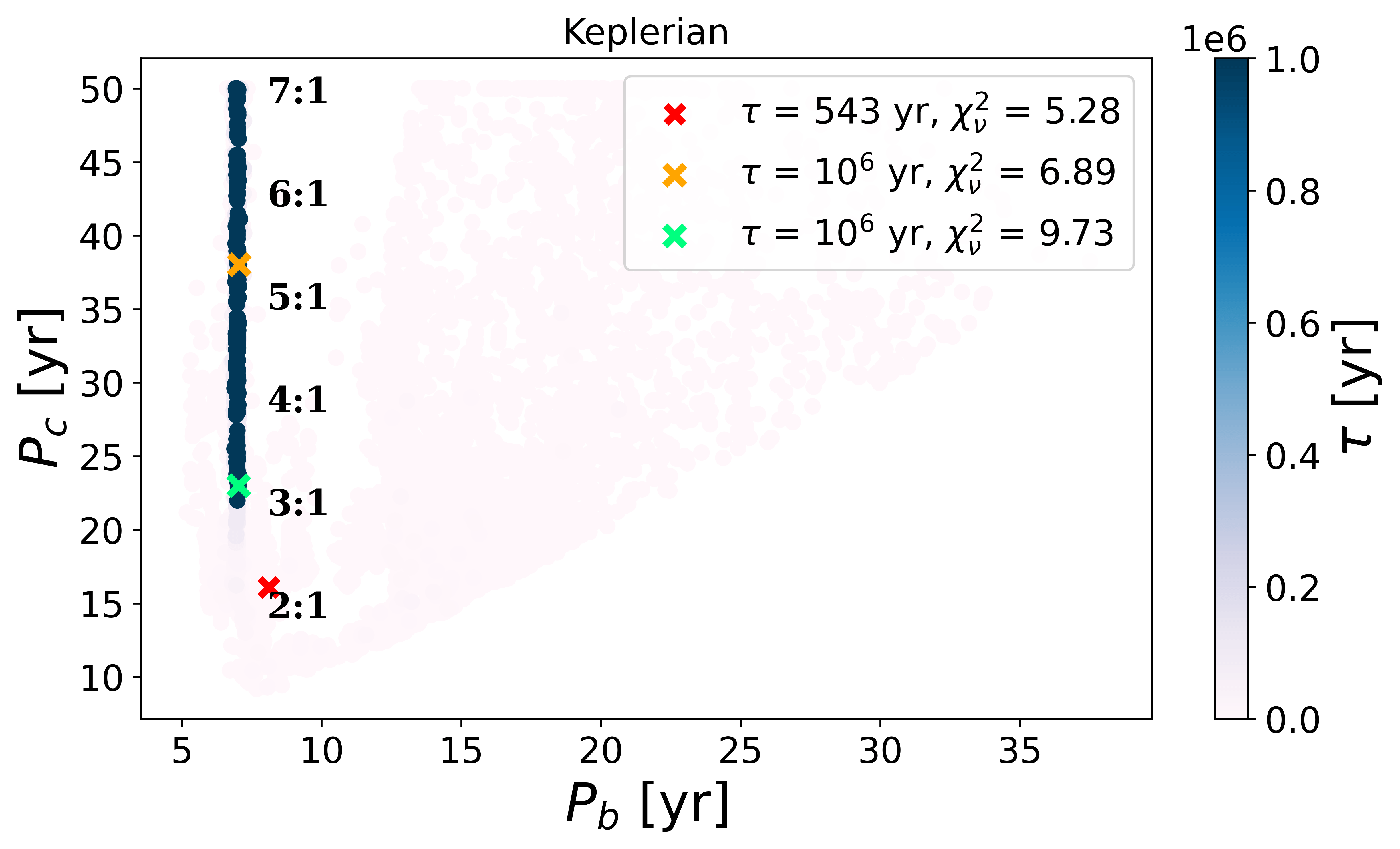}\label{fig:sbf_run1B_PiPolt_Keplerian}}\\
		\subfloat[]{\includegraphics[width=0.4\textwidth, keepaspectratio]{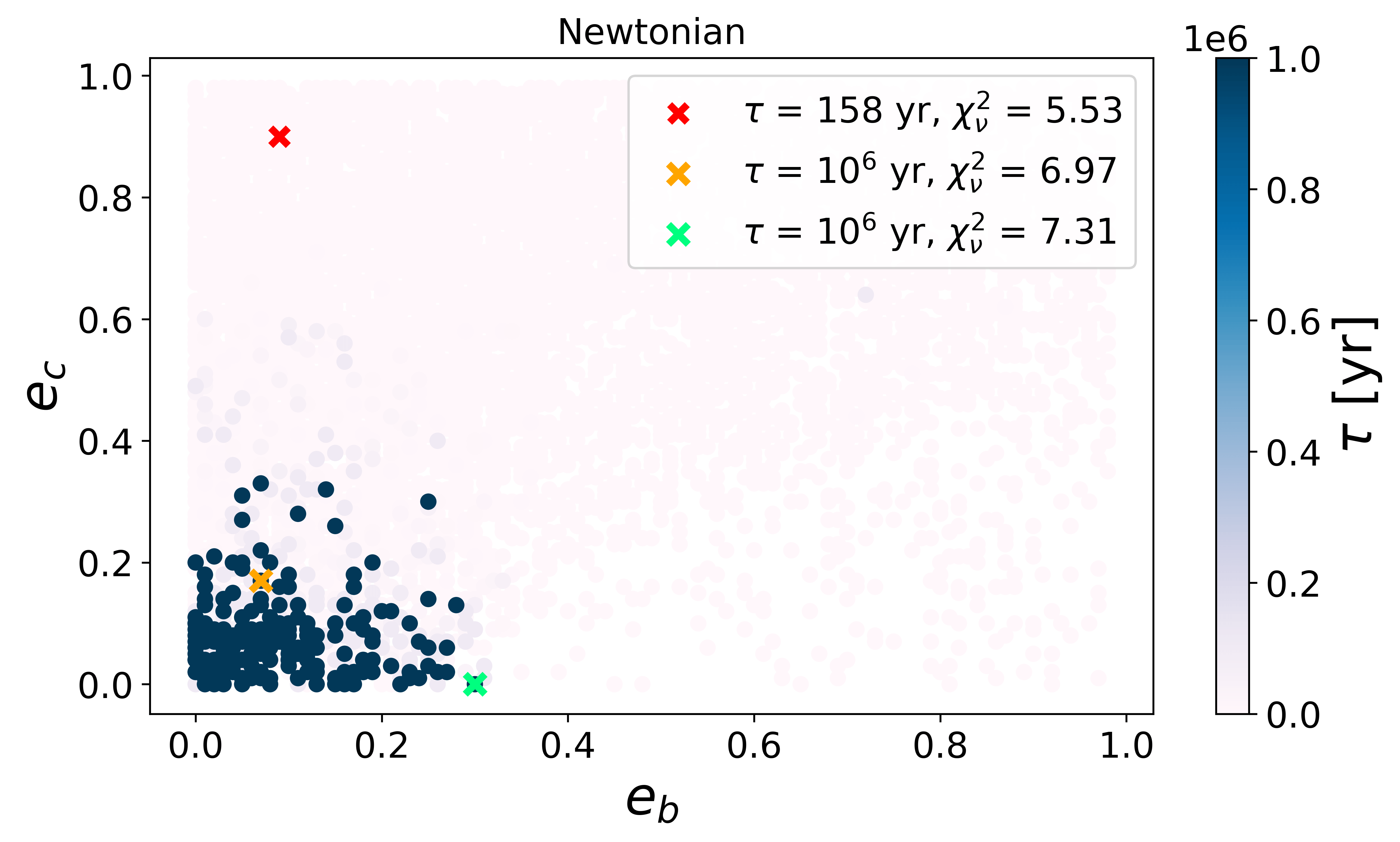}\label{fig:sbf_run1B_eieolt_Newtonian}}
		\subfloat[]{\includegraphics[width=0.4\textwidth, keepaspectratio]{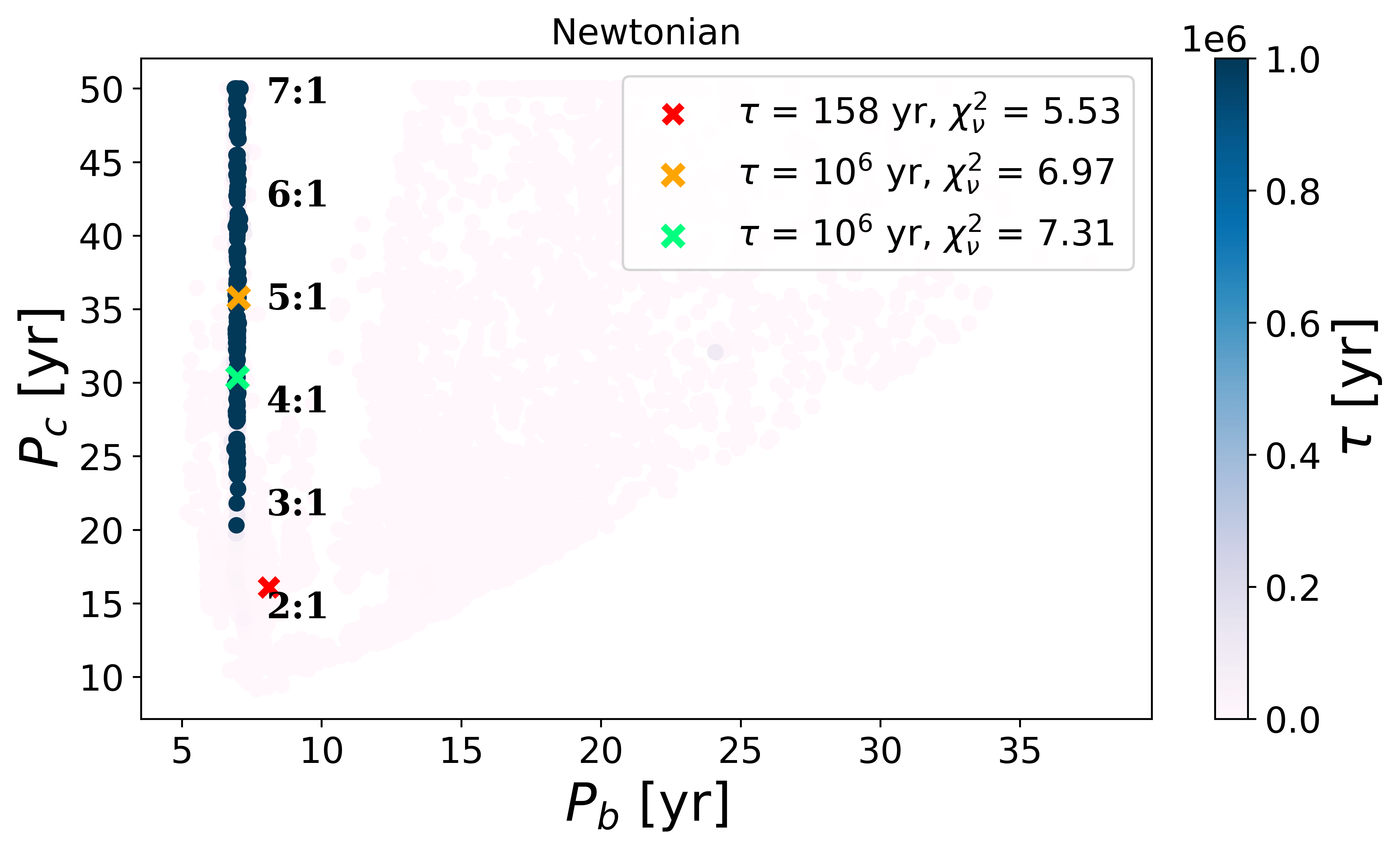}\label{fig:sbf_run1B_PiPolt_Newtonian}}
		
		\caption{Same as Figure \ref{fig:lifetime_distributions_rA1} but for 6055 Keplerian models (198 stable) within the $90\%$ confidence level of $\chi_{\nu,best}^2$ as they resulted from optimization Run 1 for dataset B (top) and for 6055 Newtonian models (171 stable, bottom).}
		\label{fig:lifetime_distributions_rB1}
	\end{figure*}
	
	\begin{figure*}[h!]
		\centering
		\subfloat[]{\includegraphics[width=0.4\textwidth, keepaspectratio]{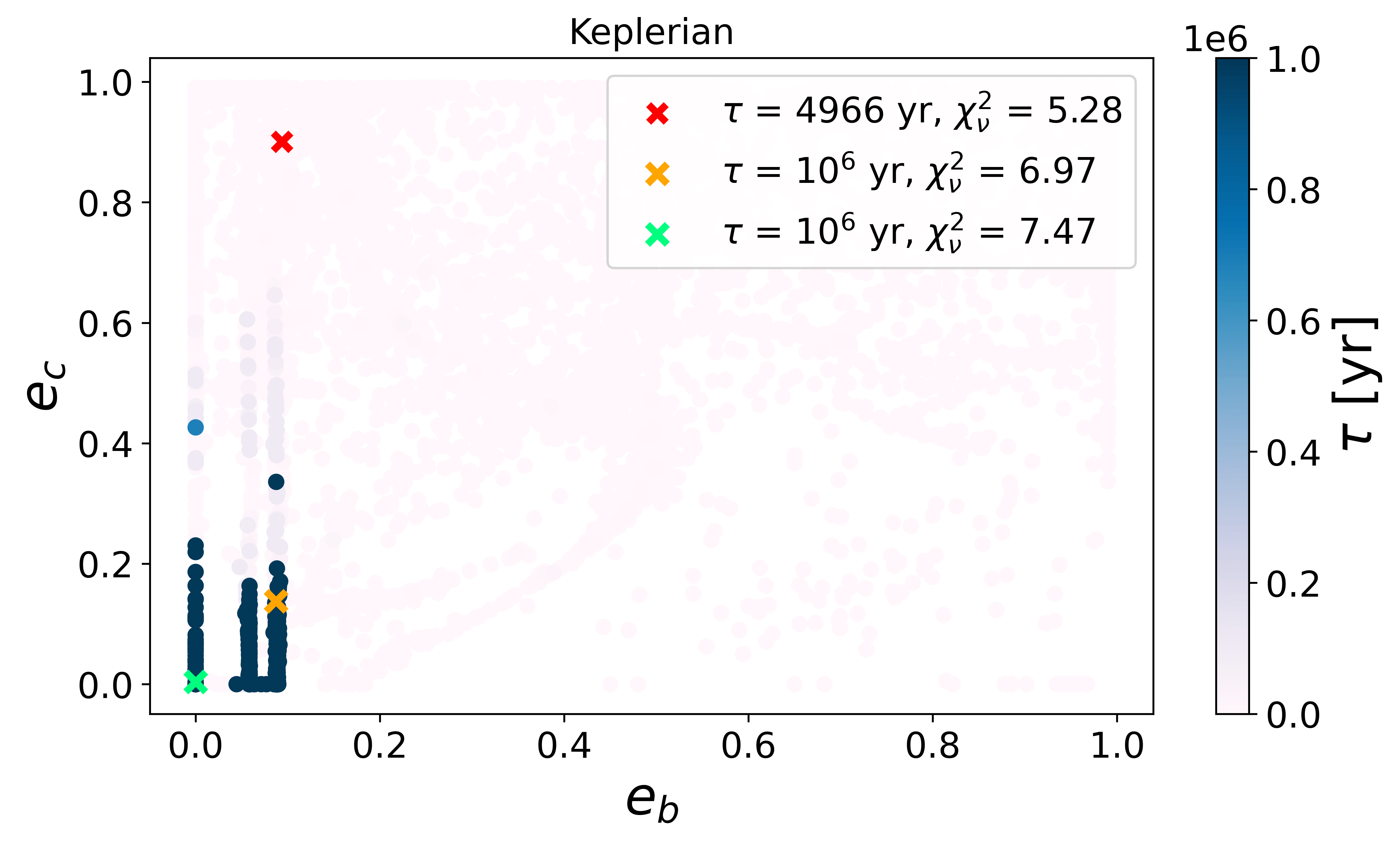}\label{fig:sbf_run2B_eieolt_Keplerian}}
		\subfloat[]{\includegraphics[width=0.4\textwidth, keepaspectratio]{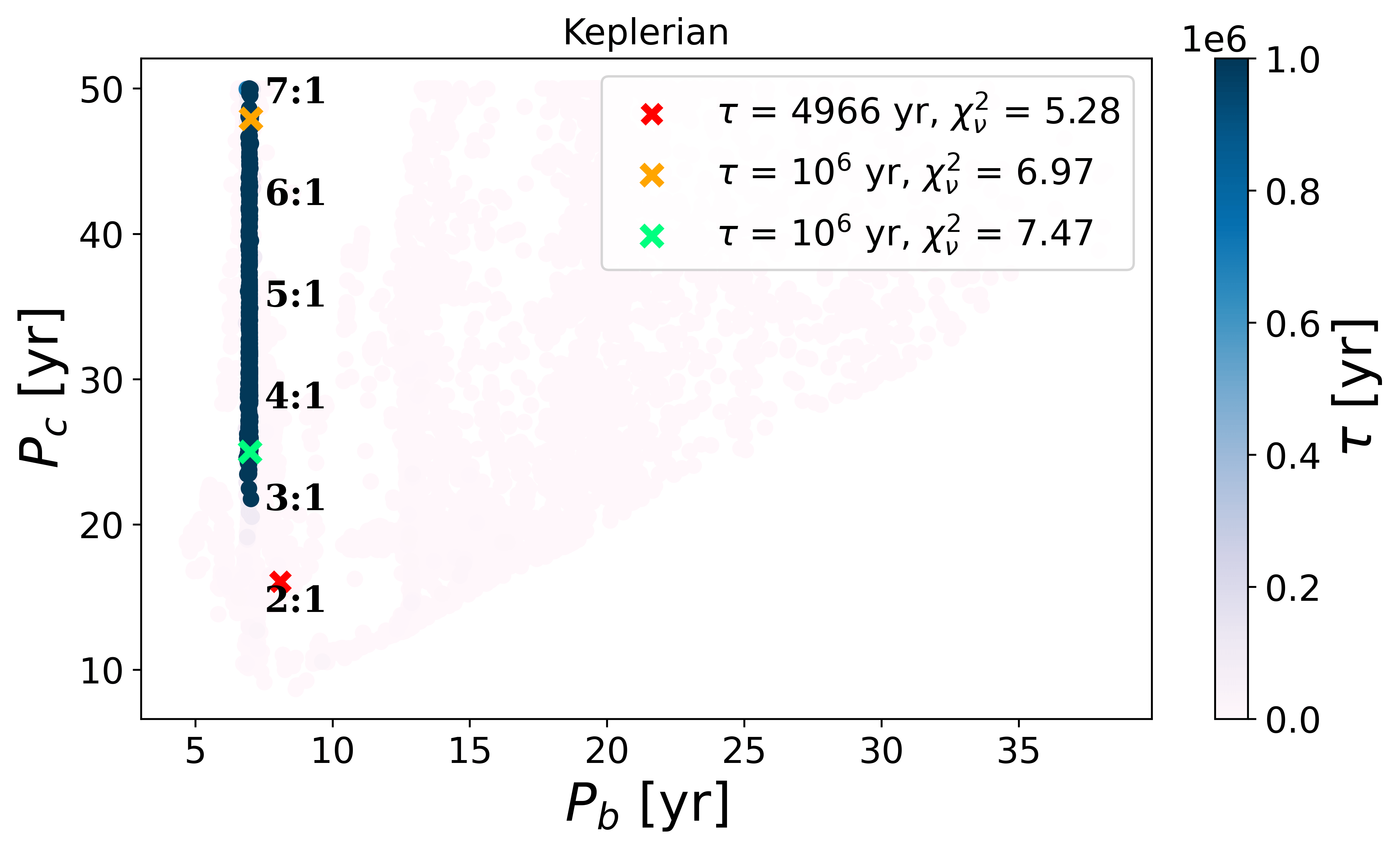}\label{fig:sbf_run2B_PiPolt_Keplerian}}\\
		\subfloat[]{\includegraphics[width=0.4\textwidth, keepaspectratio]{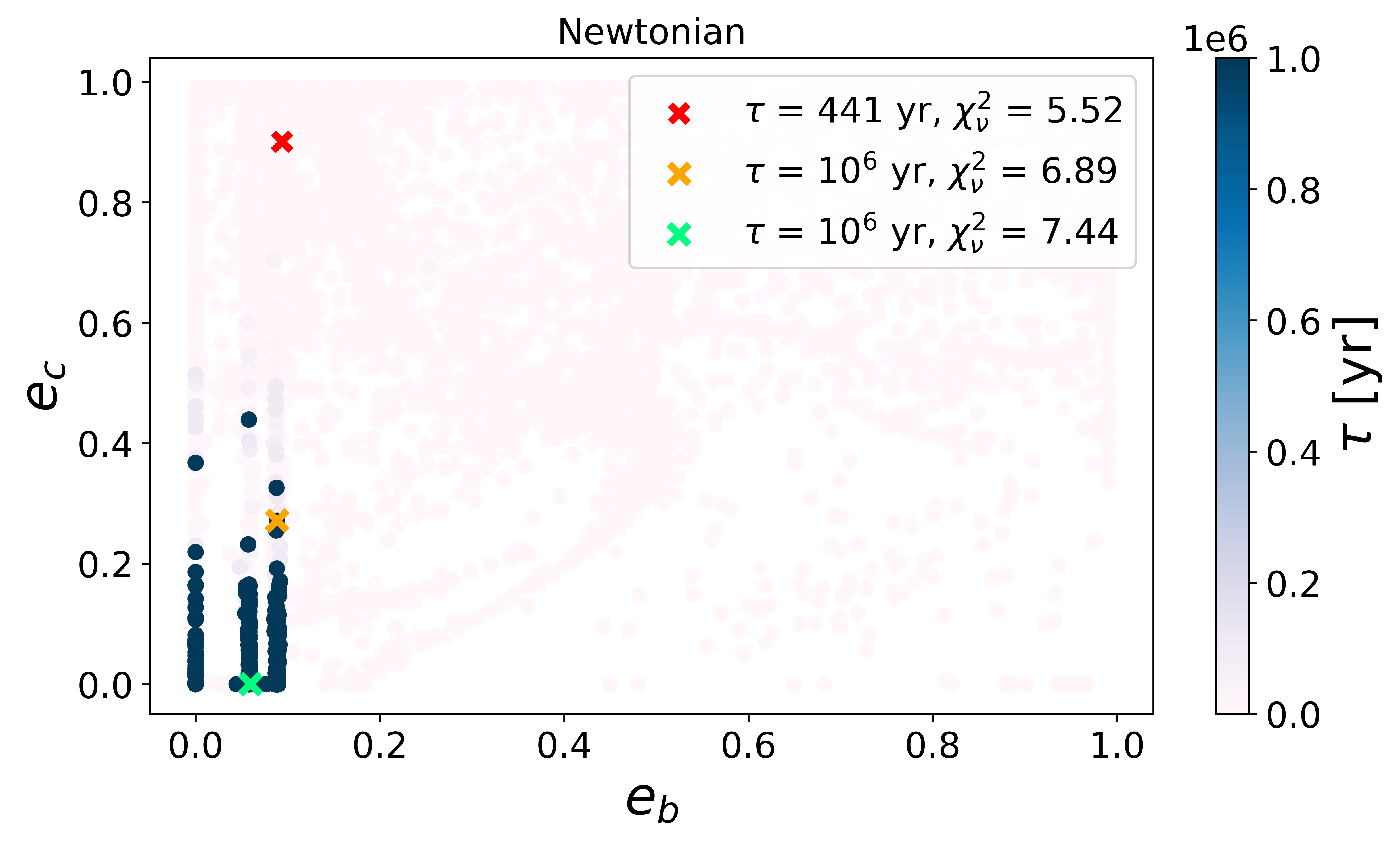}\label{fig:sbf_run2B_eieolt_Newtonian}}
		\subfloat[]{\includegraphics[width=0.4\textwidth, keepaspectratio]{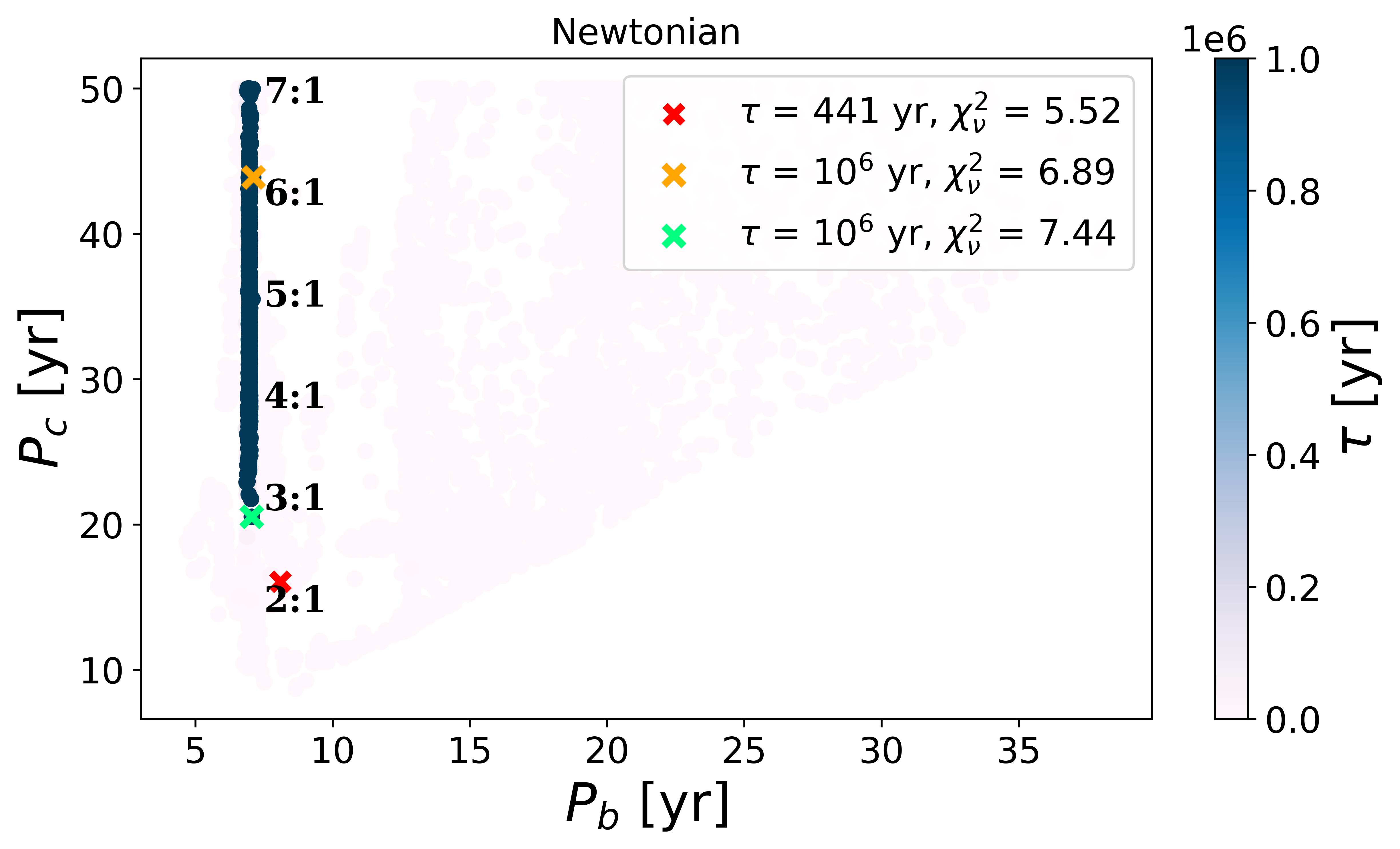}\label{fig:sbf_run2B_PiPolt_Newtonian}}
		
		\caption{Same as Figure \ref{fig:lifetime_distributions_rA1} but for 8216 Keplerian models (454 stable) within the $90\%$ confidence level of $\chi_{\nu,best}^2$ as they resulted from optimization Run 2 for dataset B (top) and for 8216 Newtonian models (423 stable, bottom).}
		\label{fig:lifetime_distributions_rB2}
	\end{figure*}
	
	\begin{figure*}
		\hspace{1cm}\subfloat[$\chi_{\nu,best}^2$ (dynamically unstable)]{\includegraphics[width=0.3\textwidth, keepaspectratio]{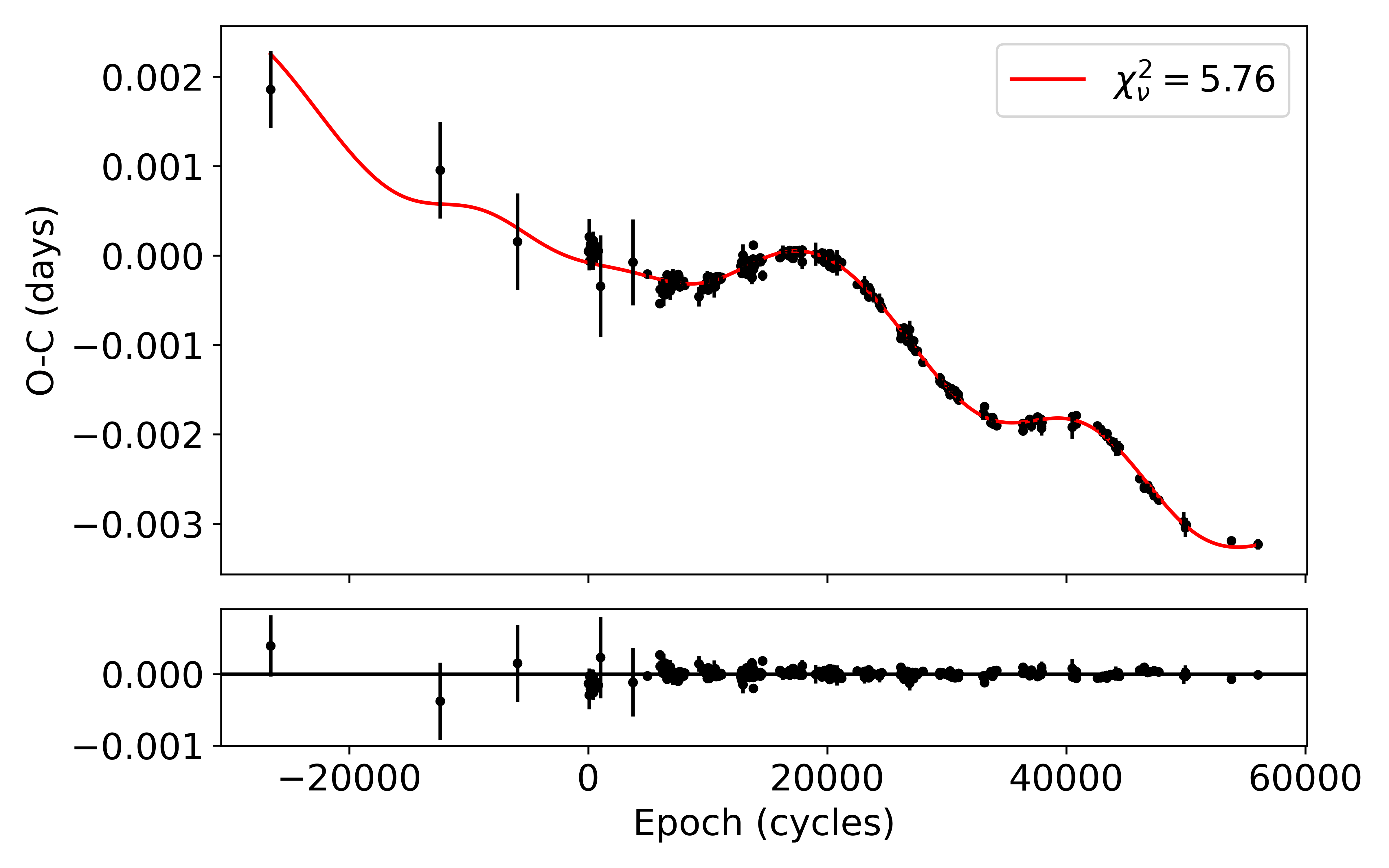}}
		\subfloat[$\chi_{\nu,min}^2$ (dynamically stable)]{\includegraphics[width=0.3\textwidth, keepaspectratio]{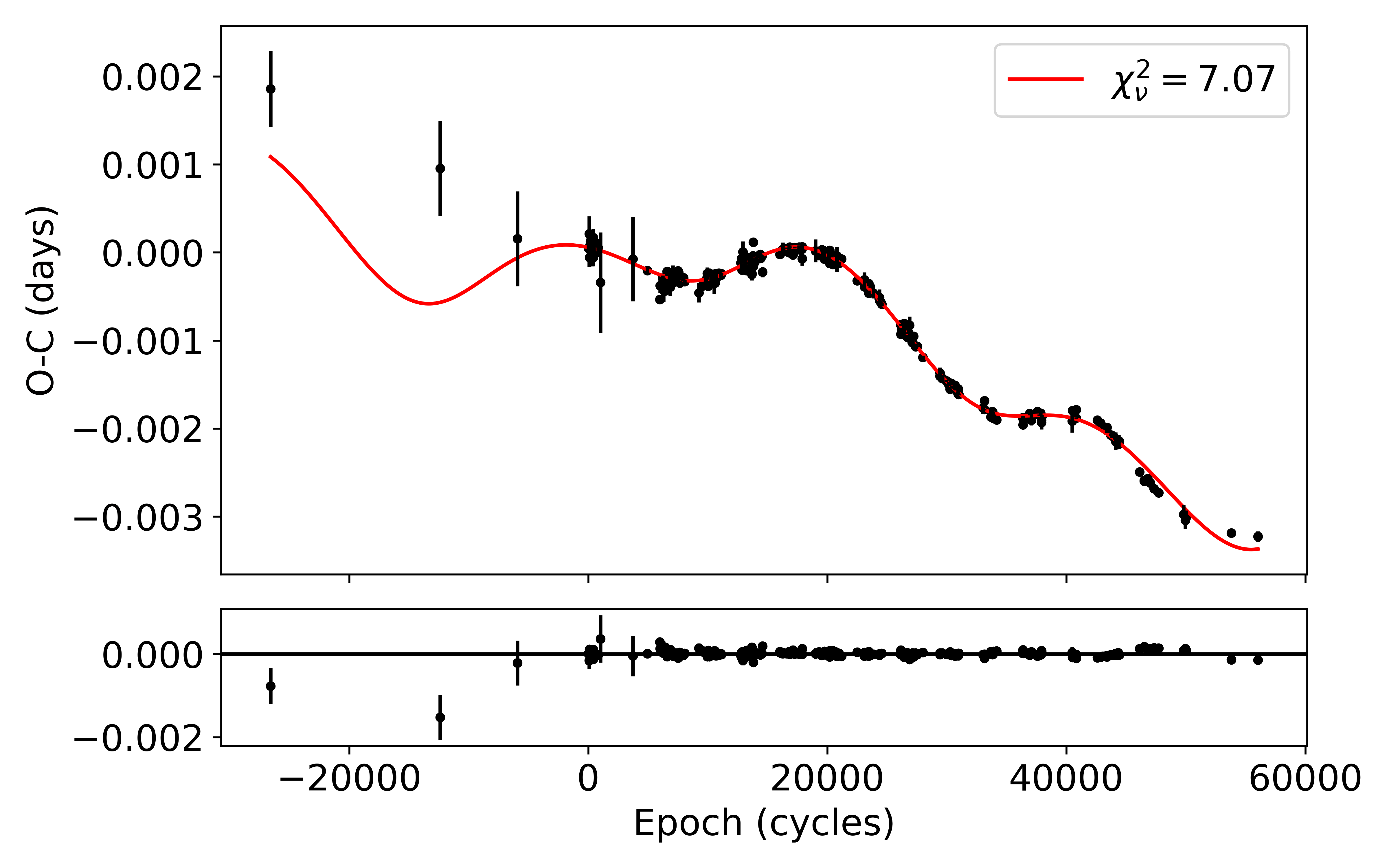}}
		\subfloat[$\chi_{\nu,max}^2$ (dynamically stable)]{\includegraphics[width=0.3\textwidth, keepaspectratio]{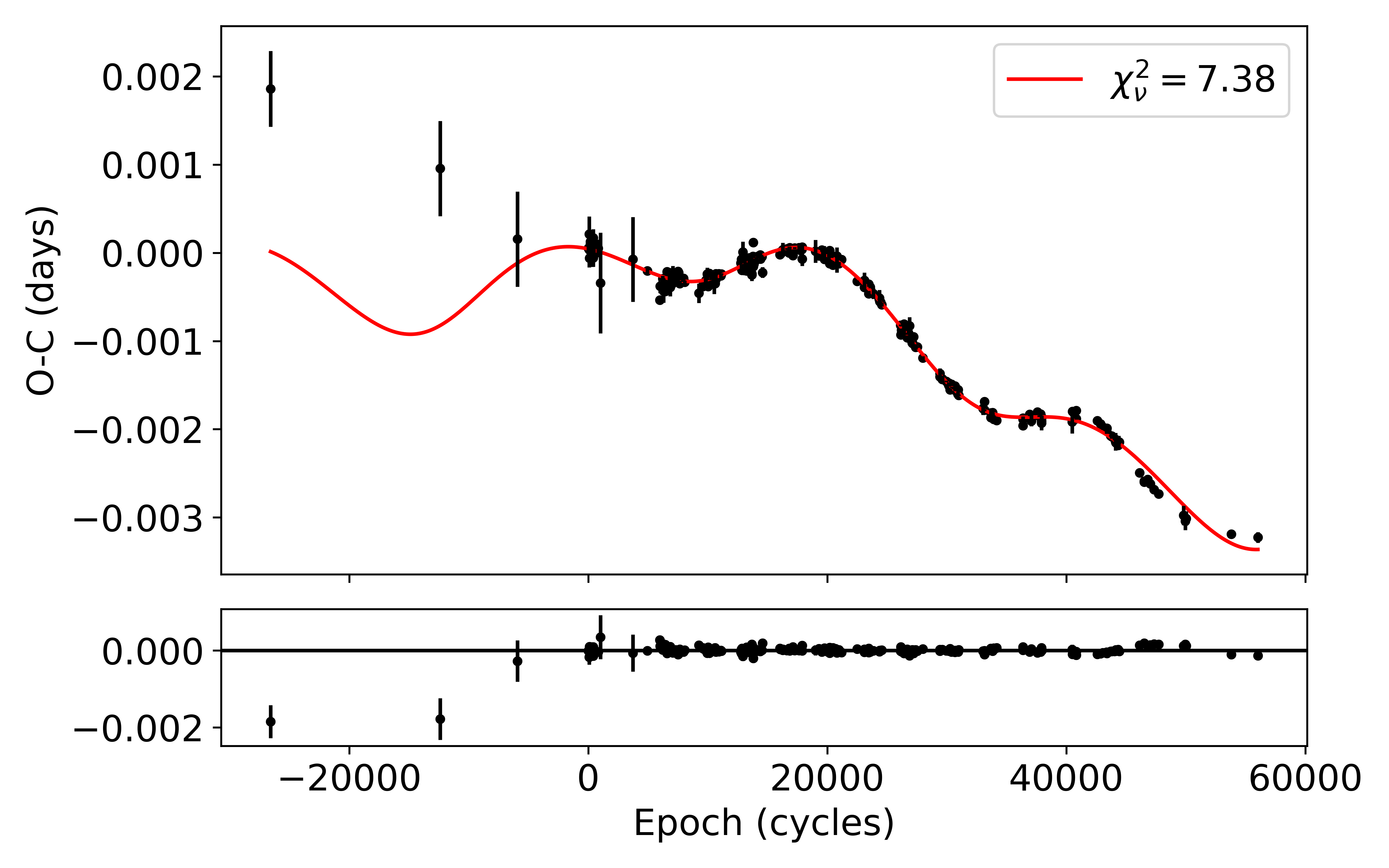}}\\
		
		\hspace{1cm}\subfloat[$\chi_{\nu,best}^2$ (dynamically unstable)]{\includegraphics[width=0.3\textwidth, keepaspectratio]{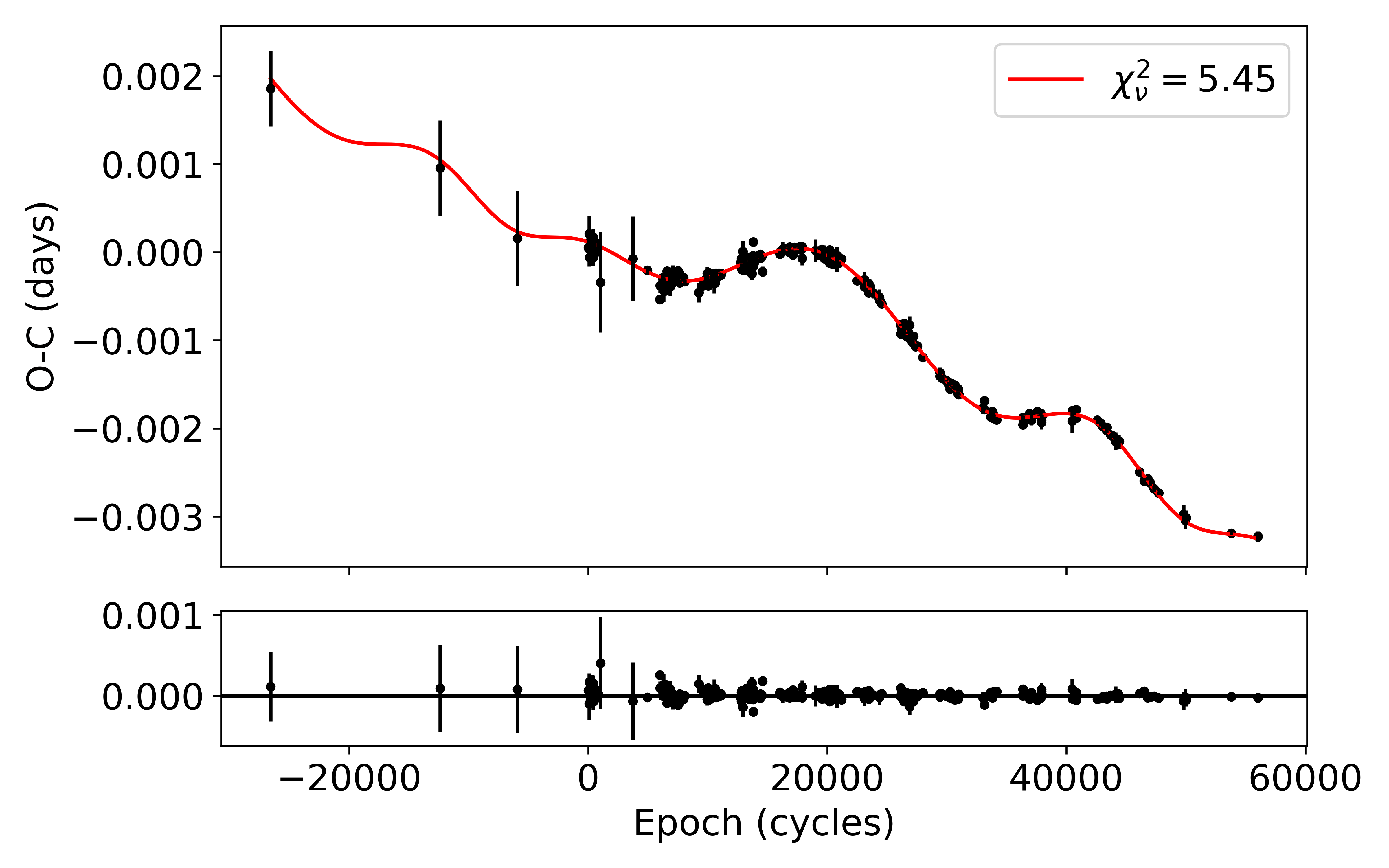}}
		\subfloat[$\chi_{\nu,min}^2$ (dynamically stable)]{\includegraphics[width=0.3\textwidth, keepaspectratio]{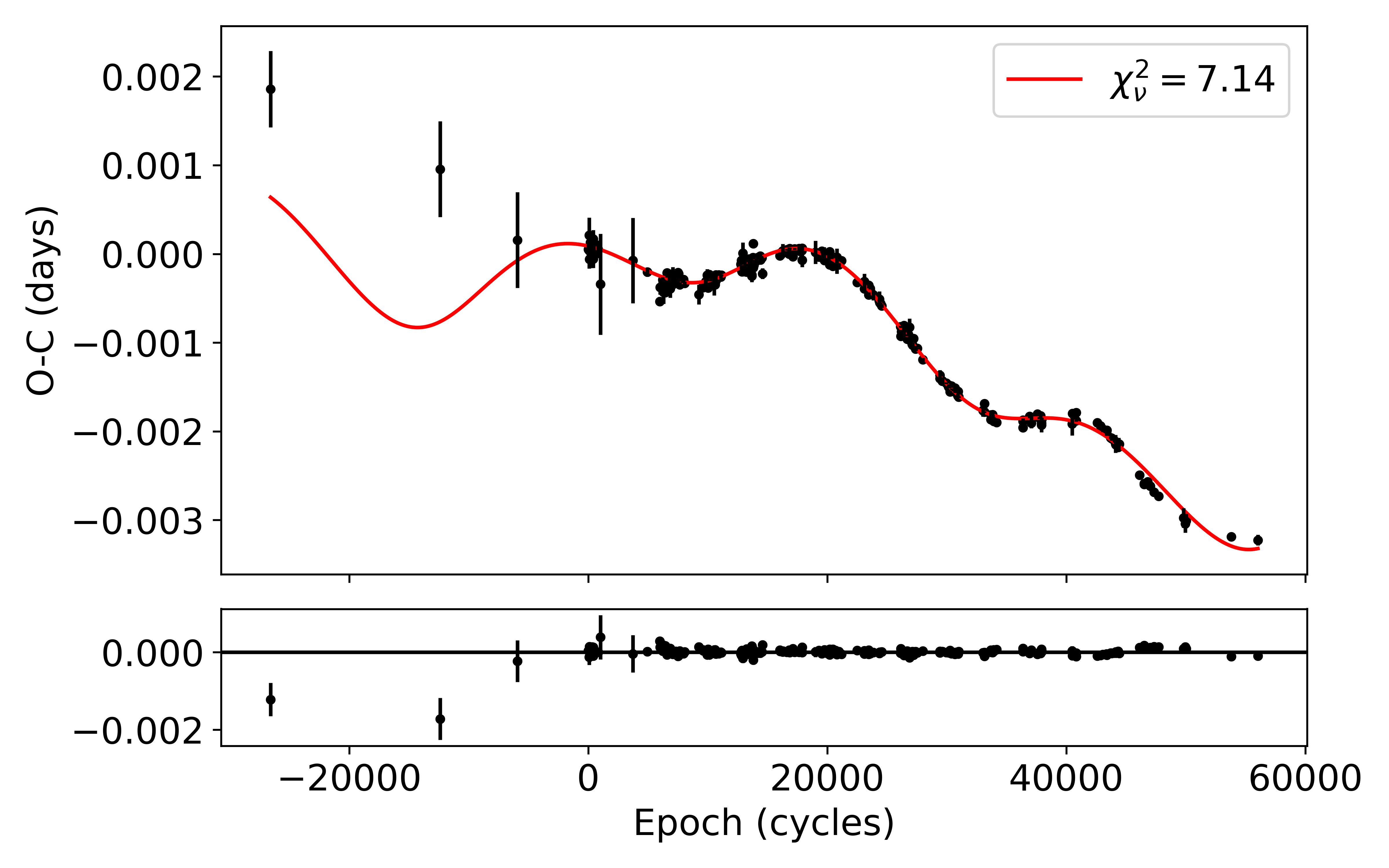}}
		\subfloat[$\chi_{\nu,max}^2$ (dynamically stable)]{\includegraphics[width=0.3\textwidth, keepaspectratio]{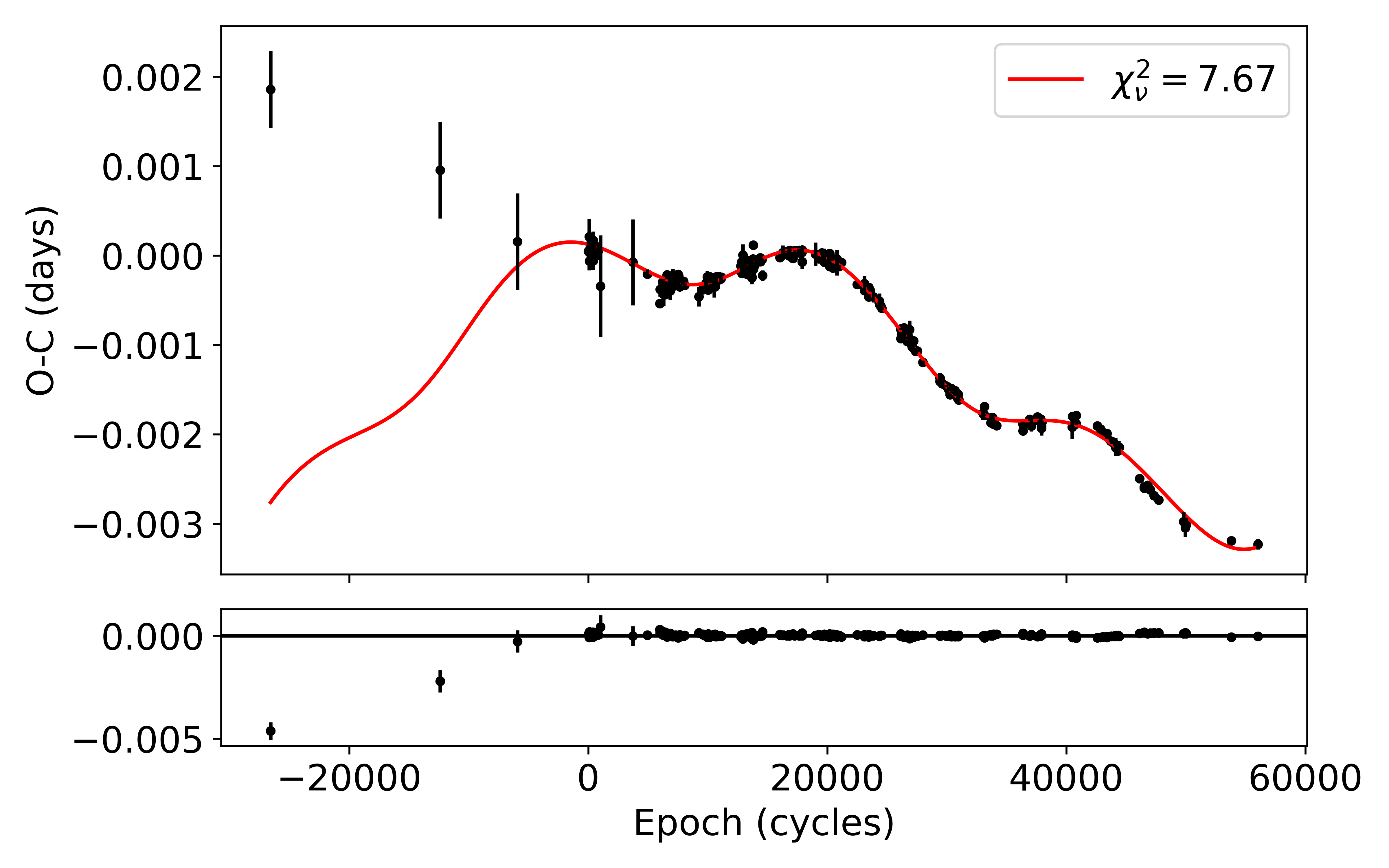}}\\
		
		\hspace{1cm}\subfloat[$\chi_{\nu,best}^2$ (dynamically unstable)]{\includegraphics[width=0.3\textwidth, keepaspectratio]{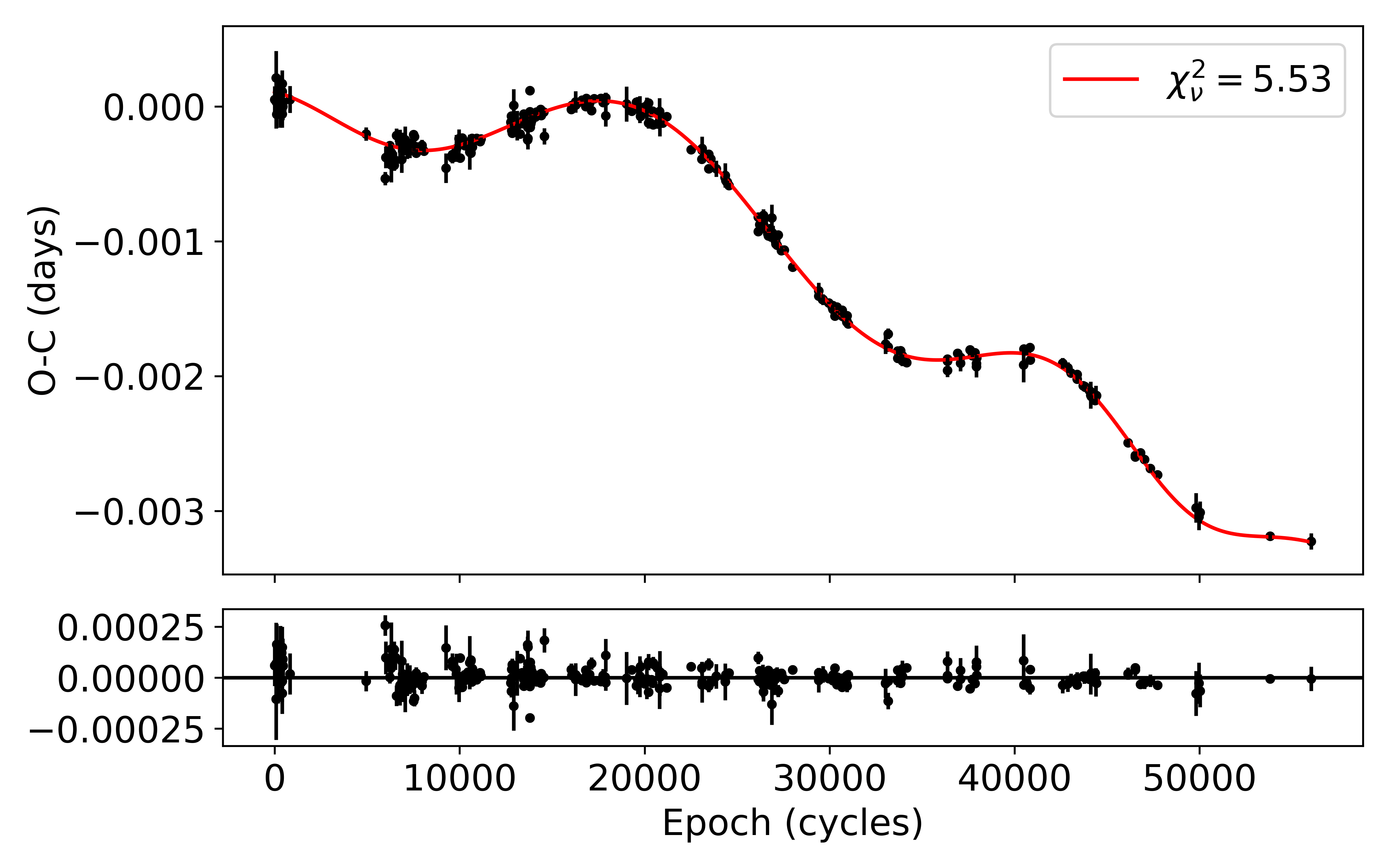}}
		\subfloat[$\chi_{\nu,min}^2$ (dynamically stable)]{\includegraphics[width=0.3\textwidth, keepaspectratio]{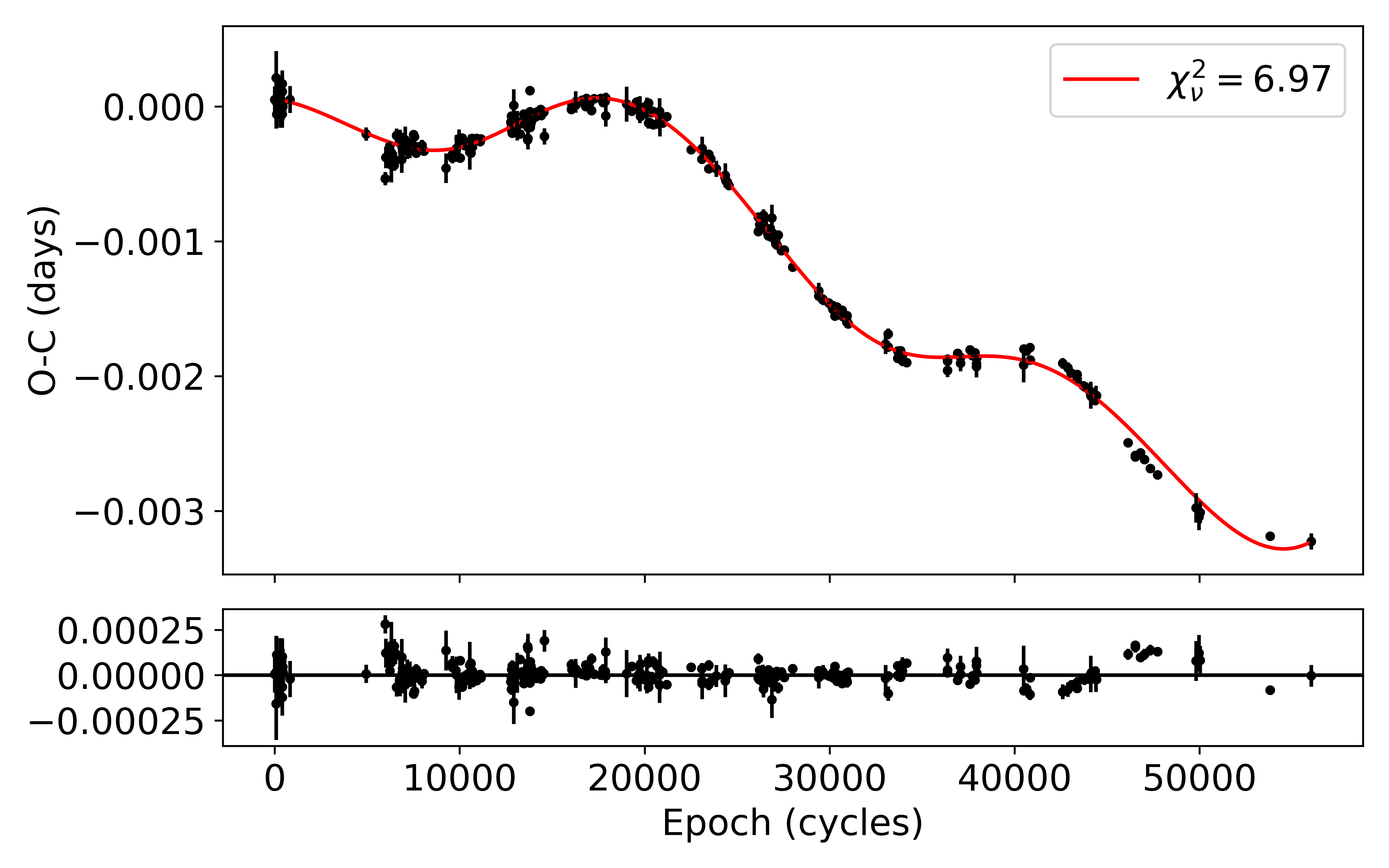}}
		\subfloat[$\chi_{\nu,max}^2$ (dynamically stable)]{\includegraphics[width=0.3\textwidth, keepaspectratio]{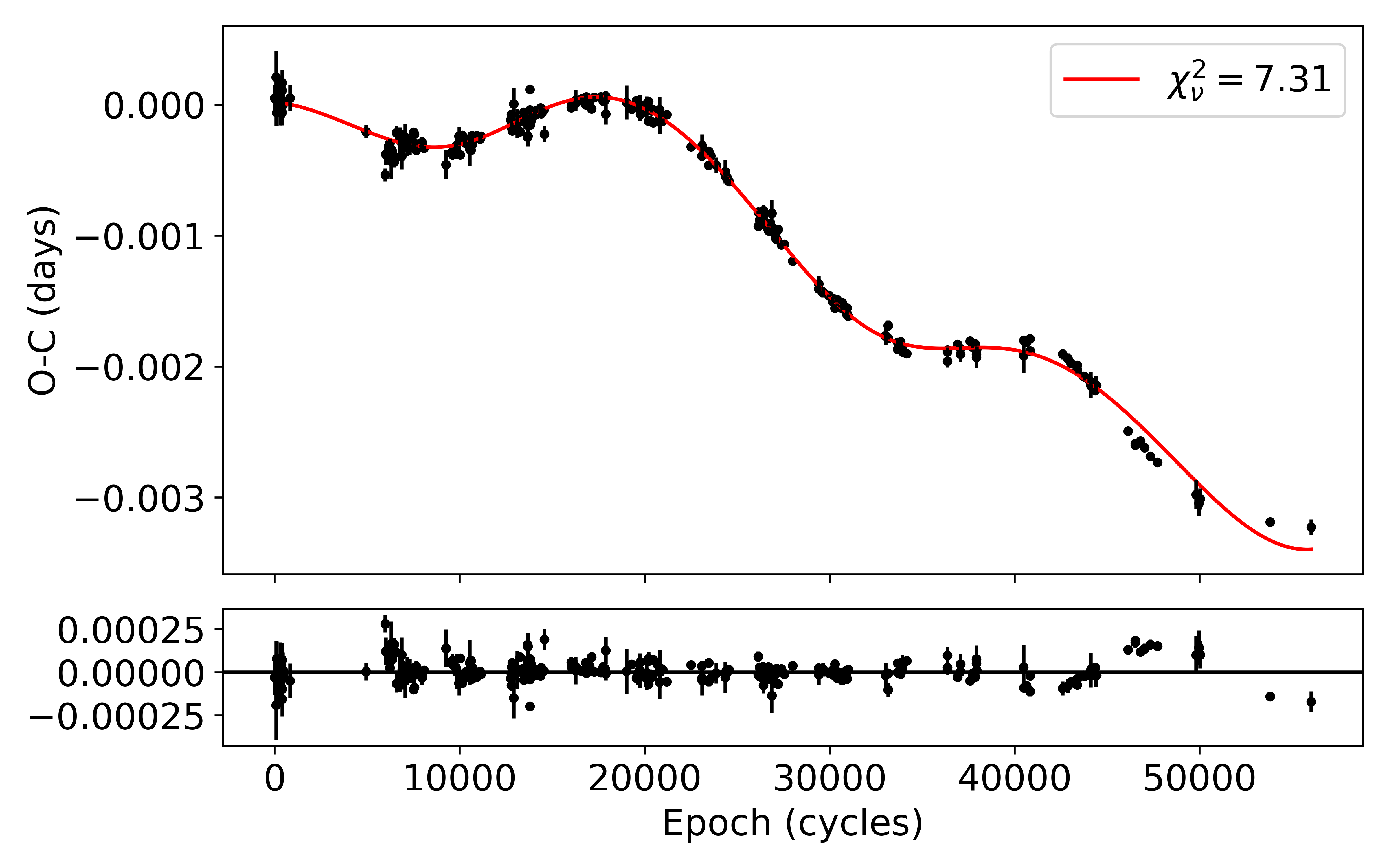}}\\
		
		\hspace{1cm}\subfloat[$\chi_{\nu,best}^2$ (dynamically unstable)]{\includegraphics[width=0.3\textwidth, keepaspectratio]{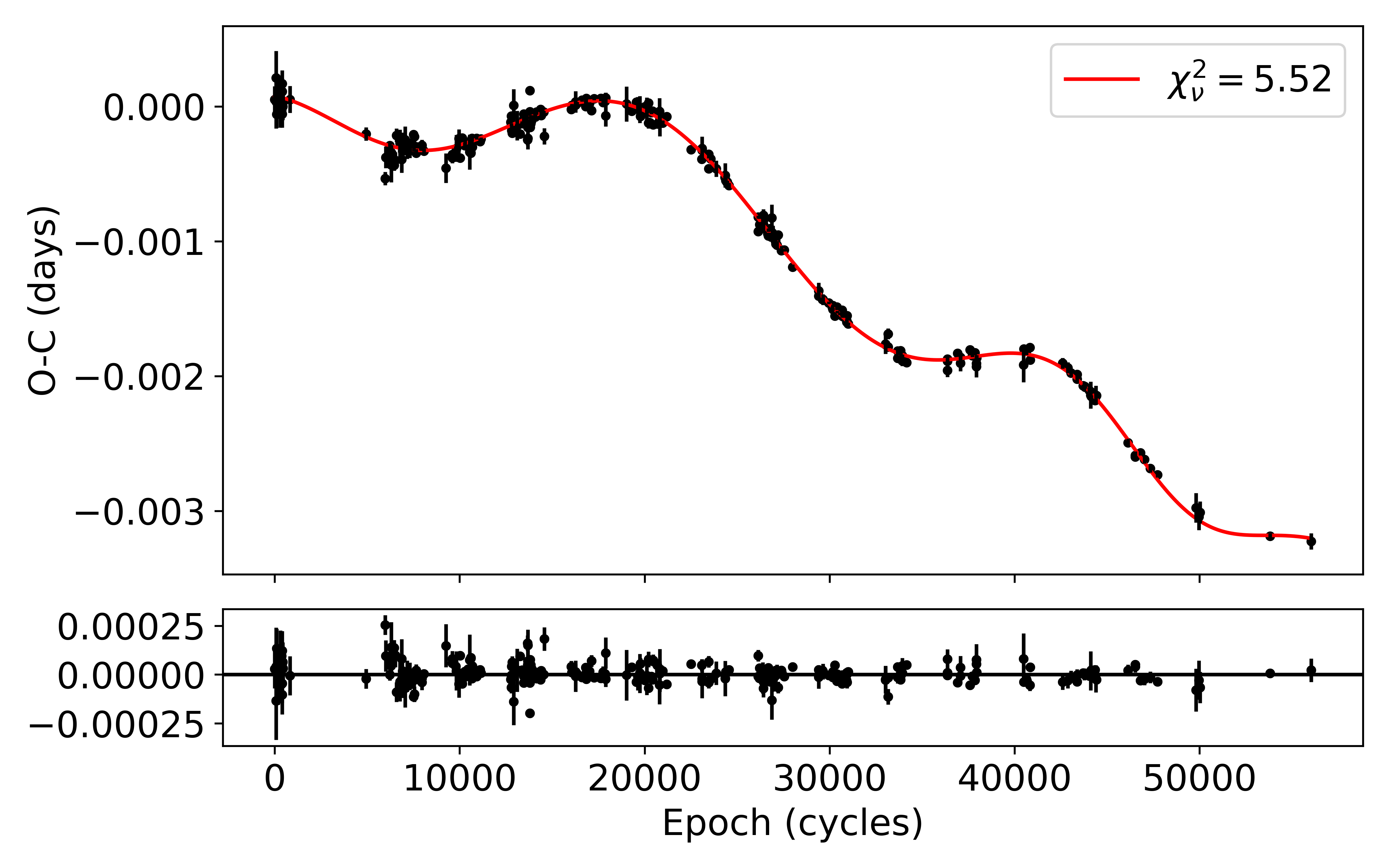}}
		\subfloat[$\chi_{\nu,min}^2$ (dynamically stable)]{\includegraphics[width=0.3\textwidth, keepaspectratio]{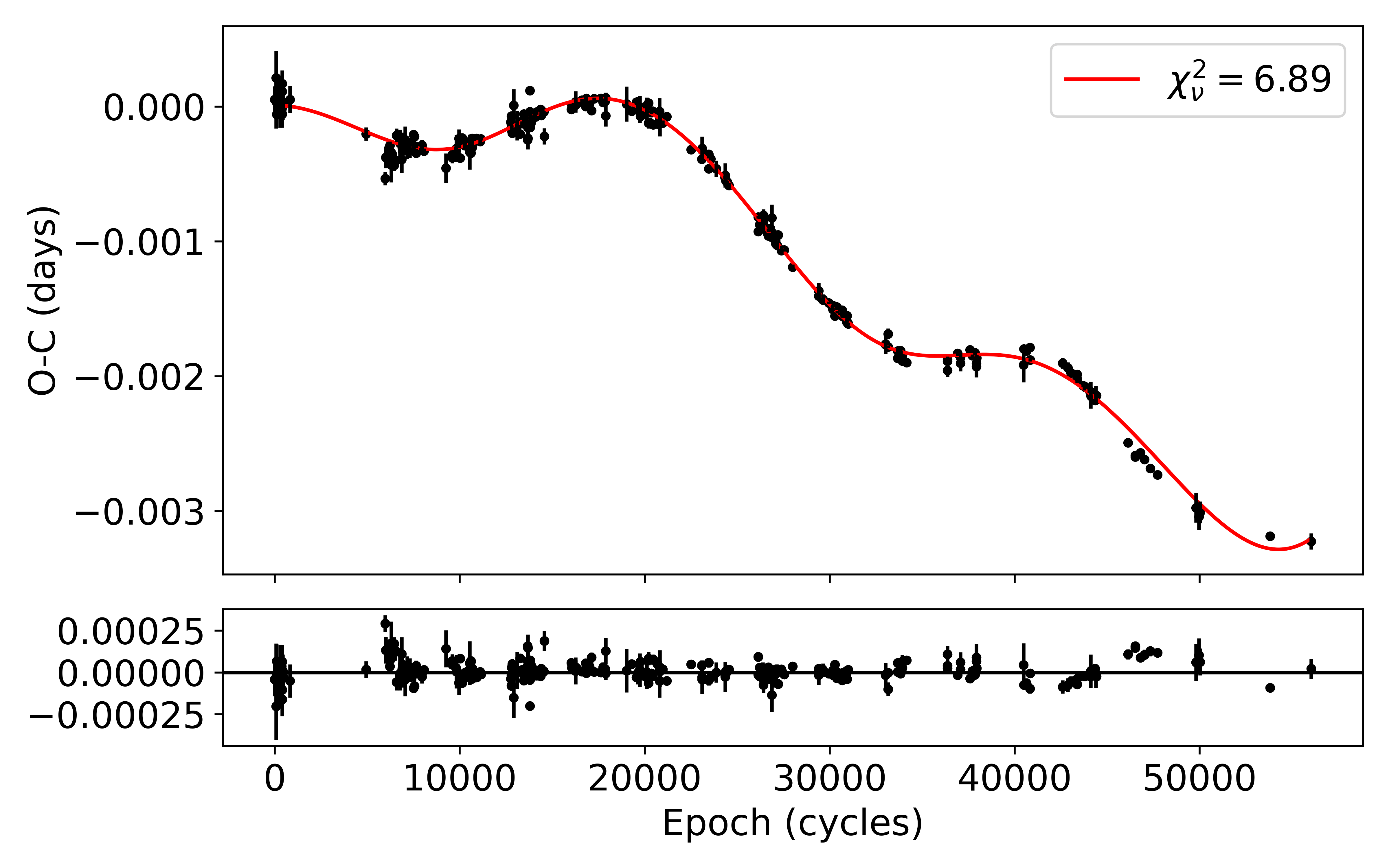}}
		\subfloat[$\chi_{\nu,max}^2$ (dynamically stable)]{\includegraphics[width=0.3\textwidth, keepaspectratio]{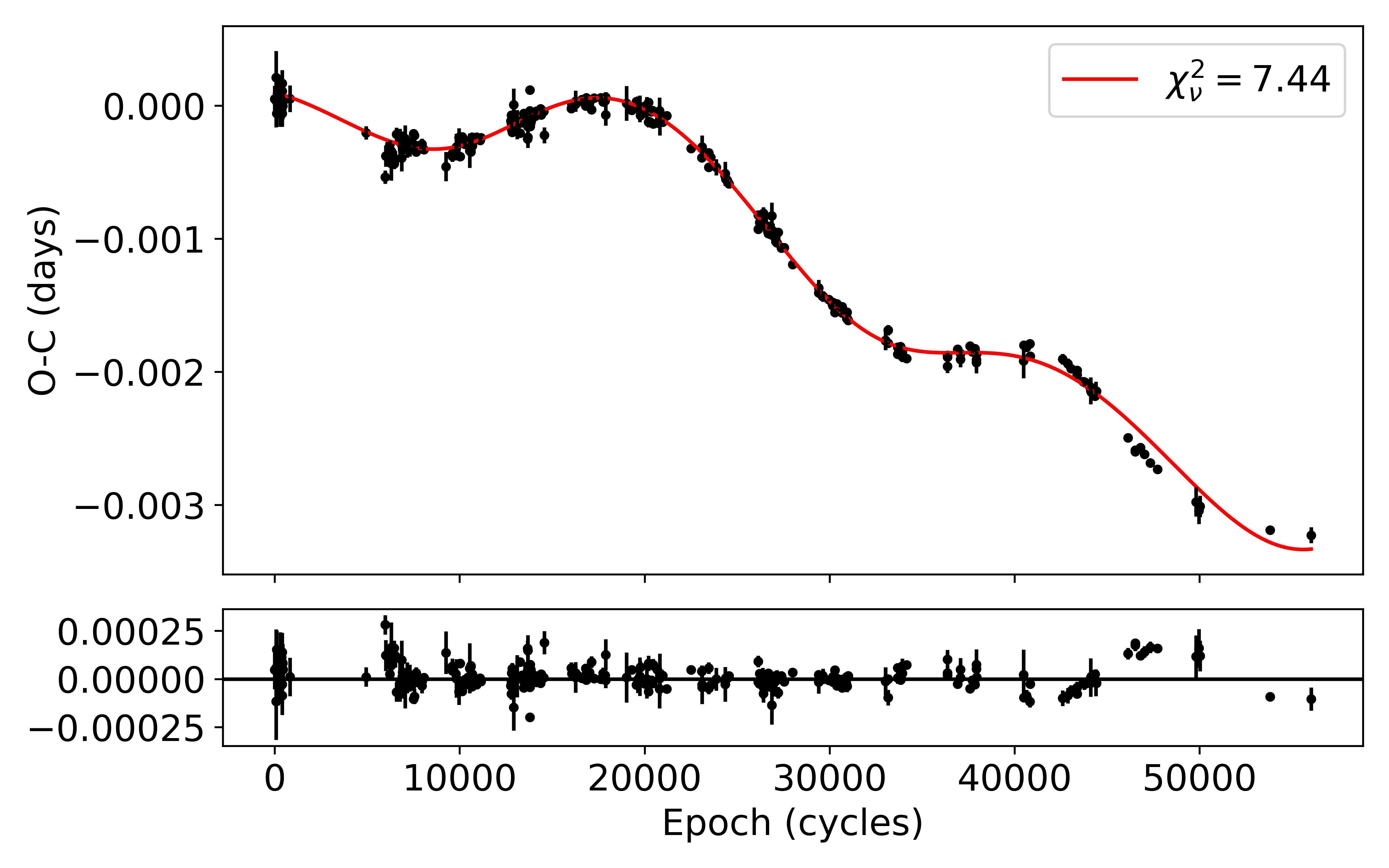}}
		\caption{Newtonian fits for both datasets [A: (a)-(f), B: (g)-(l)] and optimization runs [1: (a)-(c), (g)-(i) ,2: (d)-(f), (j)-(l)]. The best-fitting curves ($\chi_{\nu,best}^2$) result to dynamically unstable configurations, while the fitting curves of the minimum and maximum reduced chi-square value ($\chi_{\nu,min}^2$, $\chi_{\nu,max}^2$) define the $\chi_{\nu}^2$ limits among all stable configurations. The parameters of each case can be found in Table~\ref{tab:Newtonian_run1} and Table~\ref{tab:Newtonian_run2}.}
		\label{fig:Newtonian_fits}
	\end{figure*}
	
	\begin{figure*}
		\hspace{-1cm}\subfloat[]{\includegraphics[width=0.28\textwidth,keepaspectratio]{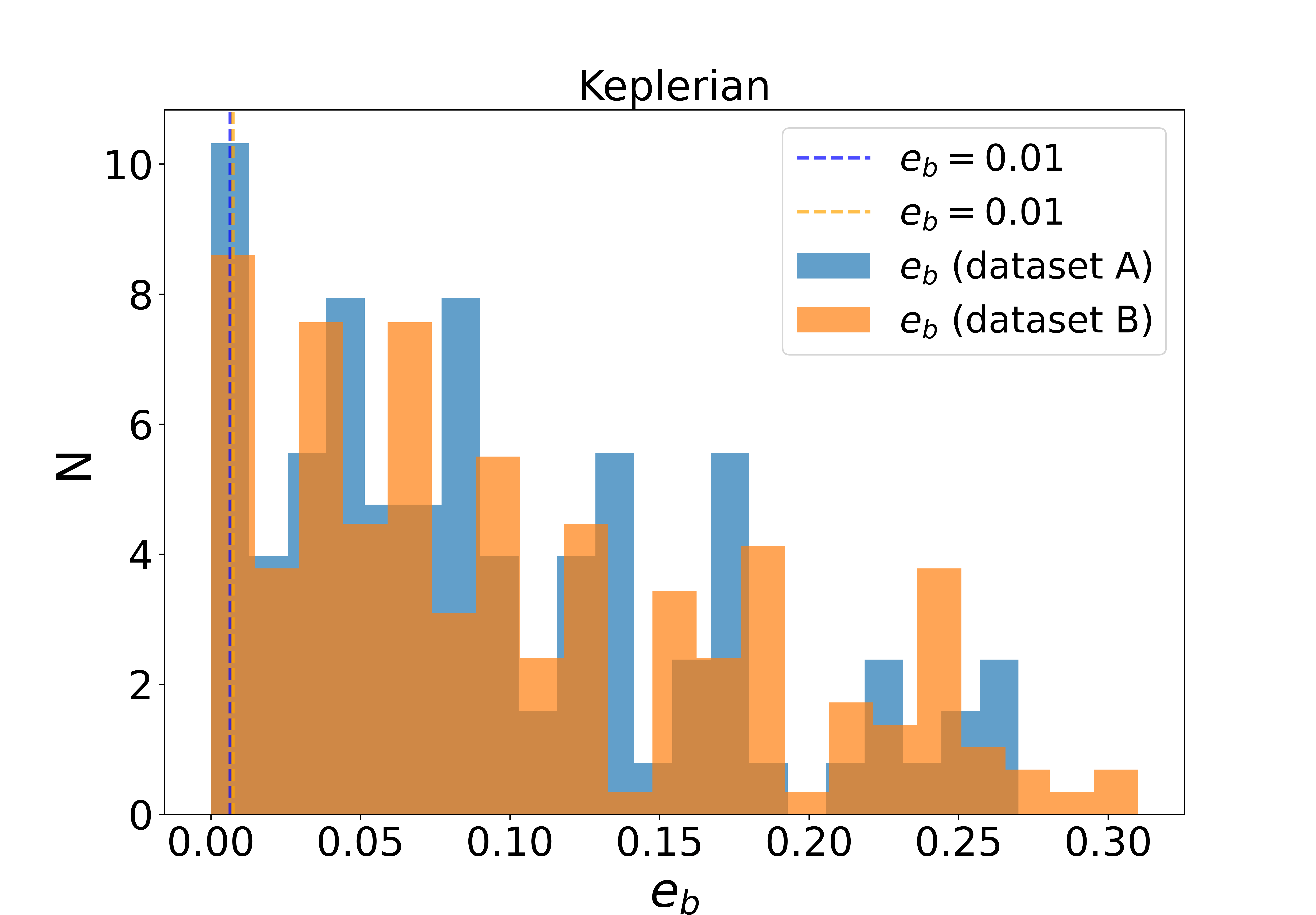}\label{fig:hist1-ei-Keplerian}}
		\subfloat[]{\includegraphics[width=0.28\textwidth, keepaspectratio]{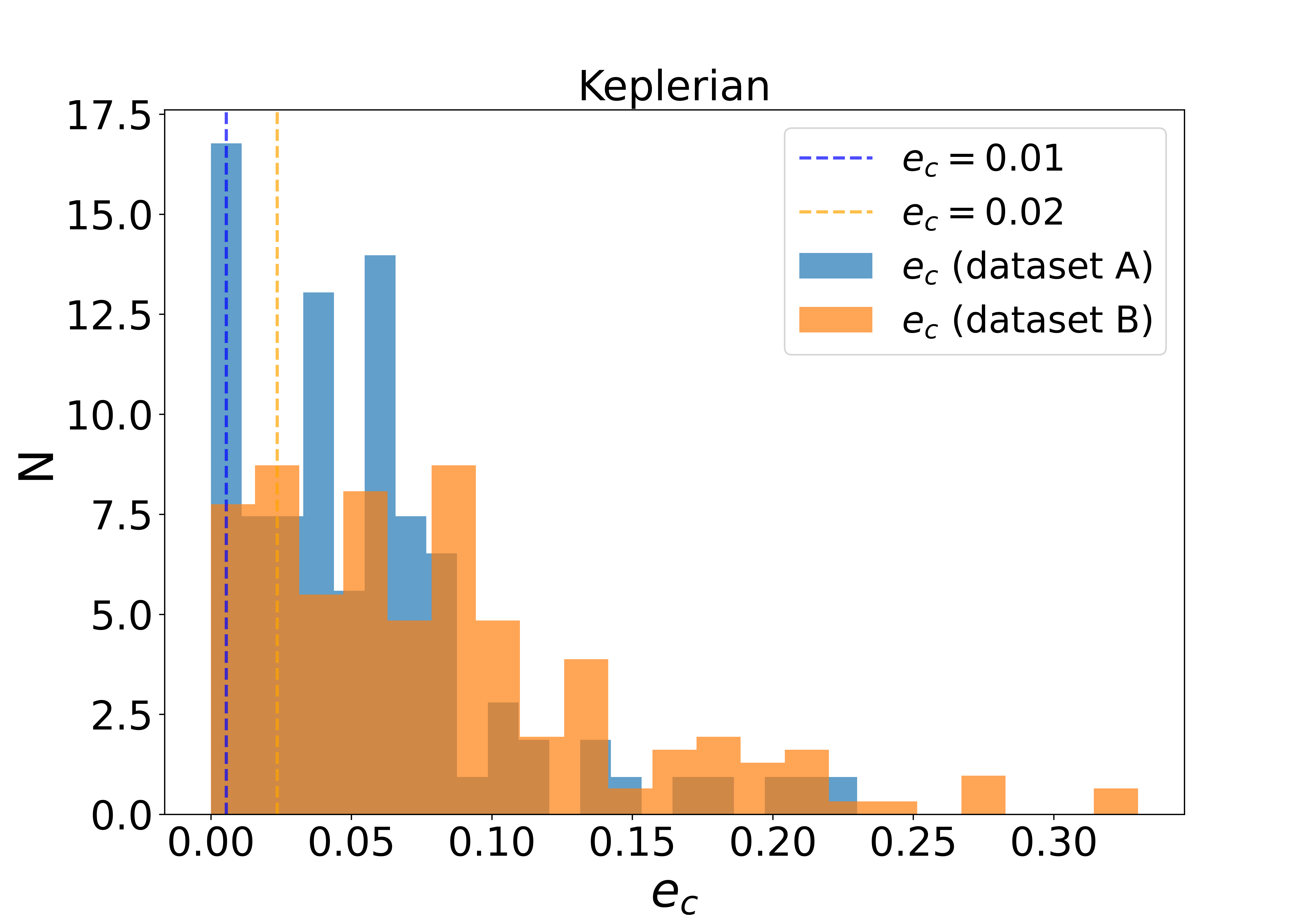}\label{fig:hist1-eo-Keplerian}}
		\subfloat[]{\includegraphics[width=0.28\textwidth,keepaspectratio]{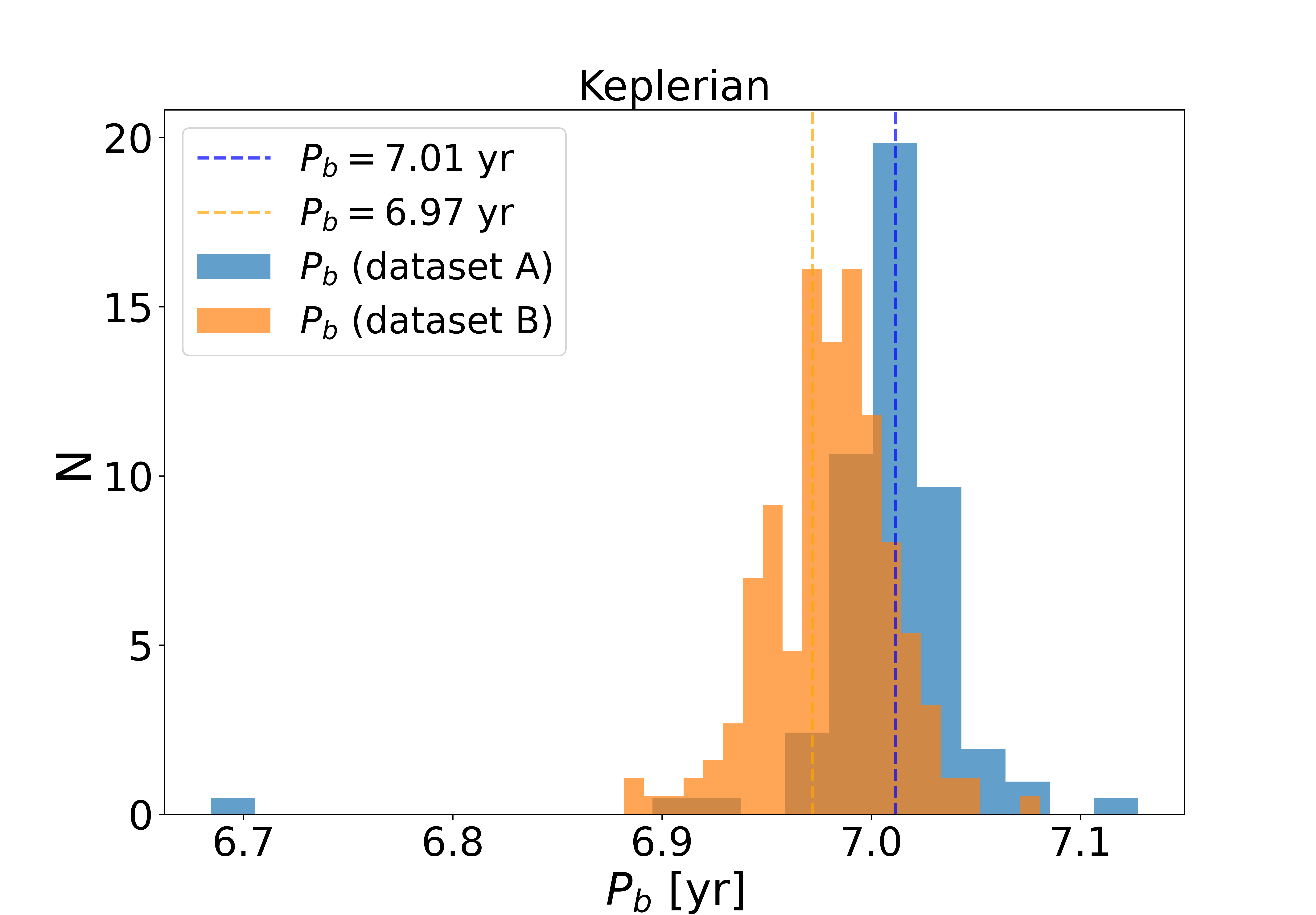}\label{fig:hist1-Pi-Keplerian}}
		\subfloat[]{\includegraphics[width=0.28\textwidth, keepaspectratio]{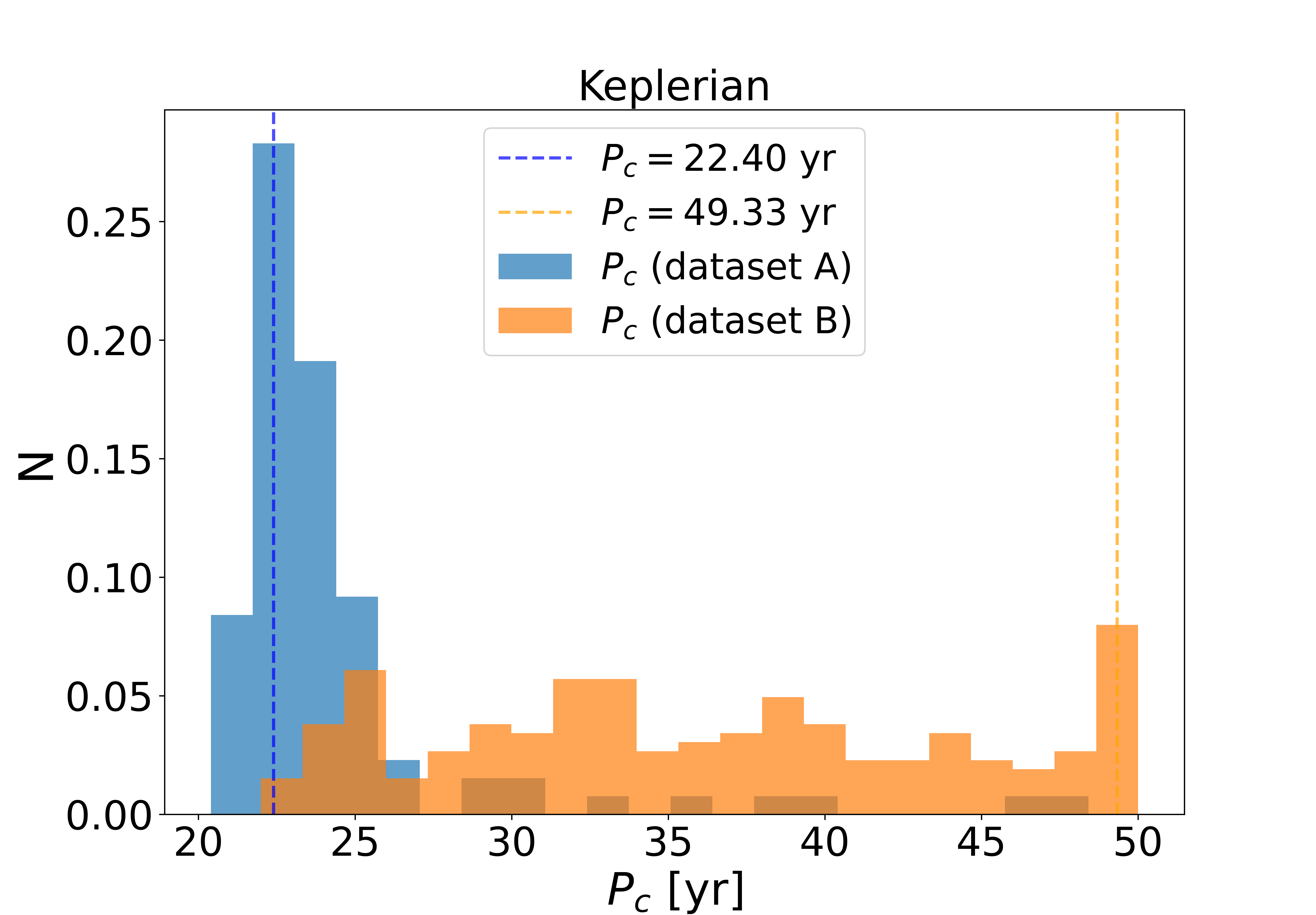}\label{fig:hist1-Po-Keplerian}}\\
		
		\hspace{-1cm}\subfloat[]{\includegraphics[width=0.28\textwidth,keepaspectratio]{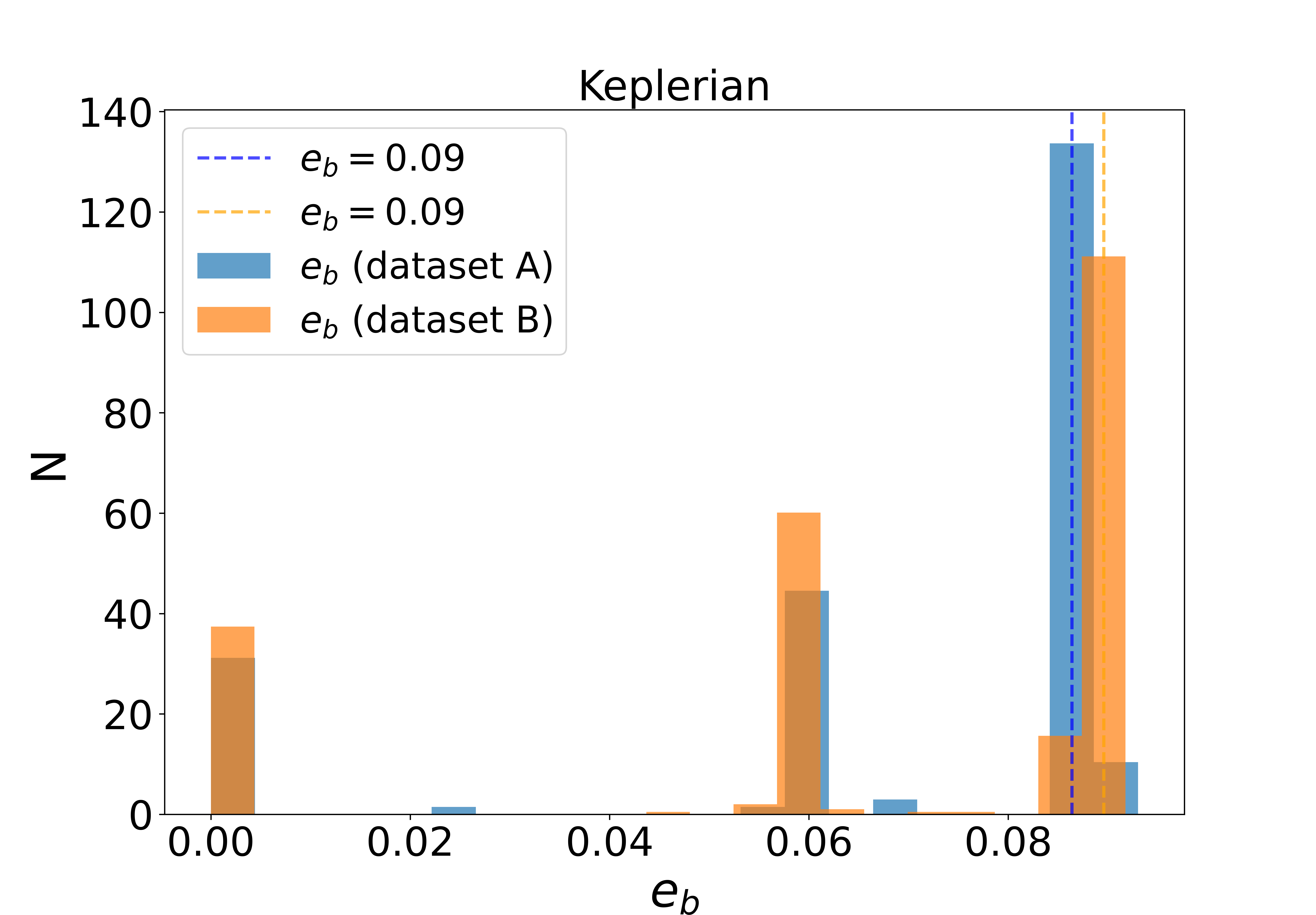}\label{fig:hist2-ei-Keplerian}}
		\subfloat[]{\includegraphics[width=0.28\textwidth, keepaspectratio]{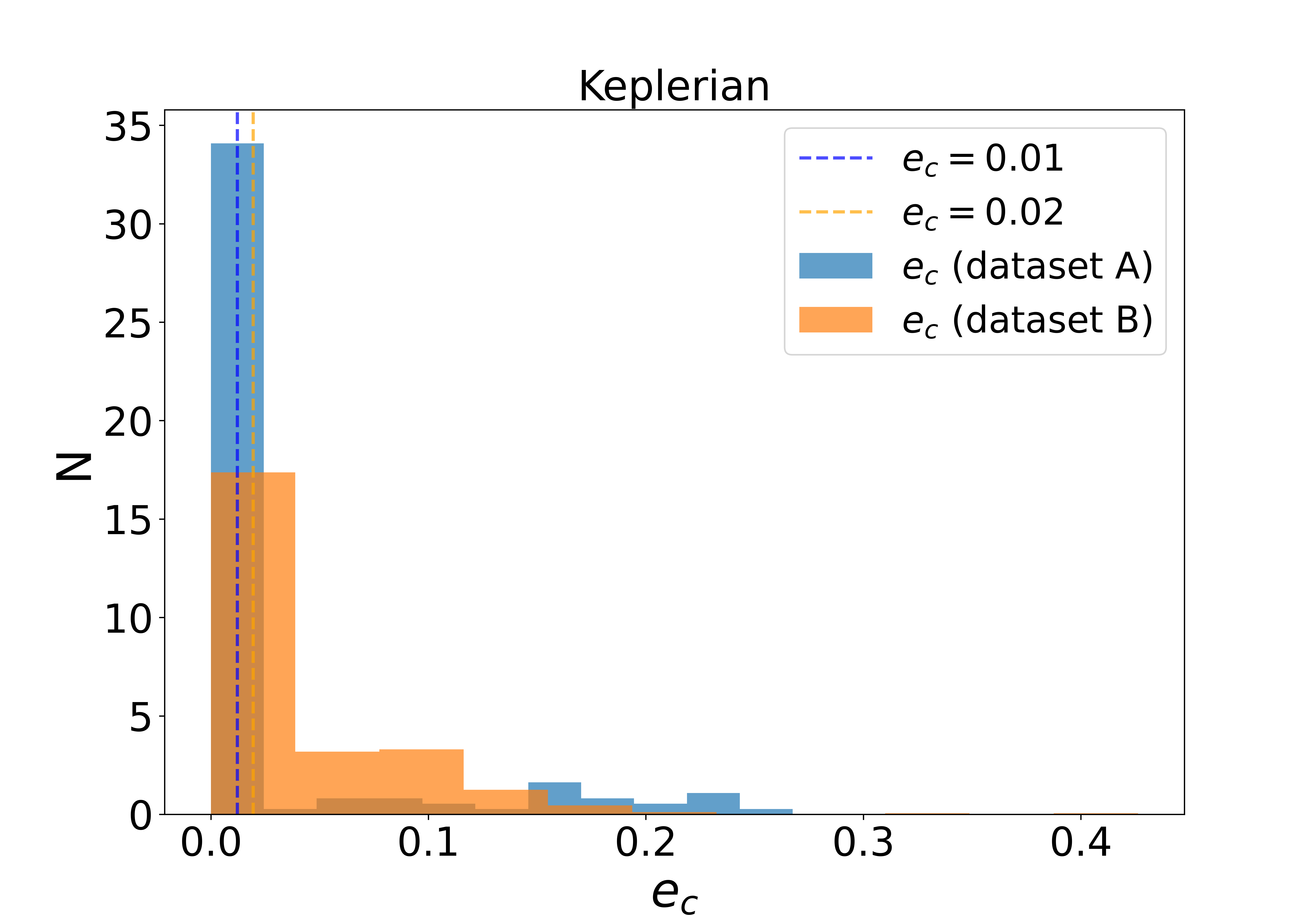}\label{fig:hist2-eo-Keplerian}}
		\subfloat[]{\includegraphics[width=0.28\textwidth,keepaspectratio]{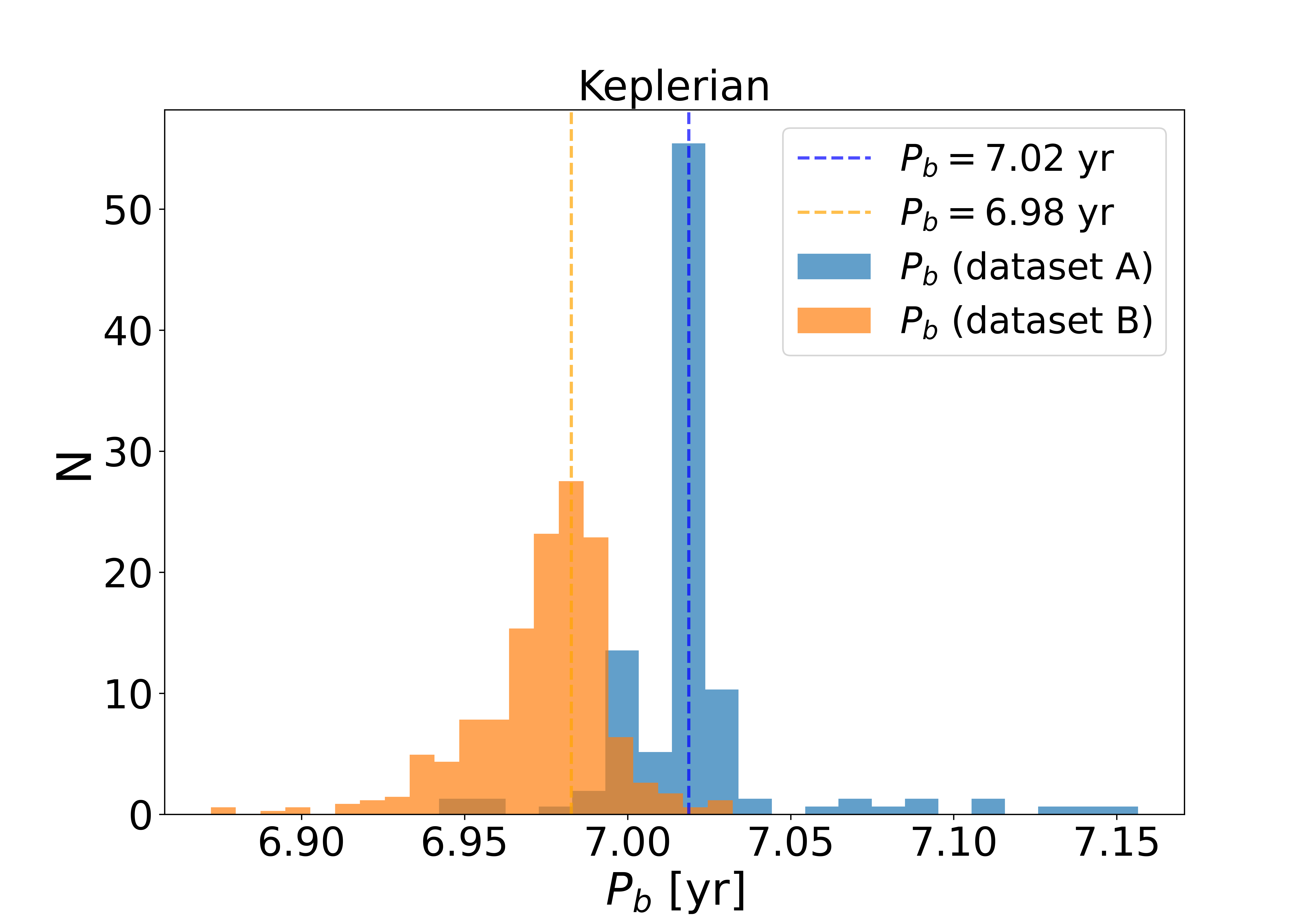}\label{fig:hist2-Pi-Keplerian}}
		\subfloat[]{\includegraphics[width=0.28\textwidth, keepaspectratio]{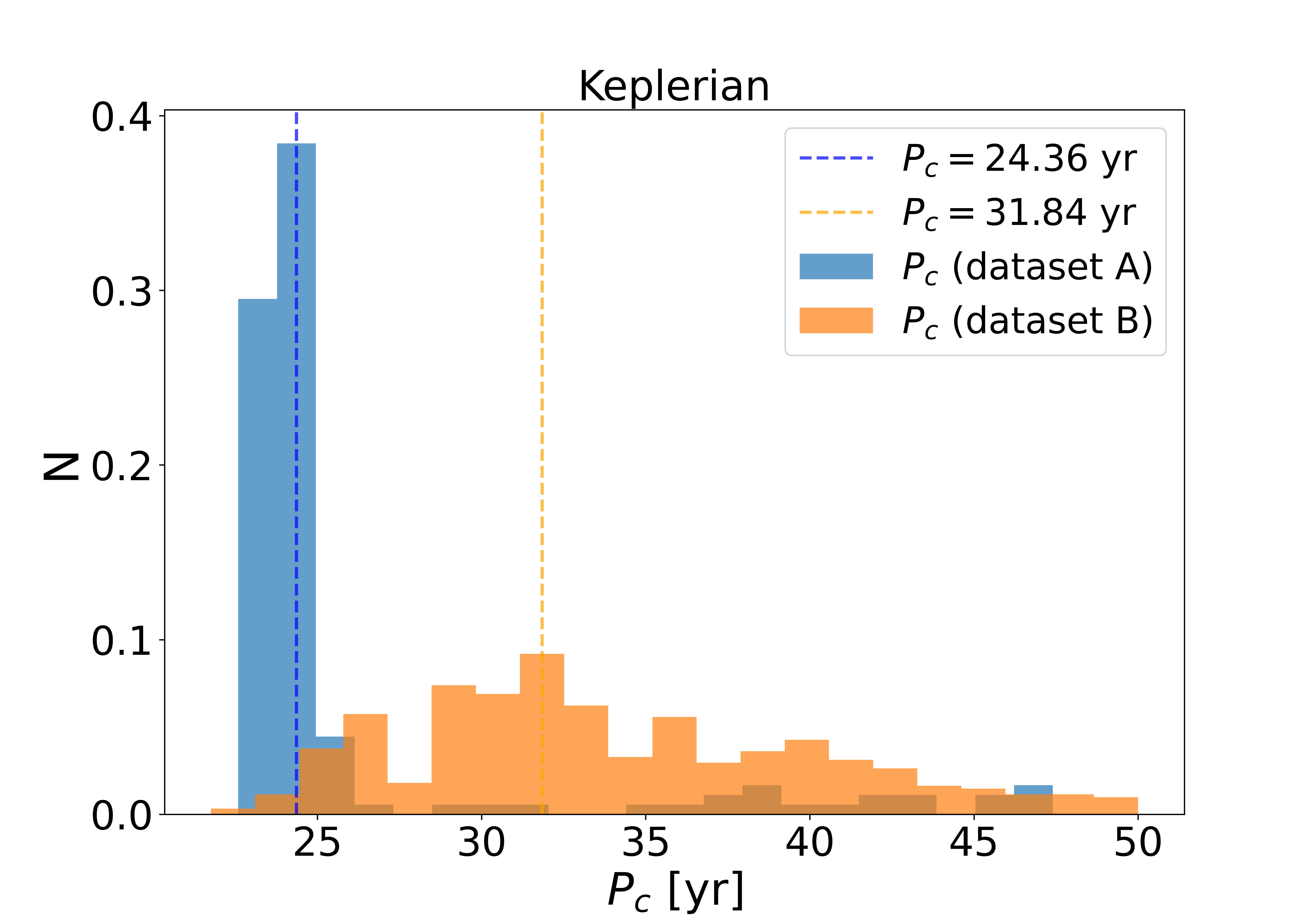}\label{fig:hist2-Po-Keplerian}}\\
		
		\caption{Histograms resulted from optimization runs. Top: Normalized histograms of $e_b$, $e_c$, $P_b$, $P_c$ for the Keplerian models as they resulted from the optimization Run 1 for both datasets A, B. The dashed lines indicate the center value of the bin with the maximum frequency. Bottom: Same as top panel but for optimization Run 2.}
		\label{fig:hist_Keplerian}
	\end{figure*}
	
	\begin{figure*}
		\hspace{-1cm}\subfloat[]{\includegraphics[width=0.28\textwidth,keepaspectratio]{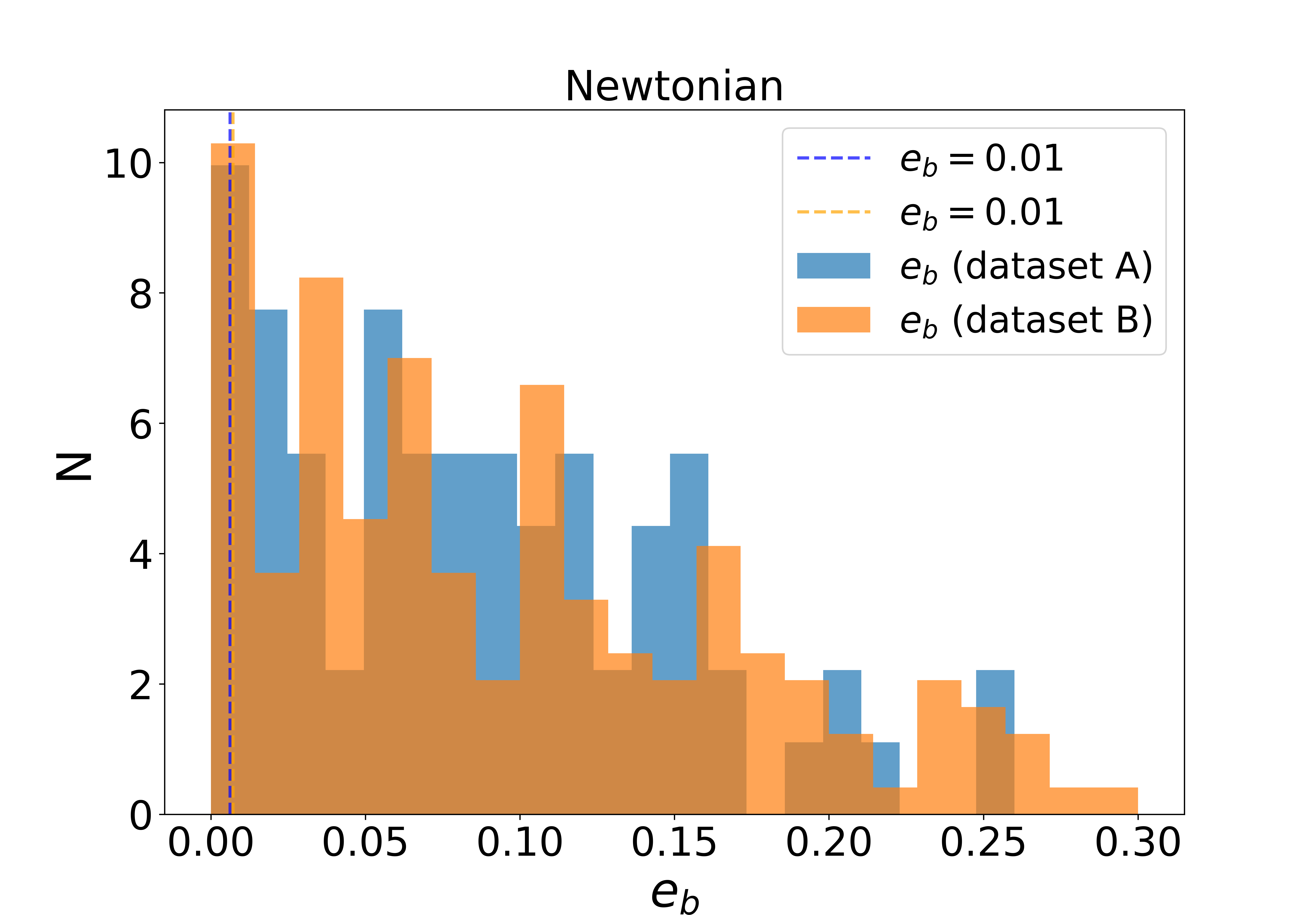}\label{fig:hist1-ei-Newtonian}}
		\subfloat[]{\includegraphics[width=0.28\textwidth, keepaspectratio]{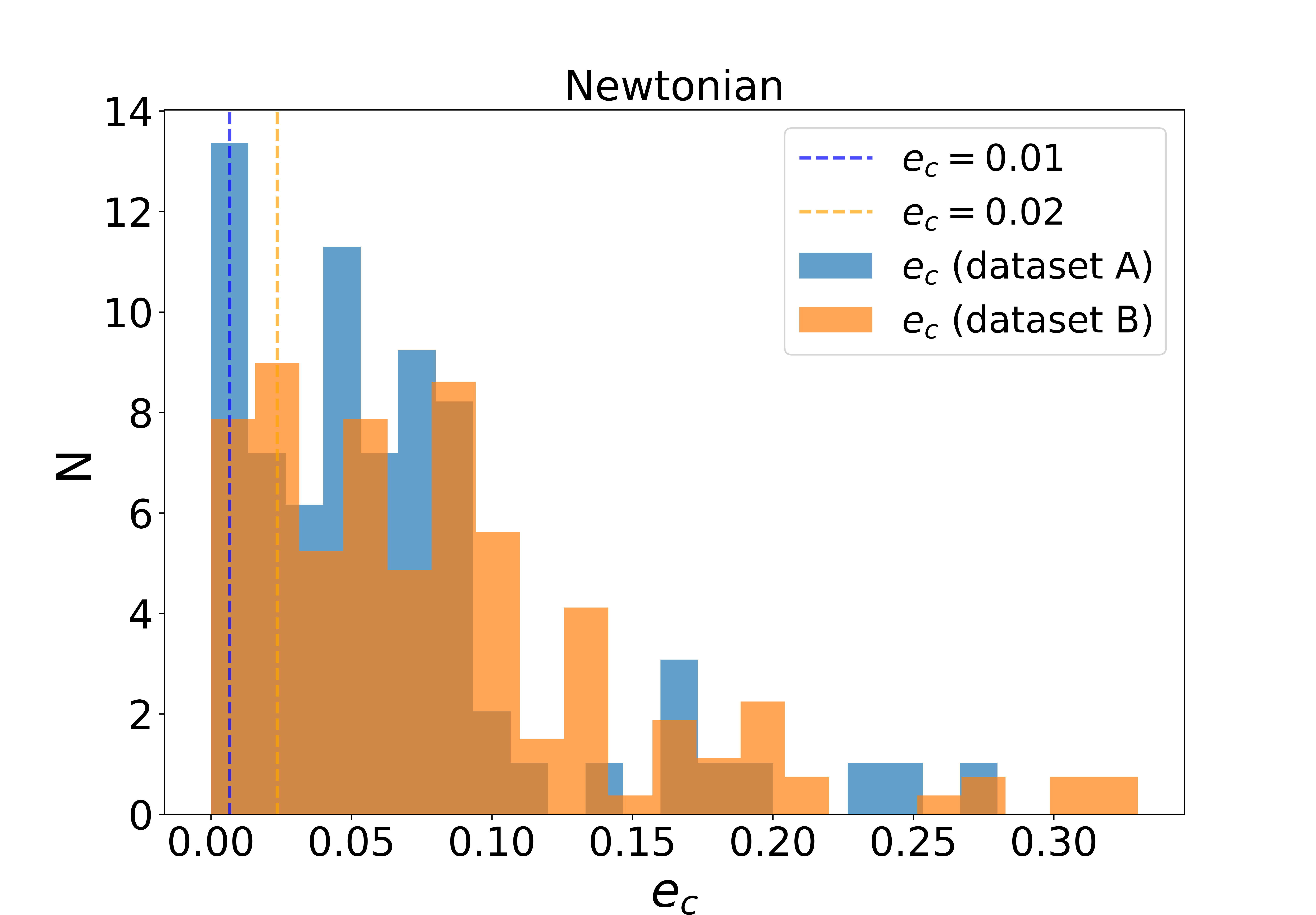}\label{fig:hist1-eo-Newtonian}}
		\subfloat[]{\includegraphics[width=0.28\textwidth,keepaspectratio]{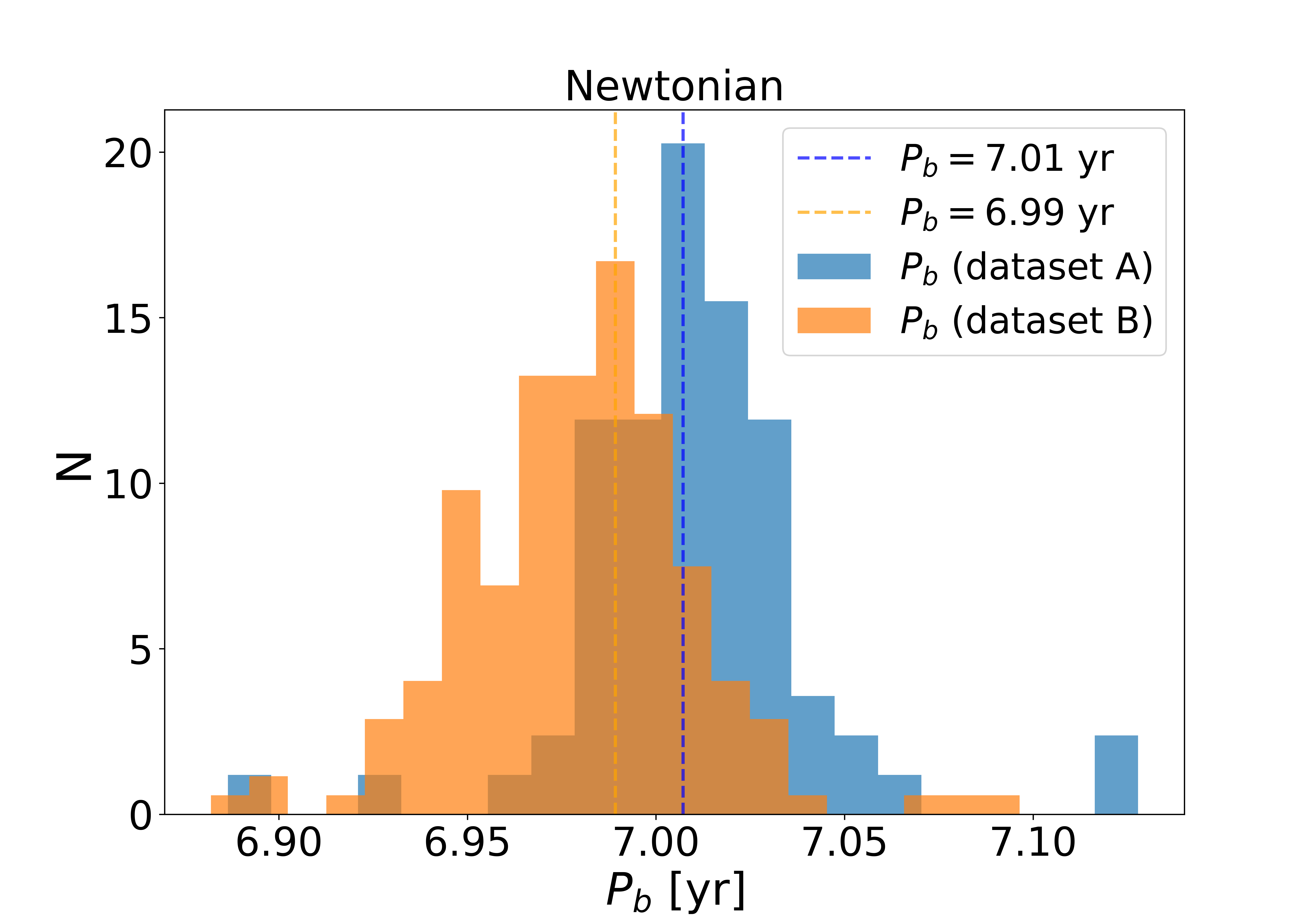}\label{fig:hist1-Pi-Newtonian}}
		\subfloat[]{\includegraphics[width=0.28\textwidth, keepaspectratio]{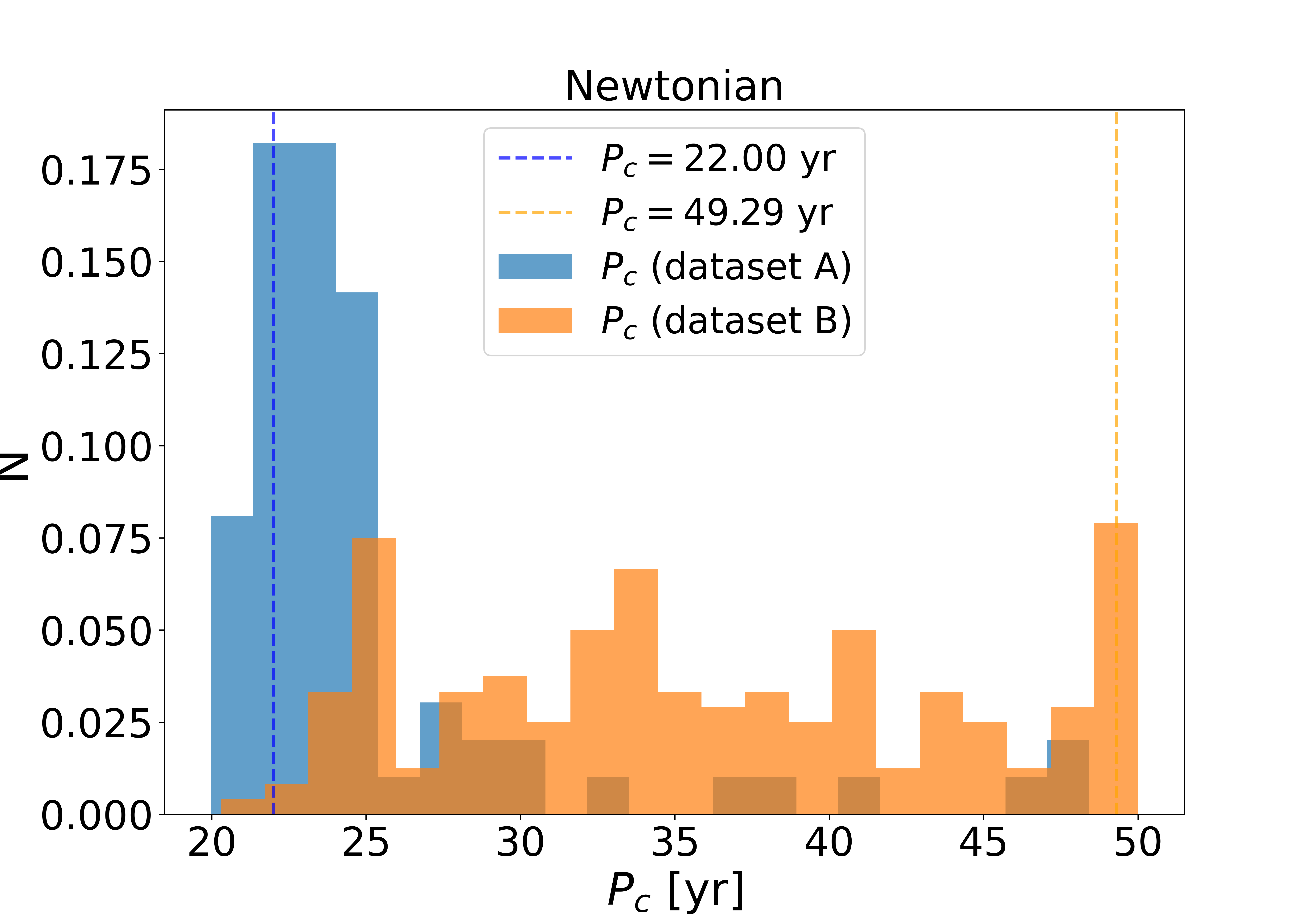}\label{fig:hist1-Po-Newtonian}}\\
		
		\hspace{-1cm}\subfloat[]{\includegraphics[width=0.28\textwidth,keepaspectratio]{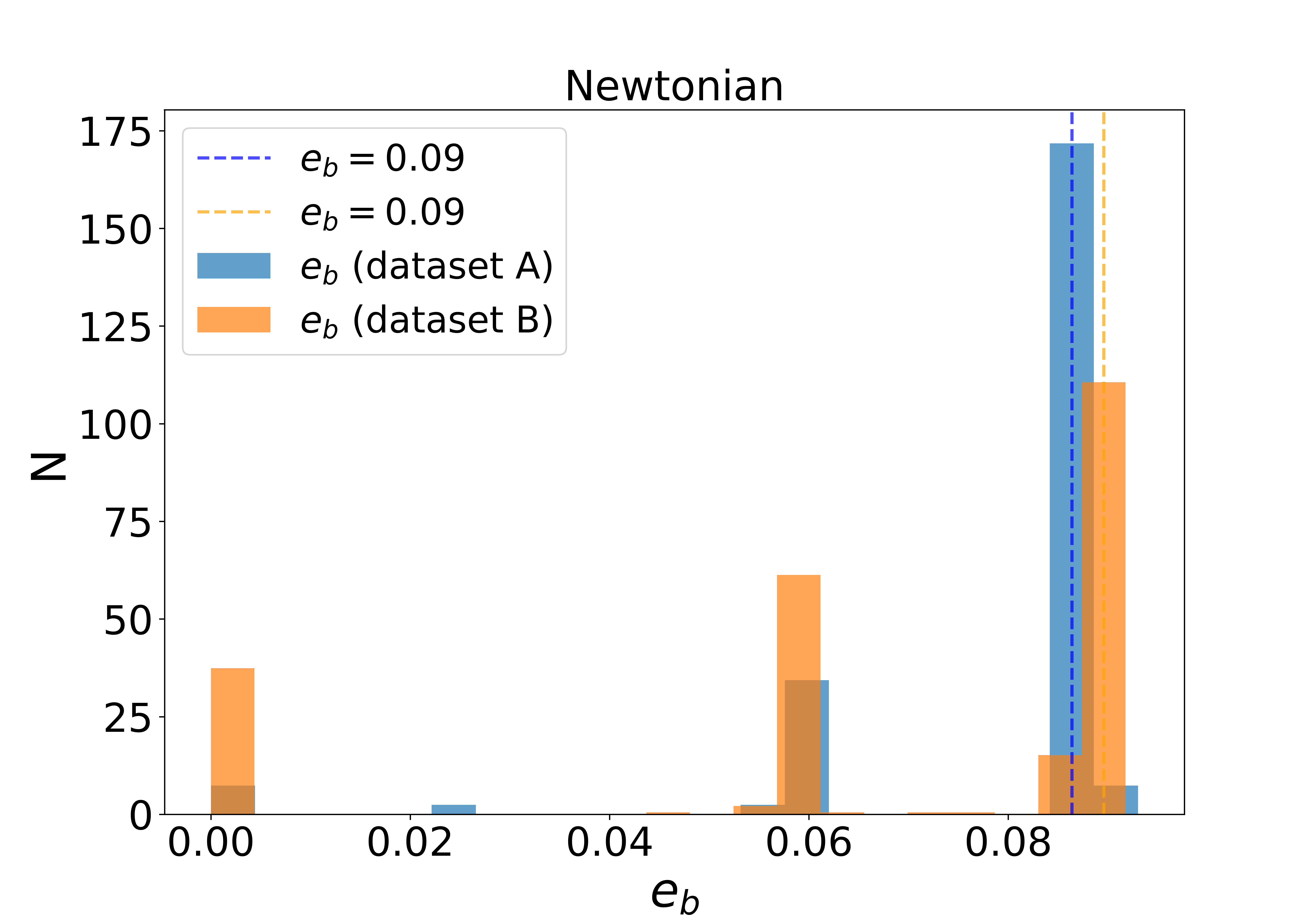}\label{fig:hist2-ei-Newtonian}}
		\subfloat[]{\includegraphics[width=0.28\textwidth, keepaspectratio]{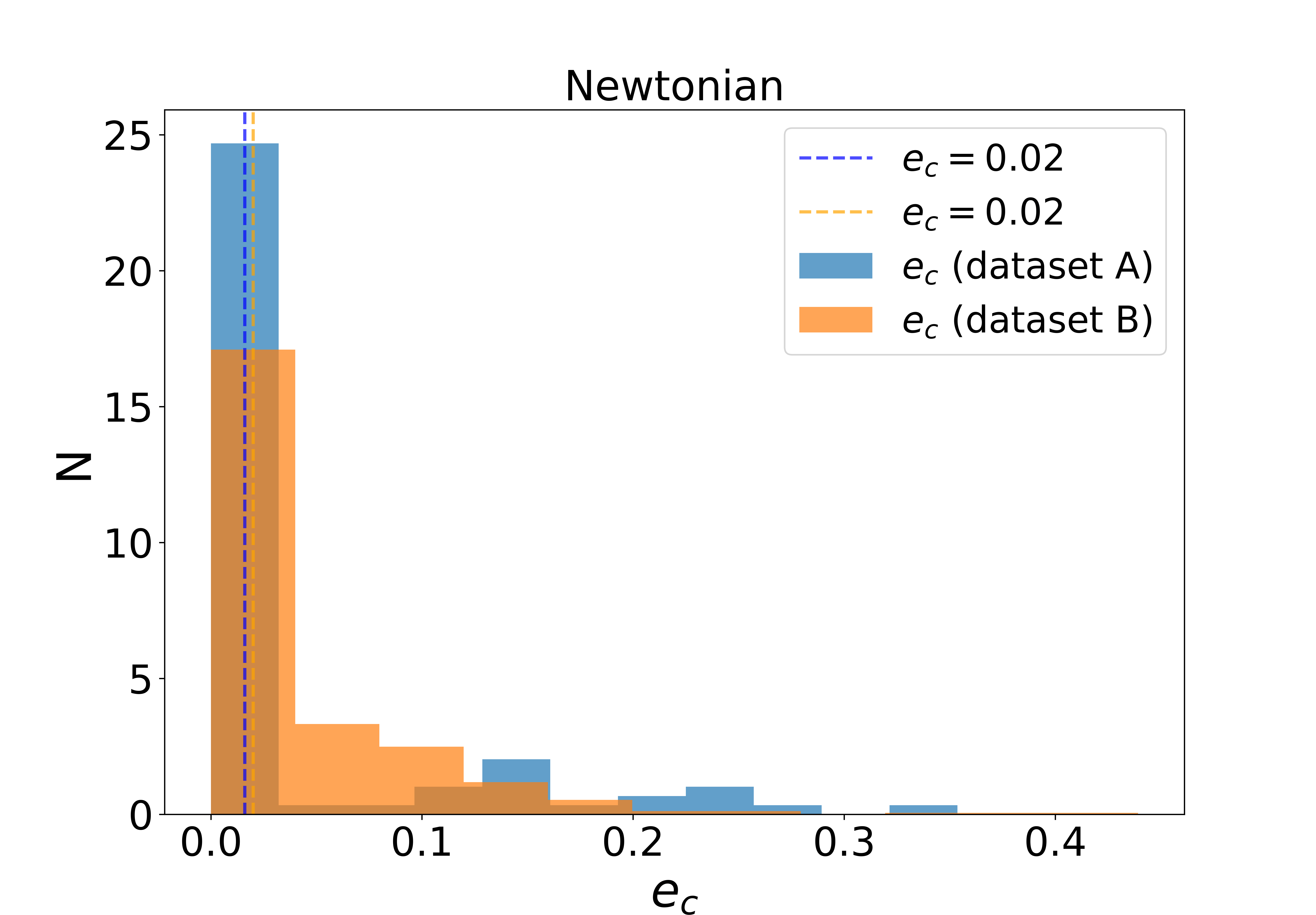}\label{fig:hist2-eo-Newtonian}}
		\subfloat[]{\includegraphics[width=0.28\textwidth,keepaspectratio]{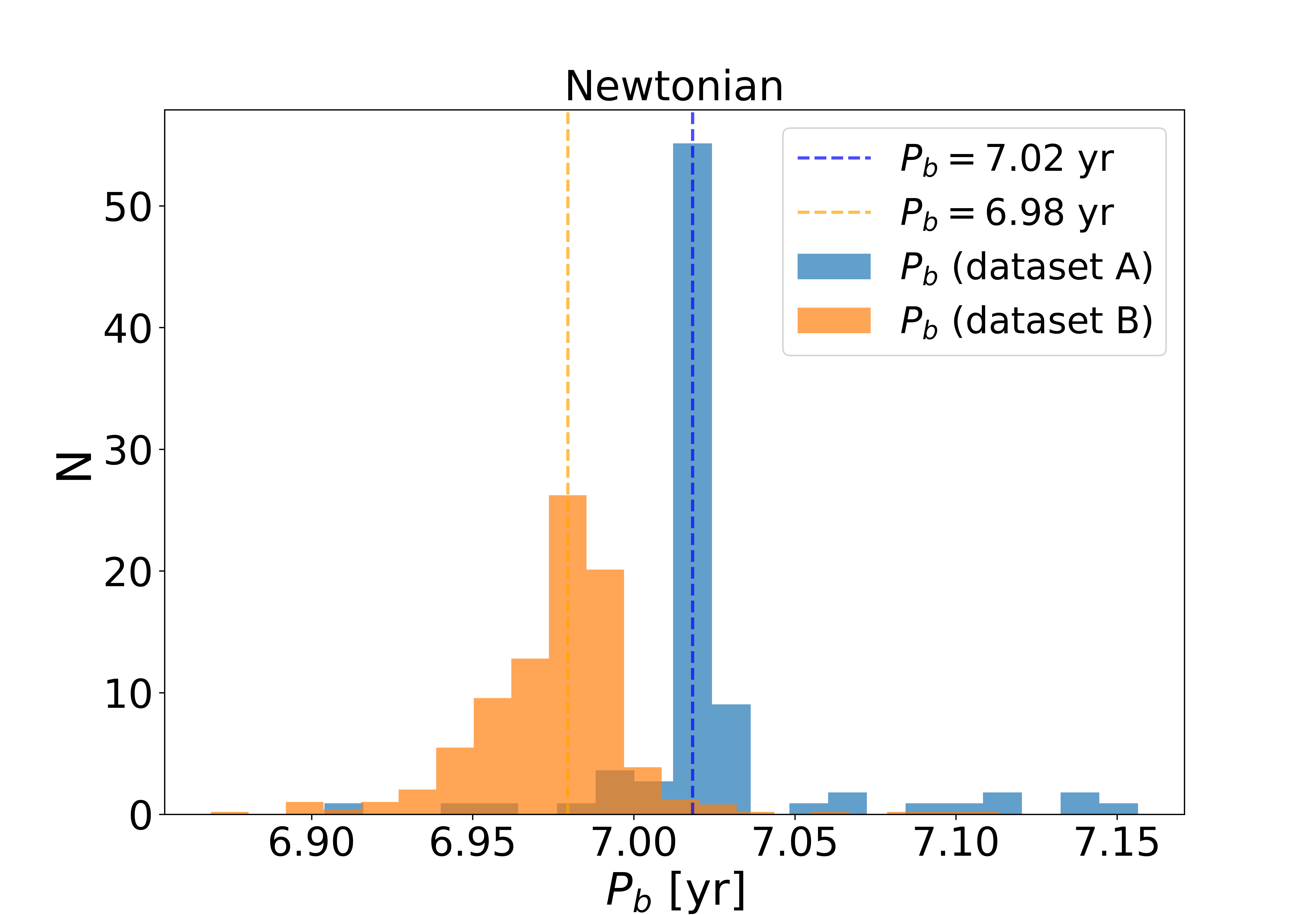}\label{fig:hist2-Pi-Newtonian}}
		\subfloat[]{\includegraphics[width=0.28\textwidth, keepaspectratio]{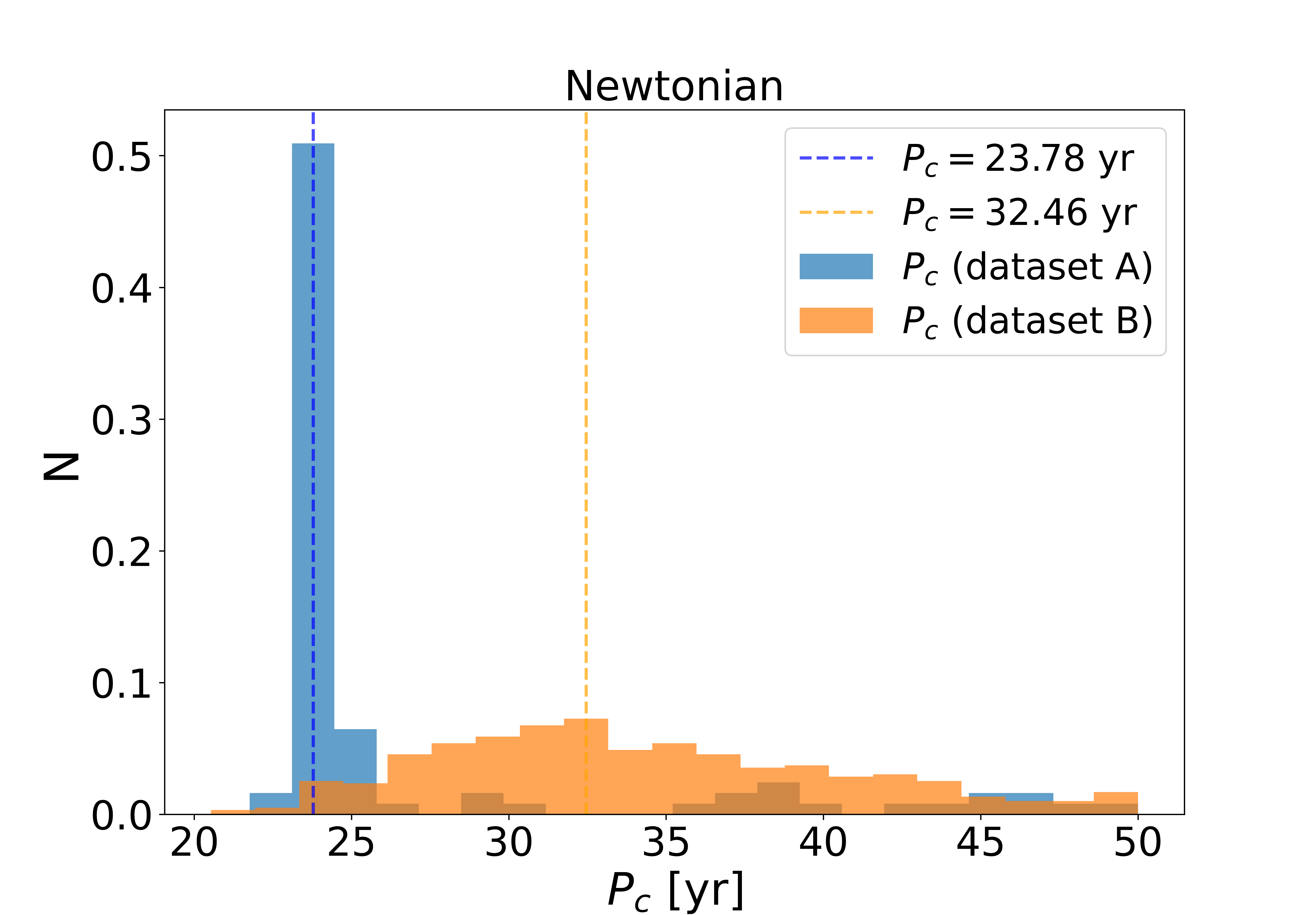}\label{fig:hist2-Po-Newtonian}}\\
		
		\caption{Same as Figure \ref{fig:hist_Keplerian} but for the Newtonian models.}
		\label{fig:hist_Newtonian}
	\end{figure*}
	
	\FloatBarrier

\section{Times of minima of NSVS 14256825}
\label{sub:ALL}
Times of minima (1999-2024) of NSVS 14256825 (literature and new).

\onecolumn
\begin{longtable}{ccccccccc}
\caption{Times of minima (1999-2024) of NSVS 14256825 (literature and new).} \label{tab:long} \\

\hline \multicolumn{1}{c}{\text{BJD-2400000}} & \multicolumn{1}{c}{$\rm Error (d)$} & \multicolumn{1}{c}{Ref} & \multicolumn{1}{c}{\text{BJD-2400000}} & \multicolumn{1}{c}{$\rm Error (d)$} & \multicolumn{1}{c}{Ref} & \multicolumn{1}{c}{\text{BJD-2400000}} & \multicolumn{1}{c}{$\rm Error (d)$} & \multicolumn{1}{c}{Ref} \\ \hline 
\endfirsthead

\multicolumn{9}{c}%
% {{\bfseries \tablename\ \thetable{} -- continued from previous page}} \\
{{\tablename\ \thetable{} -- continued from previous page}} \\
\hline \multicolumn{1}{c}{\text{BJD-2400000}} & \multicolumn{1}{c}{$\rm Error (d)$} & \multicolumn{1}{c}{Ref} & \multicolumn{1}{c}{\text{BJD-2400000}} & \multicolumn{1}{c}{$\rm Error (d)$} & \multicolumn{1}{c}{Ref} & \multicolumn{1}{c}{\text{BJD-2400000}} & \multicolumn{1}{c}{$\rm Error (d)$} & \multicolumn{1}{c}{Ref} \\ \hline
\endhead

\hline \multicolumn{9}{c}{{continued on next page}} \\ \hline
\endfoot

\hline \hline
\endlastfoot

51339.80327$^{*}$   &	0.00043	&	2	&    55165.03824	&	0.00002	&	6	&	55802.44920 &	0.00004	&	2	\\
52906.67390$^{*}$	&	0.00054	&	2	&    55296.38336	&	0.00011	&	6	&	55805.31896	&	0.00002 &	2	\\
53619.57978$^{*}$	&	0.00054	&	2	&	 55333.35878	&	0.00004	&	6	&	55805.42932 &	0.00002	&	2	\\
54274.20880	        &	0.00010	&	1	&	 55334.35217	&	0.00001	&	6	&	55808.29907	&	0.00002	&	2	\\
54282.15590	        &	0.00020	&	1	&	 55355.54403	&	0.00004	&	5	&	55826.29008	&	0.00002	&	3	\\
54282.26610	        &	0.00020	&	1	&	 55358.46897	&	0.00005	&	7	&	55829.27017	&	0.00002	&	3	\\
54286.12910	        &	0.00010	&	1	&	 55358.46897	&	0.00010	&	2	&	55837.32752	&	0.00002	&	5	\\
54293.19320	        &	0.00010	&	1	&	 55373.42474	&	0.00001	&	7	&	55861.27873	&	0.00002	&	3	\\
54294.07620	        &	0.00010	&	1	&	 55373.42474	&	0.00007	&	2	&	55863.26542	&	0.00002	&	3	\\
54294.18660	        &	0.00010	&	1	&	 55380.15742	&	0.00002	&	6	&	55863.59656	&	0.00003	&	2	\\
54295.17990	        &	0.00010	&	1	&	 55392.40910	&	0.00001	&	2	&	55864.25879	&	0.00001	&	3	\\
54309.19730	        &	0.00010	&	1	&	 55392.40910	&	0.00002	&	7	&	55872.09538	&	0.00002	&	6	\\
54310.08040	        &	0.00010	&	1	&	 55408.74442	&	0.00002	&	4	&	55879.04895	&	0.00001	&	6	\\
54314.16420	        &	0.00010	&	1	&	 55409.62744	&	0.00002	&	4	&	55883.02224	&	0.00006	&	6	\\
54316.15090	        &	0.00010	&	1	&	 55427.72877	&	0.00001	&	4	&	56042.84422	&	0.00003	&	4	\\
54318.02740	        &	0.00010	&	1	&	 55437.11053	&	0.00013	&	6	&	56047.36959	&	0.00001	&	6	\\
54319.02060	        &	0.00010	&	1	&	 55438.10395	&	0.00007	&	6	&	56069.33405	&	0.00004	&	6	\\
54319.13120	        &	0.00010	&	1	&	 55443.40192	&	0.00001	&	5	&	56069.38926	&	0.00008	&	6	\\
54323.10450	        &	0.00010	&	1	&	 55444.17447	&	0.00002	&	6	&	56078.82624	&	0.00000	&	4	\\
54324.09790	        &	0.00010	&	1	&	 55445.05746	&	0.00002	&	6	&	56107.41317	&	0.00001	&	7	\\
54366.04010	        &	0.00010	&	1	&	 55449.25176	&	0.00005	&	3	&	56129.37758	&	0.00001	&	5	\\
54386.56930$^{*}$	&	0.00057	&		&    55449.36215	&	0.00002	&	3	&	56129.43282	&	0.00002	&	5	\\
54686.67690$^{*}$	&	0.00048	&		&	 55450.02436	&	0.00005	&	6	&	56130.37098	&	0.00001	&	5	\\
54818.90501	        &	0.00005	&	6	&	 55452.23189	&	0.00002	&	3	&	56132.35772	&	0.00002	&	5	\\
54933.36268	        &	0.00005	&	6	&	 55453.11482	&	0.00002	&	6	&	56132.41291	&	0.00003	&	5	\\
54936.34294	        &	0.00008	&	6	&	 55479.27356	&	0.00001	&	5	&	56133.35111	&	0.00001	&	7	\\
54961.39796	        &	0.00003	&	6	&	 55499.03051	&	0.00002	&	6	&	56152.66655	&	0.00001	&	4	\\
54963.38463	        &	0.00007	&	6	&	 55504.49405	&	0.00001	&	4	&	56164.14541	&	0.00003	&	6	\\
54968.24111	        &	0.00003	&	6	&	 55682.91400	&	0.00001	&	2	&	56179.37713	&	0.00001	&	7	\\
54969.23445	        &	0.00008	&	6	&	 55686.88745	&	0.00001	&	2	&	56219.11183	&	0.00002	&	6	\\
54969.28957	        &	0.00013	&	6	&	 55686.94270	&	0.00003	&	2	&	56222.31268	&	0.00002	&	5	\\
54972.38013	        &	0.00002	&	6	&	 55688.32227	&	0.00004	&	6	&	56234.01231	&	0.00003	&	6	\\
54986.23200	        &	0.00004	&	6	&	 55692.29572	&	0.00003	&	6	&	56246.26385	&	0.00002	&	5	\\
54994.28935	        &	0.00006	&	6	&	 55700.35324	&	0.00012	&	6	&	56248.96791	&	0.00008	&	6	\\
55000.24974	        &	0.00005	&	6	&	 55717.57146	&	0.00001	&	3	&	56249.02323	&	0.00004	&	6	\\
55021.44152	        &	0.00010	&	7	&	 55721.32418	&	0.00011	&	6	&	56373.41486	&	0.00013	&	6	\\
55031.26470	        &	0.00010	&	6	&	 55737.32837	&	0.00001	&	6	&	56404.54032	&	0.00002	&	5	\\
55034.46565	        &	0.00010	&	7	&	 55749.69036	&	0.00001	&	4	&	55802.33887	&	0.00002	&	2	\\
55034.57601	        &	0.00010	&	7	&	 55750.68372	&	0.00000	&	4	&	55802.44920	&	0.00004	&	2	\\
55037.33534	        &	0.00002	&	2	&	 55760.83818	&	0.00003	&	2	&	55805.31896	&	0.00002	&	2	\\
55050.91137	        &	0.00001	&	7	&	 55760.89340	&	0.00001	&	2	&	55805.42932	&	0.00002	&	2	\\
55050.91137	        &	0.00002	&	2	&	 55760.94855	&	0.00002	&	2	&	55808.29907	&	0.00002	&	2	\\
55053.45005	        &	0.00010	&	7	&	 55762.93532	&	0.00002	&	2	&	55826.29008	&	0.00002	&	3	\\
55065.31521	        &	0.00008	&	5	&	 55765.47387	&	0.00004	&	2	&	55829.27017	&	0.00002	&	3	\\
55069.34387	        &	0.00004	&	5	&	 55768.89549	&	0.00001	&	2	&	55837.32752	&	0.00002	&	5	\\
55071.44100	        &	0.00001	&	5	&	 55778.49806	&	0.00003	&	4	&	55861.27873	&	0.00002	&	3	\\
55080.38128	        &	0.00007	&	2	&	 55778.82915	&	0.00001	&	2	&	55863.26542	&	0.00002	&	3	\\
55082.36800	        &	0.00002	&	2	&	 55783.35435	&	0.00004	&	6	&	55863.59656	&	0.00003	&	2	\\
55089.43200	        &	0.00002	&	5	&	 55784.12698	&	0.00008	&	6	&	55864.25879	&	0.00001	&	3	\\
55102.34584	        &	0.00002	&	5	&	 55784.34781	&	0.00006	&	2	&	55872.09538	&	0.00002	&	6	\\
55104.11174	        &	0.00004	&	6	&	 55784.34781	&	0.00001	&	7	&	55879.04895	&	0.00001	&	6	\\
55108.30603	        &	0.00004	&	5	&	 55793.84006	&	0.00001	&	2	&	55883.02224	&	0.00006	&	6	\\
55118.01888	        &	0.00001	&	6	&	 55796.37871	&	0.00001	&	7	&	56042.84422	&	0.00003	&	4	\\
55118.07402	        &	0.00003	&	6	&	 55796.48924	&	0.00001	&	7	&	56047.36959	&	0.00001	&	6	\\
55118.12921	        &	0.00002	&	6	&	 55798.14458	&	0.00002	&	6	&	56069.33405	&	0.00004	&	6	\\
55146.05390	        &	0.00001	&	6	&	 55800.35217	&	0.00001	&	2	&	56069.38926	&	0.00008	&	6	\\
55152.23486	        &	0.00002	&	5	&	 55800.46251	&	0.00001	&	2	&	56078.82624	&	0.00000	&	4	\\
55153.00750	        &	0.00004	&	6	&	 55802.33887	&	0.00002	&	2	&	56107.41317	&	0.00001	&	7	\\ 

56129.37758    &   0.00001 &   5   &    57187.25783 &   0.00008 &   6   &     58011.14475 &   0.00002 &   6   \\
56129.43282    &   0.00002 &   5   &    57189.52055 &   0.00002 &   5   &     58019.53317 &   0.00001 &   6   \\
56130.37098    &   0.00001 &   5   &    57190.51385 &   0.00001 &   5   &     58019.58832 &   0.00004 &   6   \\
56132.35772    &   0.00002 &   5   &    57190.56907 &   0.00008 &   5   &     58019.64351 &   0.00001 &   6   \\
56132.41291    &   0.00003 &   5   &    57215.45834 &   0.00001 &   5   &     58044.25693 &   0.00001 &   7   \\
56133.35111    &   0.00001 &   7   &    57219.32142 &   0.00001 &   5   &     58287.79745 &   0.00005 &   6   \\
56152.66655    &   0.00001 &   4   &    57219.43184 &   0.00001 &   5   &     58287.85272 &   0.00002 &   6   \\
56164.14541    &   0.00003 &   6   &    57219.54221 &   0.00001 &   7   &     58288.73570 &   0.00002 &   6   \\
56179.37713    &   0.00001 &   7   &    57224.50902 &   0.00001 &   5   &     58308.38229 &   0.00009 &   8   \\
56219.11183    &   0.00002 &   6   &    57230.35889 &   0.00001 &   7   &     58308.49276 &   0.00008 &   8   \\
56222.31268    &   0.00002 &   5   &    57237.31240 &   0.00002 &   5   &     58337.41069 &   0.00006 &   8   \\
56234.01231    &   0.00003 &   6   &    57240.34783 &   0.00010 &   7   &     58338.51443 &   0.00008 &   8   \\
56246.26385    &   0.00002 &   5   &    57244.48675 &   0.00001 &   5   &     58349.33117 &   0.00002 &   7   \\
56248.96791    &   0.00008 &   6   &    57247.41166 &   0.00006 &   5   &     58366.05278 &   0.00006 &   6   \\
56249.02323    &   0.00004 &   6   &    57255.30338 &   0.00002 &   5   &     58366.10801 &   0.00004 &   6   \\
56373.41486    &   0.00013 &   6   &    57262.36730 &   0.00002 &   5   &     58423.28188 &   0.00001 &   7   \\
56404.54032    &   0.00002 &   5   &    57263.36068 &   0.00003 &   5   &     58437.18898 &   0.00001 &   7   \\
56431.47168    &   0.00002 &   5   &    57265.23701 &   0.00002 &   5   &     58451.97914 &   0.00002 &   6   \\
56440.30158    &   0.00007 &   6   &    57271.30758 &   0.00003 &   5   &     58460.97453 &   0.00008 &   6   \\
56453.43608    &   0.00004 &   5   &    57278.04048 &   0.00003 &   6   &     58461.02974 &   0.00005 &   6   \\
56453.49127    &   0.00010 &   5   &    57297.35584 &   0.00001 &   7   &     58462.02315 &   0.00002 &   6   \\
56456.30576    &   0.00003 &   6   &    57299.34258 &   0.00003 &   5   &     58702.41808 &   0.00004 &   8   \\
56482.57488    &   0.00001 &   7   &    57314.24309 &   0.00001 &   5   &     58702.52845 &   0.00005 &   8   \\
56494.44007    &   0.00010 &   7   &    57365.12545 &   0.00002 &   5   &     58705.39813 &   0.00005 &   8   \\
56494.49526    &   0.00001 &   7   &    57520.47690 &   0.00006 &   5   &     58705.50847 &   0.00003 &   8   \\
56504.42899    &   0.00002 &   5   &    57520.53205 &   0.00002 &   5   &     58742.31827 &   0.00013 &   9   \\
56504.48403    &   0.00004 &   5   &    57542.49648 &   0.00001 &   5   &     58744.36031 &   0.00002 &   9   \\
56511.38250    &   0.00001 &   5   &    57543.48985 &   0.00001 &   5   &     58759.37118 &   0.00003 &   9   \\
56514.14177    &   0.00001 &   6   &    57546.46995 &   0.00004 &   5   &     58779.68005 &   0.00003 &   9   \\
56521.42654    &   0.00001 &   5   &    57580.57554 &   0.00001 &   5   &     58781.33557 &   0.00001 &   7   \\
56529.37349    &   0.00001 &   7   &    57600.33248 &   0.00001 &   5   &     58975.26294 &   0.00004 &   9   \\
56533.12611    &   0.00003 &   6   &    57603.42295 &   0.00001 &   5   &     59006.16767 &   0.00004 &   9   \\
56557.07731    &   0.00003 &   6   &    57604.41631 &   0.00001 &   5   &     59025.48311 &   0.00003 &   9   \\
56570.32230    &   0.00010 &   5   &    57605.29933 &   0.00001 &   5   &     59061.46504 &   0.00003 &   9   \\
56571.31561    &   0.00003 &   5   &    57616.33667 &   0.00001 &   7   &     59062.45844 &   0.00001 &   7   \\
56571.37078    &   0.00011 &   5   &    57629.25050 &   0.00001 &   5   &     59097.33659 &   0.00001 &   7   \\
56571.42597    &   0.00005 &   5   &    57629.47126 &   0.00002 &   5   &     59105.39390 &   0.00003 &   9   \\
56573.30232    &   0.00001 &   5   &    57630.46459 &   0.00001 &   5   &     59108.37400 &   0.00001 &   7   \\
56589.30657    &   0.00002 &   5   &    57645.36509 &   0.00001 &   7   &     59110.36073 &   0.00001 &   7   \\
56613.25781    &   0.00001 &   7   &    57658.27886 &   0.00001 &   5   &     59134.31190 &   0.00002 &   9   \\
56758.39958    &   0.00001 &   6   &    57659.27225 &   0.00001 &   5   &     59143.25217 &   0.00010 &   7   \\
56823.18914    &   0.00003 &   6   &    57659.27226 &   0.00001 &   5   &     59147.22563 &   0.00001 &   7   \\
56824.45851    &   0.00010 &   7   &    57659.38259 &   0.00001 &   5   &     59149.32277 &   0.00001 &   9   \\
56824.51371    &   0.00001 &   7   &    57659.38261 &   0.00002 &   5   &     59168.19669 &   0.00001 &   7   \\
56863.36537    &   0.00003 &   5   &    57663.24569 &   0.00001 &   5   &     59173.27394 &   0.00007 &   9   \\
56864.35863    &   0.00003 &   5   &    57664.23905 &   0.00001 &   5   &     59177.57853 &   0.00002 &   9   \\
56876.49986    &   0.00001 &   7   &    57685.21014 &   0.00001 &   5   &     59365.54537 &   0.00003 &   9   \\
56908.17717    &   0.00006 &   6   &    57688.19020 &   0.00002 &   5   &     59409.47418 &   0.00002 &   9   \\
56960.05296    &   0.00001 &   6   &    57688.19021 &   0.00001 &   5   &     59409.47419 &   0.00003 &   9   \\
56962.03971    &   0.00009 &   6   &    57689.29399 &   0.00003 &   5   &     59439.49598 &   0.00002 &   9   \\
56966.01314    &   0.00003 &   6   &    57696.24750 &   0.00001 &   5   &     59464.44049 &   0.00003 &   9   \\
56972.96670    &   0.00001 &   6   &    57696.24750 &   0.00001 &   7   &     59498.32529 &   0.00003 &   9   \\
56973.96007    &   0.00001 &   6   &    57919.47908 &   0.00007 &   10  &     59541.26079 &   0.00002 &   9   \\
56984.22484    &   0.00002 &   5   &    57919.53427 &   0.00007 &   10  &     59771.05954 &   0.00011 &   10  \\
57158.50540    &   0.00001 &   5   &    57933.82780 &   0.00004 &   6   &     59789.27121 &   0.00010 &   10  \\
57159.49866    &   0.00003 &   5   &    57933.88289 &   0.00001 &   6   &     59796.11444 &   0.00008 &   10  \\
57166.56271    &   0.00001 &   5   &    57991.60849 &   0.00001 &   6   &     60213.43895 &   0.00002 &   10  \\
57167.44565    &   0.00002 &   5   &    57991.66373 &   0.00003 &   6   &     60458.46954 &   0.00006 &   10  \\
57180.46979    &   0.00003 &   5   &    58008.16467 &   0.00002 &   6   &                 &           &       \\
\end{longtable}
\tablefoot{Minima with ($^{*}$) are not included in dataset B.}
\tablebib{
(1) \cite{2007IBVS.5800....1W}; (2) \cite{Beuermann2012B} ; (3) \cite{2012MNRAS.421.3238K}; (4) \cite{Almeida2013};(5) \cite{Nasiroglu2017}; (6) \cite{Zhu2019}; (7) \cite{Wolf2021}; (8) \cite{NehirBulut2022}; (9) \cite{Pulley2022}; (10) This work (Table~\ref{tab:TOMs}).}
\twocolumn

\end{appendix}

\end{document}